 \documentclass[aps,prd,nofootinbib,floatfix,showkeys,preprintnumbers]{revtex4}

\usepackage{epsfig}
\usepackage{latexsym}
\usepackage{graphicx}

\usepackage{amsmath}
\usepackage{amsfonts}   
\usepackage{amssymb}    

\newcommand{\newc}{\newcommand}

\newc{\hone}{h_1} \newc{\mhone}{m_{\hone}}
\newc{\htwo}{h_2} \newc{\mhtwo}{m_{\htwo}}
\newc{\hthree}{h_3} \newc{\mhthree}{m_{\hthree}}
\newc{\aone}{a_1} \newc{\maone}{m_{\aone}}
\newc{\atwo}{a_2} \newc{\matwo}{m_{\atwo}}

\newc{\msew}{m_{S}}
\newc{\mssq}{m_{S}^{2}}
\newc{\alambda}{A_{\lambda}} \newc{\aappa}{A_{\kappa}}

\def\e3{$\epsilon_3$}

\def\ch2{$\chi^2$}

\def\co#1{{\ifmmode{\cal O}_{#1}\else${\cal O}_{#1}$\fi}}

\newdimen\unit
\def\point#1 #2 #3{\vbox to0pt{\kern-#2\unit
  \hbox{\kern#1\unit#3}\vss}
 \nointerlineskip}

\newcommand{\be}{\begin{equation}}
\newcommand{\ee}{\end{equation}}
\newcommand{\bea}{\begin{eqnarray}}
\newcommand{\eea}{\end{eqnarray}}

\newcommand{\gev}{\mbox{ GeV}}
\newcommand{\tev}{\mbox{ TeV}}

\newcommand{\cl}{\text{CL}}

\newcommand{\alphas}{\alpha_s(M_Z)^{\overline{MS}}}

\newcount\hour
\newcount\minute
\newtoks\amorpm
\hour=\time\divide\hour by60 \minute=\time{\multiply\hour by60
\global\advance\minute by- \hour}
\edef\standardtime{{\ifnum\hour<12 \global\amorpm={am}%
    \else\global\amorpm={pm}\advance\hour by-12 \fi
    \ifnum\hour=0 \hour=12 \fi
    \number\hour:\ifnum\minute<100\fi\number\minute\the\amorpm}}
\edef\militarytime{\number\hour:\ifnum\minute<100\fi\number\minute}
\def\bold#1{\setbox0=\hbox{$#1$}%
     \kern-.025em\copy0\kern-\wd0
     \kern.05em\copy0\kern-\wd0
     \kern-.025em\raise.0433em\box0 }

\newc\eg{{\rm {e.g.}}}  \newc\etal{{\rm {et al.}}} \newc\ie{{\rm i.e.}}
\newc\etc{{\rm {etc}}}
\newcommand\lsim{\mathrel{\rlap{\lower4pt\hbox{\hskip1pt$\sim$}}
    \raise1pt\hbox{$<$}}}
\newcommand\gsim{\mathrel{\rlap{\lower4pt\hbox{\hskip1pt$\sim$}}
    \raise1pt\hbox{$>$}}}

\newc{\mhalf}{m_{1/2}}      \newc{\mzero}{m_0}
\newc{\tanb}{\tan\beta}
\newc{\azero}{A_0}
\newc{\at}{A_t} \newc{\ab}{A_b} \newc{\atau}{A_\tau}
\newc{\bmu}{B\mu}           \newc{\sgn}{{\rm sgn}}
\newc{\mone}{M_1}           \newc{\mtwo}{M_2}

 \newc{\hu}{H_u}       \newc{\hd}{H_d}
 \newc{\mhu}{m_{H_u}}       \newc{\mhd}{m_{H_d}}
 \newc{\mhuew}{m^{\ast}_{H_u}}       \newc{\mhdew}{m^{\ast}_{H_d}}
 \newc{\mhuewsq}{m^{\ast\, 2}_{H_u}}       \newc{\mhdewsq}{m^{\ast\, 2}_{H_d}}
 \newc{\mhuast}{m^{\ast}_{H_u}}       \newc{\mhdast}{m^{\ast}_{H_d}}

\newc{\charone}{\chi_1^\pm} \newc{\mcharone}{m_{\chi_1^\pm}}

\newc{\hl}{h}               \newc{\mhl}{m_{\hl}}   \newc{\gammahl}{\Gamma_{\hl}}
\newc{\hh}{H}               \newc{\mhh}{m_{\hh}}   \newc{\gammahh}{\Gamma_{\hh}}
\newc{\ha}{A}               \newc{\mha}{m_{\ha}}   \newc{\gammaha}{\Gamma_{\ha}}
\newc{\hpm}{H^{\pm}}        \newc{\mhpm}{m_{\hpm}} \newc{\gammahpm}{\Gamma_{\hpm}}
\newc{\hp}{H^{+}} \newc{\mhp}{m_{\hp}} \newc{\hm}{H^{-}}
\newc{\mhm}{m_{\hm}}
\newc{\xt}{X_{t}}           \newc{\xb}{X_{b}}

\newc{\qzero}{Q_0}          \newc{\qstop}{Q_{\widetilde t}}
\newc{\amu}{a_{\mu}}        \newc{\amususy}{a_{\mu}^{\text{SUSY}}}
\newc{\amuexpt}{a_{\mu}^{\text{expt}}}        \newc{\amusm}{a_{\mu}^{\text{SM}}}
\newc{\deltaamususy}{\delta a_{\mu}^{\text{SUSY}}}
\newc\gmtwo{(g-2)_{\mu}} 
\newc\deltagmtwo{\delta (g-2)_{\mu}} 
\newc\deltaamu{\Delta a_{\mu}}
\newc{\msbar}{\overline{MS}} \newc{\drbar}{\overline{DR}}
\newc{\yt}{h_t} \newc{\yb}{h_b} \newc{\ytau}{h_{\tau}}

\newc{\mtop}{m_t}               \newc{\mtpole}{M_t}
\newc{\mtaupole}{m_{\tau}^{\text{pole}}}
\newc{\mtmtsmmsbar}{m_t(m_t)^{\msbar}_{{\text{SM}}}}
\newc{\mtmtsmdrbar}{m_t(m_t)^{\drbar}_{{\text{SM}}}}
\newc{\mtmtmssmdrbar}{m_t(m_t)^{\drbar}_{{\text{SUSY}}}}

\newc{\mbmbmsbar}{m_b(m_b)^{\msbar} }

\newc{\mbmbsmmsbar}{m_b(m_b)^{\msbar}_{{\text{SM}}}}
\newc{\mbmzsmmsbar}{m_b(\mz)^{\msbar}_{{\text{SM}}}}
\newc{\mbmzsmdrbar}{m_b(\mz)^{\drbar}_{{\text{SM}}}}
\newc{\mbmzmssmdrbar}{m_b(\mz)^{\drbar}_{{\text{SUSY}}}}

\newc{\mtaumzsmmsbar}{m_{\tau}(\mz)^{\msbar}_{{\text{SM}}}}
\newc{\mtaumzsmdrbar}{m_{\tau}(\mz)^{\drbar}_{{\text{SM}}}}
\newc{\mtaumzmssmdrbar}{m_{\tau}(\mz)^{\drbar}_{{\text{SUSY}}}}

\newc{\mgut}{M_{\rm GUT}}
\newc{\mplanck}{M_{\rm P}}      \newc{\mpl}{M_{\text{Pl}}}
\newc{\msusy}{M_{\rm SUSY}}      \newc{\ms}{M_{\text{S}}}
\newc{\jxf}{J({\xf})}
\newc{\jxfexact}{J_{\rm exact}({\xf})}  \newc{\jxfexp}{J_{\rm exp}({\xf})}
\newc{\VEV}[1]{\langle #1 \rangle}
\newc{\xf}{x_f}
\newc\vrel{v_{\rm rel}}
\newcommand\mchi{m_{\chi}}              
\newc\sell{{\widetilde e}_L}      \newc\msell{m_{\sell}}
\newc\selr{{\widetilde e}_R}      \newc\mselr{m_{\selr}}
\newc\snue{{\widetilde \nu}_e}      \newc\msnue{m_{\snue}}
\newc\snutau{{\widetilde \nu}_\tau}      \newc\msnutau{m_{\snutau}}
\newc\supl{{\widetilde u}_L}      \newc\msupl{m_{\supl}}
\newc\supr{{\widetilde u}_R}      \newc\msupr{m_{\supr}}
\newc\sdl{{\widetilde d}_L}      \newc\msdl{m_{\sdl}}
\newc\sdr{{\widetilde d}_R}      \newc\msdr{m_{\sdr}}

\newcommand\stauone{{\widetilde \tau}_1}

\newc\sfermion{\tilde f}  \newc\msfermion{m_{\sfermion}}
\newc\cmeter{{\rm cm}} \newc\meter{{\rm m}} \newc\kmeter{{\rm km}}
\newc\second{{\rm sec}}

\newc\sr{{\rm sr}}

\newc{\gstar}{g_\ast}           \newc{\gsstar}{g_{s\ast}}
\newc{\geff}{g_{\rm eff}}
\newcommand\mz{m_{Z}}

\newc{\sthw}{\sin\theta_W}              \newc{\cthw}{\cos\theta_W}
\newc{\bino}{\widetilde B}              \newc{\wino}{\widetilde W_30}
\newc{\higgsinob}{{\widetilde H}^0_b}   \newc{\higgsinot}{{\widetilde H}^0_t}
\newc{\abund}{\Omega h^2}
\newc{\abundchi}{\Omega_\chi h^2}
\newc{\abundcdm}{\Omega_{\text{CDM}} h^2}
\newc{\omegam}{\Omega_{M}}       \newc{\abundm}{\Omega_{M} h^2}
\newc{\omegab}{\Omega_{b}}       \newc{\abundb}{\Omega_{b} h^2}
\newc{\omegacdm}{\Omega_{CDM}}
\newc{\omegatot}{\Omega_{TOT}}
\newc{\rhocrit}{\rho_{crit}}
\newc{\rhochi}{\rho_{\chi}}
\newcommand\pb{\,\mbox{pb}} 

\newc\pc{\,\mbox{pc}} \newc\kpc{\,\mbox{kpc}}
\newc\mpc{\,\mbox{Mpc}} \newc\gpc{\,\mbox{Gpc}}

\newc\BR{BR}
\newc\bsgamma{b\rightarrow s \gamma }
\newc\bxsgamma{\overline{B}\rightarrow X_{s}\gamma}
\newc\brbsgamma{\BR(\overline{B}\rightarrow X_s\gamma)}

\newcommand\bsmumu{\overline{B}_s\to\mu^+\mu^-}
\newcommand\brbsmumu{\BR(\overline{B}_s\to\mu^+\mu^-)}

\newcommand\brbtaunu{\BR(\overline{B}_u\to \tau \nu)}

\newc{\beq}{\begin{equation}}
\newc{\eeq}{\end{equation}}

\renewcommand\){\right)}

\newc\stoponetwo{{\widetilde t}_{1,2}}
\newc\sbotonetwo{{\widetilde b}_{1,2}}
\newc\stauonetwo{{\widetilde \tau}_{1,2}}

\newc{\sigsip}{\sigma^{SI}_{p}} \newc{\sigsin}{\sigma^{SI}_{n}}
\newc{\sigsiN}{\sigma^{SI}_{N}}
\newc{\sigsdp}{\sigma^{SD}_{p}} \newc{\sigsdn}{\sigma^{SD}_{n}}
\newc{\sigsiA}{\sigma^{SI}_{A}}

\newc{\pbar}{\bar{p}}

\newc{\egamma}{E_{\gamma}}
\newc{\flux}[1]{\Phi_{#1}}
\newc{\dfluxde}[1]{\frac{d\Phi_{#1}}{d E_{#1}}}

\newc{\fluxg}{\Phi_{\gamma}}
\newc{\dfluxgde}{\frac{d\Phi_{\gamma}}{d\egamma}}
\newc{\dfluxgdetext}{ d\Phi_{\gamma} / d\egamma}

\newc{\eplus}{e^+}
\newc{\epos}{E_{\eplus}}
\newc{\eps}{\varepsilon}

\newc{\npos}{n_{\eplus}} \newc{\Npos}{N_{\eplus}}
\newc{\dnposde}{\frac{d n_{\eplus}}{d\epos}}
\newc{\dnposdeps}{\frac{d n_{\eplus}}{d\eps\phantom{_{\eplus}}}}
\newc{\dnposdepstext}{ d n_{\eplus} / d\eps}
\newc{\fluxpos}{\Phi_{\eplus}}  \newc{\fluxelec}{\Phi_{e^{-}}}
\newc{\dfluxposde}{\frac{d\Phi_{\eplus}}{d\epos}}
\newc{\dfluxposdetext}{ d\Phi_{\eplus} / d\epos}

\newc{\nfwc}{{\text{NFW+ac}}} \newc{\moorec}{{\text{Moore+ac}}}

\newc{\chisq}{\chi^2}  \newc{\chisqred}{\chi^2_{\text{red}}}

\newc{\abb}{BR\(a_1\to b\anti{b}\)}
\newc{\hbb}{BR\(h_1\to b\anti{b}\)}
\newc{\haa}{BR\(h_1\to a_1 a_1\)}
\newc{\cbeffii}{ C^{2b}_\text{eff} }
\newc{\cbeffiv}{ C^{4b}_\text{eff} }
\newc{\cvi}{ C_{V}(1)}
\newc{\cvii}{C_{V}(2)}
\newc{\cvisq}{ |C_{V}(1)|^2}
\newc{\cviisq}{|C_{V}(2)|^2}
\newc{\akappa}{A_{\kappa} }
\def\wtil{\widetilde}
\def\hi{h_1}
\def\ai{a_1}
\def\hii{h_2}

\def\hiii{h_3}
\def\mhi{m_{\hi}}
\def\mai{m_{\ai}}
\def\caibb{C_{\ai b\anti b}}
\def\mhii{m_{\hii}}
\def\anti{\overline}
\def\br{BR}
\def\bit{\begin{itemize}}
\def\eit{\end{itemize}}
\def\epem{e^+e^-}
\def\mupmum{\mu^+\mu^-}
\def\cnone{\chi}
\def\mcnone{m_{\cnone}}
\def\sigpsi{\sigma_p^{SI}}
\def\cta{\cos\theta_A}
\def\sta{\sin\theta_A}
\def\mhii{m_{\hii}}
\def\hp{h^+}
\def\mhp{m_{\hp}}

\def\fbi{~\mbox{fb}^{-1}}

\def\alam{A_\lambda}
\def\akap{A_\kappa}

\newc\xilim{\xi_{\rm lim}} 
\newc\tlim{t_{\rm lim}} 
\newc\zetalim{\zeta_{\rm lim}} 

\newc\zetah{\zeta_h}
\newc{\relprobone}[1]{p({#1} \vert d)}
\newc{\relprobtwo}[2]{p({#1},{#2} \vert d)}

\long\def\begincomment#1\endcomment{%
        \begingroup\sf\baselineskip12pt#1\endgroup}

\newcommand{\squishlist}{
   \begin{list}{$\bullet$}
    { \setlength{\itemsep}{0pt}      \setlength{\parsep}{3pt}
      \setlength{\topsep}{3pt}       \setlength{\partopsep}{0pt}
      \setlength{\leftmargin}{1.em} \setlength{\labelwidth}{1em}
      \setlength{\labelsep}{0.5em} } }
\newcommand{\squishend}{
    \end{list}  }
        
\def    \be            {\begin{equation}}
\def    \ee            {\end{equation}}
\def    \bea           {\begin{eqnarray}}
\def    \eea           {\end{eqnarray}}

\def \ie{{\it i.e.}}
\def \eg{{\it e.g.}}
\def \etal{{\it et al.}}
\def\omhsq{\Omega h^2}

\newcommand{\data}{d}
\newcommand{\nuis}{\psi}
\newcommand{\params}{\theta}
\newcommand{\basis}{m}
\newcommand{\derived}{\xi}

\def\ds@jhep{\def\@journal{jhep}}

\def\ds@jhep{\def\@journal{jcap}}

\def\ds@plb{\def\@journal{plb}}

\def\ds@prep{\def\@journal{prep}}

\def\ds@cpc{\def\@journal{cpc}}

\def\ds@ijmpa{\def\@journal{ijmpa}}

\def\ds@hepph{\def\@journal{hepph}}

\def\ds@app{\def\@journal{app}}

\begin{document}

\title{Next-to-Minimal Supersymmetric Model Higgs Scenarios for
  Partially Universal GUT Scale Boundary Conditions}

\author{John F. Gunion$^1$, Daniel E. L\'opez-Fogliani$^{2,3}$, Leszek Roszkowski$^{3,4}$
  Roberto Ruiz de Austri$^5$, and Tom A. Varley$^{3}$}
\affiliation{$^1$ Department of Physics, University of California at
  Davis, 
Davis, CA, USA\\
$^2$Laboratoire de Physique Th\'eorique, Universit\'e Paris-Sud, F-91405 Orsay, France.\\
$^3$Department of Physics and Astronomy, The University of Sheffield,
Sheffield S3 7RH, England\\
$^4$The Andrzej Soltan Institute for Nuclear Studies, Warsaw, Poland\\
$^5$Instituto de F\'isica Corpuscular, IFIC-UV/CSIC, Valencia, Spain\\
}

\begin{abstract}
  We examine the extent to which it is possible to realize the NMSSM
  ``ideal Higgs'' models espoused in several papers by Gunion \etal\
  in the context of partially universal GUT scale boundary conditions.
  To this end we use the powerful methodology of nested sampling. We
  pay particular attention to whether ideal-Higgs-like points not only
  pass LEP constraints but are also acceptable in terms of the
  numerous constraints now available, including those from the
  Tevatron and $B$-factory data, $(g-2)_\mu$ and the relic density
  $\abund$. In general for this particular methodology and range of
  parameters chosen, very few points corresponding to said previous
  studies were found, and those that were found were at best $2\sigma$
  away from the preferred relic density value. Instead, there exist a
  class of points, which combine a mostly singlet-like Higgs with a
  mostly singlino-like neutralino coannihilating with the lightest
  stau, that are able to effectively pass all implemented constraints
  in the region $80<m_h<100$.  It seems that the spin-independent
  direct detection cross section acts as a key discriminator between
  ideal Higgs points and the hard to detect singlino-like points.
 \end{abstract}

\preprint{LPT-ORSAY 10/104, UCD-HEP-TH-2010-14}

\maketitle

\section{Introduction}\label{sec:gunion_intro}

As the LHC begins its first physics runs, more scrutiny can be placed
on possible regions where Beyond the Standard Model (BSM) physics can
exist. One intriguing possibility is the ``ideal Higgs'' scenario
pointed out by Gunion \etal~\cite{gunion1,gunion2,gunion3}.  In this
Next-to-Minimal Supersymmetric Model (NMSSM) scenario, the lightest
Higgs, $\hi$, has mass $\mhi\sim 100\gev$ and SM-like couplings to $WW,ZZ$, but
decays in such a way that LEP limits are obeyed. In general, this is
possible if the  $\hi$ decays primarily to a
pair of the lightest pseudoscalar Higgs bosons, $\hi\to\ai\ai$,
(resulting in a small branching ratio for $\hi\to b\anti b$). Because
$\mai<2 m_B$, the $\ai$ then subsequently decays to $2\tau$, $2g$, or
$2c$, the precise mixture depending on $\tanb$, with $2\tau$ being
dominant at high $\tan\beta$ while all states are important at low
$\tan\beta$. This allows these points to sidestep the LEP results
\cite{lhwg} on $\cbeffii$ and $\cbeffiv$, defined as:
\begin{eqnarray}
C^{2b}_\text{eff} &=& |C_V(1)|^2 \times BR(\hi \to b \anti{b}),\nonumber\\
\cbeffiv &\equiv& |C_V(1)|^2 \times \haa \times \left[\abb \right]^2,
\label{eqn:defs}
\end{eqnarray}
where $|C_V(1)|^2 \equiv g^2_{\hi ZZ}/g^2_{h_{SM}ZZ}$, where $h_{SM}$
denotes the Higgs boson of the Standard Model. Roughly speaking, LEP
constraints are evaded if $\br(\hi\to\ai\ai)\gsim 0.7$.

There are three main motivations for such a scenario. 
\bit
\item Precision electroweak (PEW) data is most consistent with a
  SM-like Higgs having mass below $100\gev$.
\item There is an excess in the combined LEP data for $\epem\to Z
  b\anti b$ for $M_{b\anti b}$ in the region $80-100\gev$ that is
  consistent with $\br(\hi\to b\anti b)\lsim 0.3$, as obtained for
  $\br(\hi\to \ai\ai)\gsim 0.7$.
\item A mass $\mhi\lsim 100\gev$ is consistent with a light
  superparticle spectrum for which GUT scale parameters need not be
  fine tuned in order to obtain the correct value of $m_Z$ at low
  scales.  
\eit 
One of the primary goals of this paper is to elucidate the extent to
which the emergence of ideal-Higgs-like scenarios depends on the
extent to which the pattern of soft supersymmetry breaking is highly
constrained/universal.

Indeed, a key uncertainty in both the MSSM and NMSSM is the pattern of soft
supersymmetry breaking, as described by the scalar masses $m_0$,
gaugino masses $m_{1/2}$ and trilinear couplings $A_0$. These
presumably originate from physics at some high-energy scale, e.g.,
from some supergravity or superstring theory, and then evolve down to
lower energy scale according to well-known renormalization-group
equations. What is uncertain, however, is the extent to which
universality applies to the scalar masses $m_0$ for different squark,
slepton and Higgs fields, the gaugino masses $m_{1/2}$ for the
$SU(3)$, $SU(2)$ and $U(1)$ gauginos, and the trilinear couplings
$A_0$ corresponding to different Yukawa couplings. Certain types of
universality are much better motivated than others.

The suppression of flavour-changing neutral interactions suggests that
the $m_0$ may be universal for different matter fields with the same
quantum numbers, e.g., the different squark and slepton generations.
However, there is no very good reason to postulate universality
between, say, the spartners of left- and right-handed quarks, or
between squarks and sleptons. In Grand Unified Theories (GUTs), there
must be universality between fields in the same GUT multiplet, e.g.,
$u_L, d_L, u_R$ and $e_R$ in a ${\mathbf 10}$ of $SU(5)$, and this
would extend to all matter fields in a ${\mathbf 16}$ of
$SO(10)$. However, there is less reason to postulate universality
between these and the Higgs fields. Nevertheless, this extension of
universality to the Higgs masses (UHM) is often assumed, resulting in
what is commonly termed the constrained MSSM (CMSSM) or constrained
NMSSM (CNMSSM).  Alternatively, there may be non-universal Higgs
masses (NUHM) in the more general MSSM and NMSSM.  As regards the $A$
parameters, in the MSSM or CMSSM context it is primarily the 3rd
generation $A_t$ that matters and so universality or lack thereof does
not have significant phenomenological impact. However, in the NMSSM
there are two new $A$ parameters: one, $\alam$, associated with the
singlet-Higgs-Higgs interaction at the superfield level; and a second,
$\akap$, associated with the singlet-cubed superpotential
terms. Currently, there is no reason for these to have the same value
as $A_t$. Indeed, the limit in which $\akap$ and $\alam$ are zero at
the GUT scale is one in which the NMSSM has an extra $U(1)_R$
symmetry, independent of the GUT-scale value of
$A_t$~\cite{Dobrescu:2000yn}.  Although it can certainly be the case
that the universalities assumed in the NUHM
relaxation of the CMSSM and the NUHM plus $\akap,\alam$ relaxation of
the CNMSSM are still more restrictive than suggested by
many schemes of supersymmetry breaking, comparing results
incorporating such relaxations to those obtained using the most
restrictive  CMSSM or CNMSSM universalities provides a
simplified framework for understanding what new phenomena arise when
poorly-motivated restrictive boundary conditions are relaxed. In
particular, we will see that in the NMSSM context relaxation of the
CNMSSM boundary conditions to include NUHM and non-universality for
$\alam$ and $\akap$ relative to $A_t$ already gives rise to dramatic
new physics possibilities, including the possibilities of
ideal-Higgs-like scenarios and singlino-singlet dark matter scenarios.
If nothing else, this shows that overly restrictive universality
assumptions should be avoided in order that dramatic new physics
scenarios are not ``needlessly'' excluded.

As referred to above, previous studies in the context of the Constrained NMSSM
(CNMSSM)~~\cite{CNMSSMpaper,Balazs:2008ph}, in which all $A$ parameters and all
soft-SUSY-breaking masses-squared are unified, did not find
ideal-Higgs-like points.   In this paper, we allow the Higgs soft
masses-squared to be independent of the other (unified) soft
masses-squared and we allow the soft-SUSY-breaking $A_\kappa$
parameter associated with the singlet field to range freely,
independent of the other $A$ parameters (which are taken to be
universal).  Using the advanced scanning techniques from SuperBayeS coupled to MultiNest \cite{superbayes} we can efficiently scan for interesting regions of
parameter space and see how often points of an ideal-Higgs nature are
found and look at the many phenomenological aspects that these points
entail. In addition, the power of this approach to scan over
\emph{all} interesting parameters instead of fixing crucial ones to
some canonical value can provide us with some insights into the full
structure of the parameter space.  In particular, we will not be using
fine-tuning as the defining criteria when searching for
ideal-Higgs-like points.  This said, our choices of
parameters are influenced by \cite{gunion1,gunion2,gunion3} and the
findings therein.

\section{The Implementation of NMSSM model.}\label{sec:gunion_overNMSSM}

The NMSSM superpotential contains a new superfield $S$ which is a
singlet under the SM gauge group $SU(3)_c \times SU(2)_L \times
U(1)_Y$. (For simplicity, we use the same notation for superfields and
their respective spin-0 component fields.) 
\begin{equation}\label{2:Wnmssm}
W=
\epsilon_{ij} \left(
Y_u \, H_u^j\, Q^i \, u +
Y_d \, H_d^i\, Q^j \, d +
Y_e \, H_d^i\, L^j \, e \right)
- \epsilon_{ij} \lambda \,S \,H_d^i H_u^j +\frac{1}{3} \kappa S^3\,,
\end{equation}
where $H_d^T=(H_d^0, H_d^-)$, $H_u^T=(H_u^+, H_u^0)$, $i,j$ are
$SU(2)$ indices with $\epsilon_{12}=1$, while $\lambda$ and $\kappa$
are dimensionless couplings. With the
addition of a scalar singlet superfield field, there will be five
neutralinos and the Higgs content of the NMSSM is
extended to include three scalar Higgses, $h_{1}$, $h_{2}$ and
$h_{3}$, and two pseudoscalars, $a_{1}$ and $a_{2}$. The lightest Higgs $h_{1}$ plays an important role in the scenarios we consider here, and in particular the composition will be commented on. The state composition can be written as,
\begin{equation}
 h_{1}= S_u H_u + S_d H_d + S_s S.
\end{equation}

The superpotential in Eq.~(\ref{2:Wnmssm}) is scale invariant, and the
EW scale will only appear through the soft-SUSY-breaking terms in
$\mathcal{L}_{\text{soft}}$, which in our conventions is given by
\begin{align}\label{2:Vsoft}
-\mathcal{L}_{\text{soft}}=&\,
 {m^2_{\tilde{Q}}} \, \tilde{Q}^* \, \tilde{Q}
+{m^2_{\tilde{U}}} \, \tilde{u}^* \, \tilde{u}
+{m^2_{\tilde{D}}} \, \tilde{d}^* \, \tilde{d}
+{m^2_{\tilde{L}}} \, \tilde{L}^* \, \tilde{L}
+{m^2_{\tilde{E}}} \, \tilde{e}^* \, \tilde{e}
 \nonumber \\
&
+m_{H_d}^2 \,H_d^*\,H_d + m_{H_u}^2 \,H_u^* H_u +
m_{S}^2 \,S^* S \nonumber \\
&
+\epsilon_{ij}\, \left(
A_u \, Y_u \, H_u^j \, \tilde{Q}^i \, \tilde{u} +
A_d \, Y_d \, H_d^i \, \tilde{Q}^j \, \tilde{d} +
A_e \, Y_e \, H_d^i \, \tilde{L}^j \, \tilde{e} + \text{H.c.}
\right) \nonumber \\
&
+ \left( -\epsilon_{ij} \lambda\, A_\lambda S H_d^i H_u^j +
\frac{1}{3} \kappa \,A_\kappa\,S^3 + \text{H.c.} \right)\nonumber \\
&
- \frac{1}{2}\, \left(M_3\, \lambda_3\, \lambda_3+M_2\, \lambda_2\, \lambda_2
+M_1\, \lambda_1 \, \lambda_1 + \text{H.c.} \right) \,.
\end{align}
When the scalar component of $S$ acquires a VEV, $s=\langle S
\rangle$, an effective interaction $-\eps_{ij}\mu H^i_d H^j_u$ is generated, with
$\mu \equiv \lambda s$.

In addition to terms from $\mathcal{L}_{\text{soft}}$, the
tree-level scalar Higgs potential receives the usual $D$ and $F$ term
contributions:
\begin{align}\label{2:Vfd}
V_D = & \, \,\frac{g_1^2+g_2^2}{8} \left( |H_d|^2 - |H_u|^2 \right)^2 +
\frac{g_2^2}{2} |H_d^\dagger H_u|^2 \, , \nonumber \\
V_F = & \, \,|\lambda|^2
\left( |H_d|^2 |S|^2 + |H_u|^2 |S|^2 + |\epsilon_{ij} H_d^i H_u^j|^2 \right)
+ |\kappa|^2 |S|^4
\nonumber \\
&
-\left( \epsilon_{ij} \lambda \kappa^* H_d^{i} H_u^{j}S^{*2} + \mathrm{H.c.}
\right) \,.
\end{align}

Using the minimization equations we can re-express the soft breaking
Higgs masses in terms of $\lambda$, $\kappa$, $A_\lambda$, $A_\kappa$,
$v_d=\langle H_d^0 \rangle$, $v_u=\langle H_u^0 \rangle$ (with
$\tanb=v_u/v_d$), and $s$: 
 \begin{align}
m_{H_d}^2 = & -\lambda^2 \left( s^2 + v^2\sin^2\beta \right)
- \frac{1}{2} M_Z^2 \cos 2\beta
+\lambda s \tan \beta \left(\kappa s +A_\lambda \right) \,,
\label{2minima:mh1} 
\\
m_{H_u}^2 = & -\lambda^2 \left( s^2 + v^2\cos^2\beta \right)
+\frac{1}{2} M_Z^2 \cos 2\beta
+\lambda s \cot \beta \left(\kappa s +A_\lambda \right) \,,
\label{2minima:mh2} 
\\
\label{2minima:ms} 
m_{S}^2 = & -\lambda^2 v^2 - 2\kappa^2 s^2 + \lambda \kappa v^2
\sin 2\beta + \frac{\lambda A_\lambda v^2}{2s} \sin 2\beta -
\kappa A_\kappa s\,.
\end{align}
We are now looking at a Lagrangian that is identical in structure to
our previous work~\cite{CNMSSMpaper} but with a few important differences
in terms of unification. In contrast to ~\cite{CNMSSMpaper}, where we
took CMSSM-like boundary conditions, in the present paper, we allow
the Higgs mass parameters $\mhu$ and $\mhd$ to freely vary, in a
similar manner to~\cite{nuhm1}. In addition $A_\kappa$ is no longer
taken equal to the universal value, $A_0$, of the other $A$
parameters. This freedom, specifically allowing $|A_{\kappa}|$ to be
small, will make it possible to obtain lighter pseudoscalar and scalar
Higgs masses than in ~\cite{CNMSSMpaper} that are nonetheless still
allowed by collider constraints. This additional freedom in Higgs mass
can be seen by looking at the tree level pseudoscalar mass matrix in
the basis $(A^0,S)$ \cite{Cerdeno:2004xw}: \newline
 {\footnotesize \begin{equation}
   \mathcal{M}^2_{A} =  \left(
      \begin{array}{cc}
       \frac{2 \lambda s}{\sin 2 \beta}\left( \kappa s +A_\lambda \right) & \lambda v \left(A_\lambda -2 \kappa s \right)\\
   \lambda v \left(A_\lambda -2 \kappa s \right) &\lambda \left(2 \kappa +\frac{A_\lambda}{2 s}
\right) v^2 \sin 2 \beta -3 \kappa A_\kappa s 
    \end{array} \right).
   \label{pseumatrix}
 \end{equation}}
\newline
After diagonalization of $M^2_A$, there will be two mass eigenstates.
The lightest state will be a mixture of the CP-odd doublet state $A_{MSSM}$
that is present in the Minimal Supersymmetric Model and the new
CP-odd component, $A_S$, of the complex scalar $S$ field. We write
\begin{equation}
\ai\equiv \cos \theta_A A_{MSSM}+ \sin \theta_A A_S,
\label{aidef}
\end{equation}
where the entries in $M_A^2$ are such that the 11 entry is the MSSM
diagonal entry. From Eqs.~(\ref{pseumatrix}) and (\ref{aidef}) it is
clear that having the freedom to vary $\akappa$ is crucial if we want
to control the mass of the singlet component independently of other
parameters.  Despite these changes, just as in~\cite{CNMSSMpaper} the
minimisation equations, (\ref{2minima:mh1})-(\ref{2minima:ms}) will be
used to fix $m_S$, $\kappa$, and $s$ giving us a model with input
parameters $\mhalf$, $\mzero$, $\mhu$, $\mhd$, $\azero$, $A_\kappa$,
$\tanb$ and $\lambda$, in addition to $\sgn(\mu)$. In particular note
that in our procedure, the value of $\kappa$ is an output that depends
on $\akappa$.  It is important to emphasize that in contrast to
previous studies looking at the ideal-Higgs region, where scanning was
done at the EW scale and run up, here all of our parameters (excluding
$\lambda$) are searched over at the GUT scale and then run down. One
implication is that in the studies we present here it is impossible to
obtain the values of the gaugino masses in \cite{gunion1} for a given
parameter point as we are constrained by unification
considerations. That said, our choice of parameters, although
constrained, leads to a scan that is practicable and is a useful
starting point to perturb from in order to better satisfy, for
example, fine-tuning or phenomenology.


%

%
 We also present the neutralino sector since the lightest neutralino
 will, by assumption, play the r\^{o}le of dark matter. The mass term in
 the Lagrangian is given by

 \begin{equation}
 \mathcal{L}_{\mathrm{mass}}^{\chi^0} =
 -\frac{1}{2} (\Psi^0)^T \mathcal{M}_{\chi^0} \Psi^0 + \mathrm{H.c.}\,,
\end{equation}

 with $\mathcal{M}_{\chi^0}$ given by a $5 \times 5$ matrix in
 the basis $(\bino, \widetilde{W}, \wtil H_u, \wtil H_d, \wtil S)$,
 {\footnotesize \begin{equation}
   \mathcal{M}_{\chi^0} = \left(
     \begin{array}{ccccc}
       M_1 & 0 & -M_Z \sin \theta_W \cos \beta &
       M_Z \sin \theta_W \sin \beta & 0 \\
       0 & M_2 & M_Z \cos \theta_W \cos \beta &
       -M_Z \cos \theta_W \sin \beta & 0 \\
       -M_Z \sin \theta_W \cos \beta &
       M_Z \cos \theta_W \cos \beta &
       0 & -\lambda s & -\lambda v_u \\
       M_Z \sin \theta_W \sin \beta &
       -M_Z \cos \theta_W \sin \beta &
       -\lambda s &0 & -\lambda v_d \\
       0 & 0 & -\lambda v_u & -\lambda v_d & 2 \kappa s
     \end{array} \right),
   \label{neumatrix}
 \end{equation}}
 where $\mone$ ($\mtwo$) denotes the soft mass of the bino (wino) and $\theta_W$ denotes the weak mixing angle. After diagonalization the lightest neutralino $\chi_1$ (which we will denote as $\chi$ from now on) can be written as:
\begin{equation}
\chi = N_B \bino + N_W \widetilde{W} + N_u \wtil H_u + N_d \wtil H_d
+ N_s \wtil S.  
\end{equation}

Finally, we note that couplings of the Higgs bosons and $\chi$ to SM
particle states depend upon the compositions of the former.  In
particular, the coupling of $\hi$ to $WW,ZZ$ is given by $\cvi \equiv
g_{\hi WW}/g_{h_{SM} WW}=S_u\sin \beta + S_d\cos\beta$ (with analogous
results for $\hii$ and $\hiii$) and the
coupling of $\ai$ to $b\anti b$ is given by $C_{\ai b\anti b}\equiv
g_{\ai b\anti b}/g_{h_{SM} b\anti b}=\cta\tanb$.

\section{Outline of the method}\label{sec:gunion_Bayes}

Following the discussion of section~\ref{sec:gunion_overNMSSM}, in
this constrained version of the NMSSM the free parameters are given by
\beq
\theta = (\mhalf, \mzero, m_{H_u}, m_{H_d}, \azero, A_\kappa, \tanb,\lambda)\,.
\label{eq:nmssm}
\eeq
Without loss of generality, one can choose
$\lambda>0$~\cite{Cerdeno:2004xw}.  However, we will also (as in our
previous work) fix $\sgn(\mu)= +1$ and then $\mu=\lambda s$ implies
$s>0$. The ``nuisance'' parameters are treated in the same manner as
in our previous work~\cite{CNMSSMpaper}, and are shown in Eq.~(\ref{eq:nuis2}):
\beq
\psi = (\mtpole,m_b(m_b)^{\msbar}, \alpha_{s}(\mz)^{\msbar}).
\label{eq:nuis2}
\eeq
Using notation consistent with previous analyses we define our eleven
dimensional {\em basis parameter} set as
\beq
\basis=(\theta,\psi)\,,
\label{eq:basis}
\eeq
all of which will be simultaneously scanned over. For
each choice of $\basis$ a number of collider and cosmological observables
are calculated. These derived variables are denoted by
$\derived=(\xi_1,\xi_2,\ldots)$, which are then compared with the
relevant measured data, $\data$.

In this study we will be using the ``nested sampling''
method~\cite{skilling-nsconv} as implemented in the
MultiNest~\cite{Feroz:2007kg} algorithm to efficiently explore the
likelihood space. Generally speaking, for this study we will be looking at
points of interest irrespective of statistical
considerations. MultiNest provides an extremely efficient sampler even
for likelihood functions defined over a parameter space of large
dimensionality with a very complex structure. (See, \eg,
Refs.~\cite{tfhrr1,nuhm1}.)

\begin{table}
\centering
\begin{tabular}{|c | | c |}
 \hline
{CNMSSM parameters $\params$}  &   {SM (nuisance) parameters $\nuis$}   \\ \hline
 $10^{-2} < \mhalf < 4000 \gev$ & $160 < \mtpole < 190 \gev$\\ 
 $ 10^{-2} < \mzero < 4000 \gev$ &  $4 < m_b(m_b)^{\overline{MS}} < 5 \gev$\\
$10^{-2}  < m_{H_u} < 4000 \gev$ & $0.10 < \alphas  < 0.13$ \\
$10^{-2} < m_{H_d} < 4000 \gev$ & \\
 $|\azero| <  100 \gev$ &  \\
$|A_\kappa| <  10 \gev$  & \\
$2 < \tanb < 20$   &    \\ 
 $10^{-3}< \lambda <0.7$ & \\ \hline
\end{tabular}
\caption[]{Initial ranges for our basis parameters
$\basis=(\params,\nuis)$.  }
\label{table:prior}
\end{table}

It should be emphasised here that although nested sampling was used to
obtain the points, we will be drawing no statistical inferences from
the results. What in effect we are doing here is using the useful
properties of the technique (especially fast scans of high
dimensionality parameter spaces and the ability to simultaneously scan
in {\it all} parameters) to get a sample of representative points for
the part of the parameter space we are investigating. What is
presented below then, must be viewed with some observations in
mind. Firstly, although data is implemented in the scans, the trading
off of poor fits in one variable for good fits in another can lead to
some outlandish values for key phenomenological values. In addition as
the chain has to start somewhere, an unweighted scatter plot such as
we will be showing below will also show these initial, poorly fitting
points. To address this, for the key points we will state more
precisely what the key phenomenological values are, and in general
they correspond closely to experimental values. Also, regions that
have clearly unacceptable experimental values will be identified where
possible.

The specific region we are investigating is defined by our range
of priors which are specified in table \ref{table:prior}.  
The above choice of priors was influenced by previous work and the
preference to focus on lower values of the soft masses in order to
explore ideal-Higgs-like scenarios. This also leads us to 
choose a log prior, defined here as being flat in $\log\mhalf$,
$\log\mzero$, $\log\mhu$ and $\log\mhd$ and flat in $\azero$,
$\akappa$, $\lambda$ and $\tanb$.  For the nuisance parameters we use
flat priors (although this is not important as they are directly
constrained by measurements) and apply Gaussian likelihoods
representing the experimental observations (see table~\ref{tab:meas}),
as before~\cite{rtr1,rrt2,rrt3,nuhm1}.

The region specified by this set of priors is by no means a fair and
even-handed exploration of the parameter space, but the aim here is to
try and find points in a particular regime. In particular, allowing
$\akappa$ to be independent of $\azero$ (unlike in our previous paper)
was very important, in addition to having non-universal soft
Higgs masses ($\mhu \ne \mhd \ne \mzero$).  With this additional
flexibility compared to the CNMSSM type scan we were
able to find points with the desired characteristics, although as
displayed below these points are still in the minority.

Two alternate exploratory scans were implemented to better understand
this region, the first being to allow a much more generous range in
the parameters, and the second was to do a scan with similar
constrained priors but with the unification conditions
$\mhu=\mhd=\mzero$ and $\akappa=\azero$ enforced, as in the so-called
CNMSSM studied in \cite{CNMSSMpaper}. The objective was to see if it was
the focusing into a small region or the extra freedoms in the Higgs
sector that lead to us finding points of interest. What we found
is that both the freedom and the focus (especially in $\akappa$) seem
to be important.

We compute our mass spectra and observable quantities using the
publicly available NMSSMTools (version 2.1.1) that includes NMSPEC
with a link to Micromegas \cite{micro}; for details see Ref.~\cite{NMSPEC}. We list
the observables that the current version of NMSPEC is applying to
points found in the scan in Table~\ref{tab:measderived}. The relic
density $\abund$ of the lightest neutralino is computed with the
help of Micromegas, which is also linked to NMSPEC. We further use the
same code to compute the cross section for direct detection of dark
matter via its elastic scattering with targets in underground
detectors but do not include it in the likelihood due to large
astrophysical uncertainties.
 The likelihoods for the measured observables are taken as Gaussian
 about their mean values, $\mu$ as tabulated in
 Table~\ref{tab:measderived}, the Gaussian widths being
 determined by the experimental and theoretical errors, $\sigma$ and
 $\tau$, respectively (see the detailed
 explanation in Refs.~\cite{rtr1,rrt2}). In the case where only an
 experimental limit is available, this is given, along with the
 theoretical error.  The smearing out of bounds and combination of
 experimental and theoretical errors is handled in an identical manner
 to Refs.~\cite{rtr1,rrt2}, with the notable exception of the Higgs
 mass and LEP limits on sparticle masses, which are calculated as a
 step function with values of the cross section times branching ratio
 (in the case of the Higgs) or mass that are within two standard
 deviations of the experimental limit being accepted. Finally, any
 points that fail to provide radiative EWSB, give us tachyons or the
 LSP other than the neutralino are rejected.

The above discussion does not yet include the constraints from the
recent analysis by ALEPH for the $e^+e^-\to Z h$, $h\to aa\to 4\tau$
channel~\cite{Schael:2010aw} nor the constraints from BaBar data
on $\Upsilon(3S)\to \gamma a$ with $a\to
\tau^+\tau^-$~\cite{Aubert:2009cka}. These will be considered
ex-post-facto. In this way, we can see what the impact of these latter
constraints is upon a less biased sample of otherwise acceptable
points in parameter space.
%

\begin{table} 
 \centering
\begin{tabular}{|l | l l | l|}
\hline
SM (nuisance) parameter  &   Mean value  & \multicolumn{1}{c|}{Uncertainty} & Ref. \\
 &   $\mu$      & ${\sigma}$ (exper.)  &  \\ \hline
$\mtpole$           &  172.6 GeV    & 1.4 GeV&  \cite{topmass:mar07} \\
$m_b (m_b)^{\overline{MS}}$ &4.20 GeV  & 0.07 GeV &  \cite{pdg06} \\
$\alpha_{\text{s}}(M_Z)^{\overline{MS}}$       &   0.1176   & 0.002 &  \cite{pdg06}\\
\hline
\end{tabular}
\caption[]{Experimental mean $\mu$ and standard deviation $\sigma$
 adopted for the likelihood function for SM (nuisance) parameters,
 assumed to be described by a Gaussian distribution.
\label{tab:meas}}
\end{table}

\begin{table}
\centering
\begin{tabular}{|l | l l l | l|}
\hline
Observable &   Mean value & \multicolumn{2}{c|}{Uncertainties} & Ref. \\
 &   $\mu$      & ${\sigma}$ (exper.)  & $\tau$ (theor.) & \\\hline
$\deltagmtwo \times 10^{10}$       &  29.5 & 8.8 &  1 & \cite{cite:g-2}\\
 $\brbsgamma \times 10^{4}$ &
 3.55 & 0.26 & 0.21 & \cite{cite:g-2} \\
$\brbtaunu \times 10^{4}$ &  $1.32$  & $0.49$  & $0.38$
& \cite{cite:CDF} \\
$\abund$ &  0.1099 & 0.0062 & $0.1\,\abund$& \cite{cite:WMAP} \\\hline
   &  Limit (95\%~\cl)  & \multicolumn{2}{r|}{$\tau$ (theor.)} & Ref. \\ \hline
$\brbsmumu$ &  $ <5.8\times 10^{-8}$ & \multicolumn{2}{r|}{14\%}  & \cite{cdf-bsmumu}\\
$\mhl$  & As implemented in NMSSMTools. & & & \cite{NMSPEC} \\ 
sparticle masses  & As implemented in NMSSMTools. & & & \cite{NMSPEC} \\ \hline 
$\cbeffii$  & As implemented in NMSSMTools. & & & \cite{NMSPEC} \\ 
$\cbeffiv$  & As implemented in NMSSMTools. & & & \cite{NMSPEC} \\ \hline
\end{tabular}
\caption[Summary of the observables used in the analysis.]{Summary of the observables used in the analysis. For more details on how these are implemented, see \cite{rtr1}.
\label{tab:measderived}}
\end{table}

\section{Results}\label{sec:gunion_CNMSSMpars}

In this section we present our numerical results from global scans,
mostly in the form of scatter plots for some of the most interesting
combinations of observables. Most of the figures displayed are scatter
plots with three distinct populations visible on them. There are a
large number of grey points in the figures that follow; these
so-termed {\it full scan} points come from all the points obtained in
the scan and are thinned by a factor of one hundred for clarity. This
will also eliminate many of the outlying initial points talked about
earlier. This population of points can be thought of as a
representation of the general structure of the parameter space, the
backdrop against which we specify points of interest.

To better display the points with $\abb = 0$ and $\haa > 0.5$ we have
marked them as triangles in the relevant figures. Such points will be
termed Type~I points. In addition, there are points with $\br(\hi\to
b\anti b)<0.5$ by virtue of substantial $\haa$ but for which
$\mai>2m_B$ and $\abb\neq 0$.  These are shown by squares in the
figures.  We call these points Type~II points.  In fact, there are two
subclasses of Type~II points.  Type~IIA points are such that the light
Higgs is mainly doublet. Type~IIA points pass constraints on
$\cbeffii$ and $\cbeffiv$ considered individually, but may struggle to
pass the overall constraint implicit in LEP data on $hZ$ production
with $h\to b's$, where $b's$ represents any final state containing one
or more $b$ quarks. This overall constraint becomes important for
$m_h$ below about $110\gev$. Because of the need to consider this
overall constraint, one cannot be certain of whether or not the Type
IIA points with $\mhi\lsim 110-112\gev$ should be eliminated without submitting them to the LEP
collaborations for full analysis.  Thus, we will depict them in the
figures. However, in general the Type~I points where $\mai<2m_B$ are
more interesting. Type~IIB points are ones with $\br(\hi\to b\anti
b)<0.5$, large $\haa$ and $\mai>2m_B$ but for which the light Higgs is
mainly singlet.  These easily pass the LEP constraints and have other
interesting properties that we shall elucidate later.

There is a third class of points, Type~III, that are worth singling
out.  These have $\br(\hi\to b\anti b)>0.5$ but pass our primary
selection criteria. As for Type~II points, there are two subclasses:
Type~IIIA for which the $\hi$ is mainly doublet and $S_s^2$ is small;
and Type~IIIB for which $\hi$ is mainly singlet, $S_s^2\sim 1$.  The
latter easily pass LEP limits on $\cbeffii$ and $\cbeffiv$, by virtue
of $\hi$ being mainly singlet in composition, implying very small
$ZZ\hi$ coupling. In common with Type~IIB, the Higgs with SM-like
coupling to $WW,ZZ$ is instead the $\hii$ which must have mass and/or
decays that allow it to obey all of the bounds imposed by
LEP. Type~IIIA points pass LEP limits by virtue of $\mhi\gsim
114\gev$, and are therefore not ``special'' in any way.  Type~IIIB
points have $\mhi\in[80-100]\gev$ and are particularly interesting in
that they have a high probability of providing the correct value of
$\abund$. In fact, in what follows {\it we will include in the
  definition of Type~III points the requirement that they yield the
  correct value of $\abund$ within $2\sigma$.} This will mean that
Type~III points will have the lowest overall $\chi^2$ of all the
points in our scans. However, they cannot explain the LEP excess in
this range since, as stated before, the $\hi$ is for the most part
decoupled and it is the $\hii$ which plays the role of the SM-like
Higgs, and it has mass above $110\gev$. Finally, we also include in
our final definition of Type~III points the requirement that they satisfy
the ``ex-post-facto'' constraints from ALEPH (on the $\hi\to\ai\ai\to
4\tau$ channel) and from BaBar (from $\Upsilon(3S)\to \gamma \ai$
decays) mentioned earlier.  In short, our final definition of
Type~III points is such that they are very consistent with {\it all} available
experimental constraints. We explicitly state our selection criteria
for the various points in Table~\ref{tab:types}.

A later table will provide a fuller list of
the properties of the different classes of points.


\begin{table} 
 \centering
\begin{tabular}{|c | c | c | c|}
\hline
Point Type &   $\hbb$  & $\abb$ & $S^2_{s}$ \\ \hline
 Type~I &   $< 0.5 $    & $=0$  &  $\sim 0$\\
Type~IIA  &   $< 0.5 $  &  $\ne0$    &  $\sim 0$ \\
Type~IIB &   $<0.5$ & $\ne0$ & $\sim 1$ \\
Type~IIIA &   $> 0.5 $  &   $\ne0$   &  $\sim 0$ \\
Type~IIIB &   $> 0.5 $  &   $\ne0$   &  $\sim 1$ \\
\hline
\end{tabular}
\caption[]{Definition of various points shown in figures. In the case
  of Type~III points, we have also required that $\abund$ be within $\pm
2\sigma$ of the observed value and that they obey the ALEPH
constraints on $\hi\to \ai\ai\to 4\tau$ and the BaBar constraints on
$\Upsilon(3S)\to \gamma\ai$.
  \label{tab:types}}
\end{table}

\begin{figure}[tbh!]
\vspace*{-1.9in}
  \begin{center}
\includegraphics[width=0.78\textwidth]{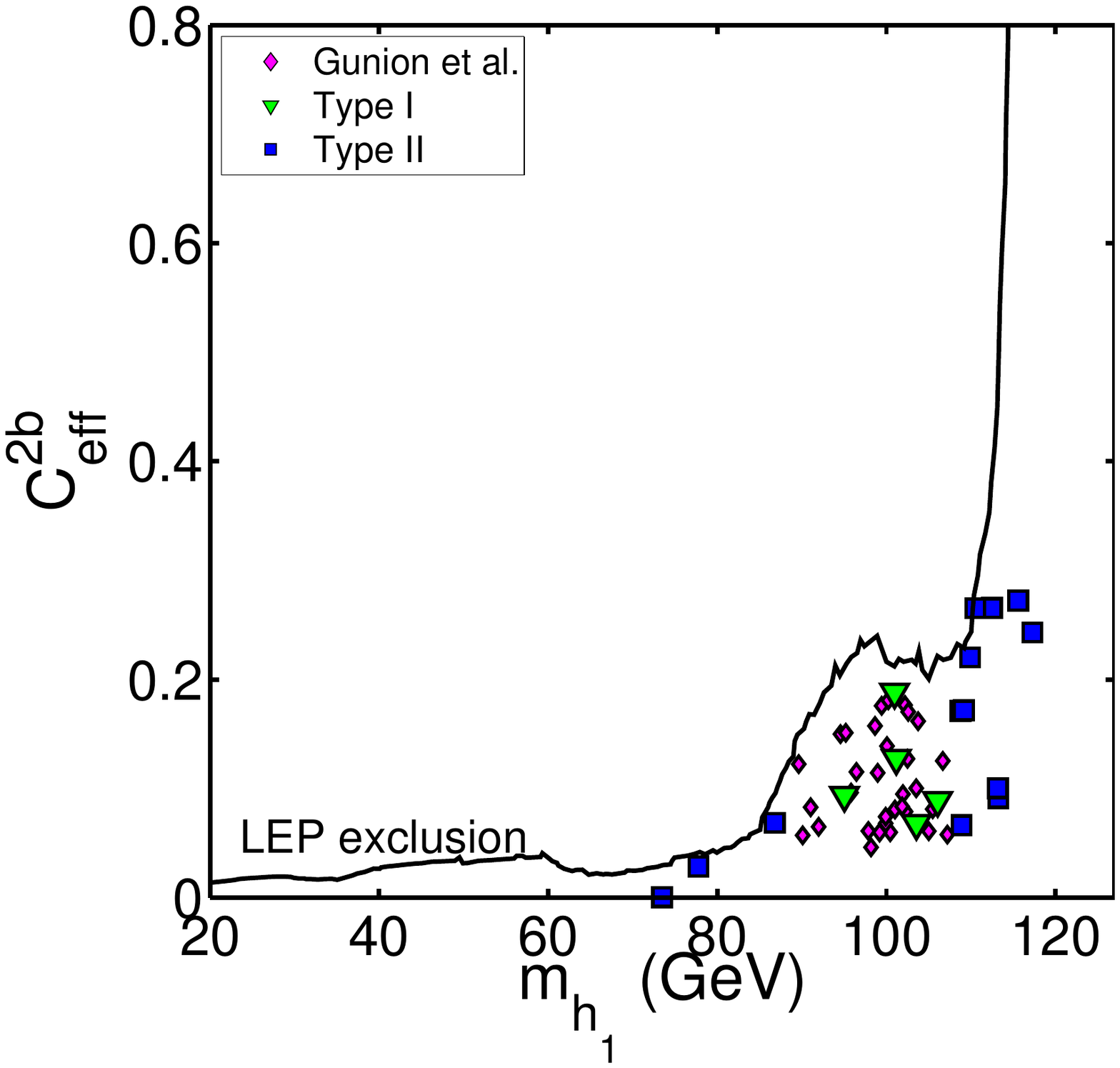}
  \end{center}
\vspace*{-1in}
\caption[A plot of the LEP limit from \cite{lhwg} in the $\cbeffii$ vs
$m_{h_{1}}$ plane.]{\label{fig:lepcomp} A plot of the LEP limit from
  \cite{lhwg} in the $\cbeffii$ vs $m_{h_{1}}$ plane,  superimposed on
  points obtained by Gunion \etal in \cite{gunion1} and so called
  ``Type~I'' points ($m_{a_1} < 2m_B$) and``Type~II'' points ($m_{a_1}
  > 2m_B$) from this study. }

  \end{figure}

  Given the motivation for searching in those regions of parameter
  space that might yield ideal-Higgs scenarios, we first look at the
  $\hi$ branching ratios and $\cbeffii$.  The first result is shown in
  Fig.~\ref{fig:lepcomp}, where we plot $\cbeffii$ as a function of
  $m_{h_{1}}$ for our study, points from previous work by Gunion
  \etal~and the experimental limits from LEP.  This plot already shows
  that the {\it only} points in our scans that reproduce exactly the
  desired qualities ($\hbb \to 0$ with $\abb=0$) while passing all LEP
  constraints are precisely those with a Higgs mass in the range
  $80\gev\lsim \mhi\lsim 100\gev$.  

  In Fig.~\ref{fig:lepcomp}, one also clearly sees the Type~IIA points
  with $\mhi\gsim 108\gev$ and the Type~IIB points with $\mhi\lsim
  90\gev$. As stated above, those of the Type~IIA points that have
  $\mhi\lsim 114\gev$, especially those with $\mhi\lsim 110\gev$,
  might be ruled out by a combined $Z+b's$ LEP analysis, even if not
  ruled out by the $\cbeffii$ and $\cbeffiv$ separate limits.

  \begin{figure}[tbh!]
\vspace*{-1.3in}   \begin{center}\hspace*{-.55in}
 \begin{tabular}{c c}
 \includegraphics[width=0.6\textwidth]{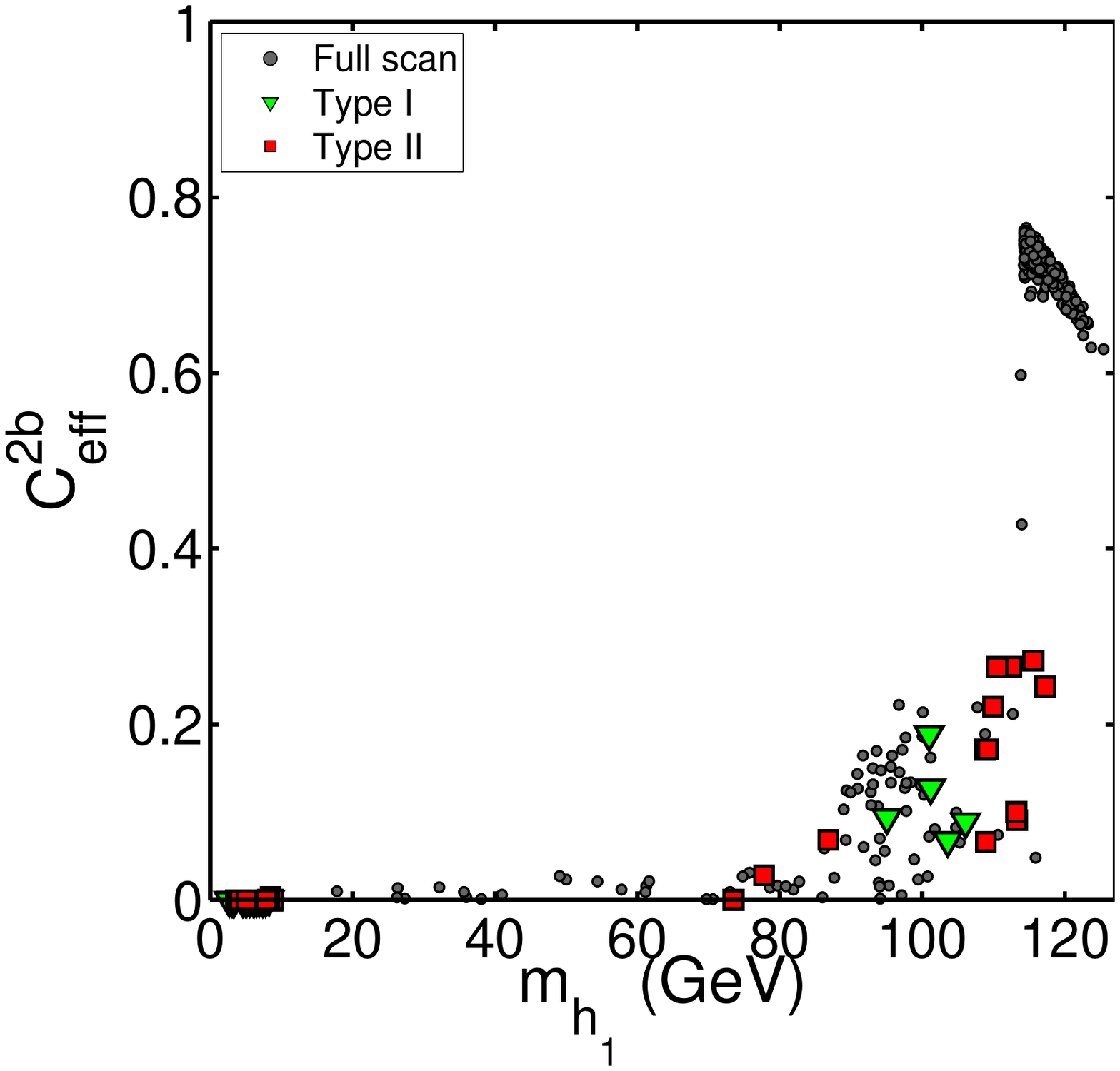}
 &
 \hspace*{-.5in}\includegraphics[width=0.6\textwidth]{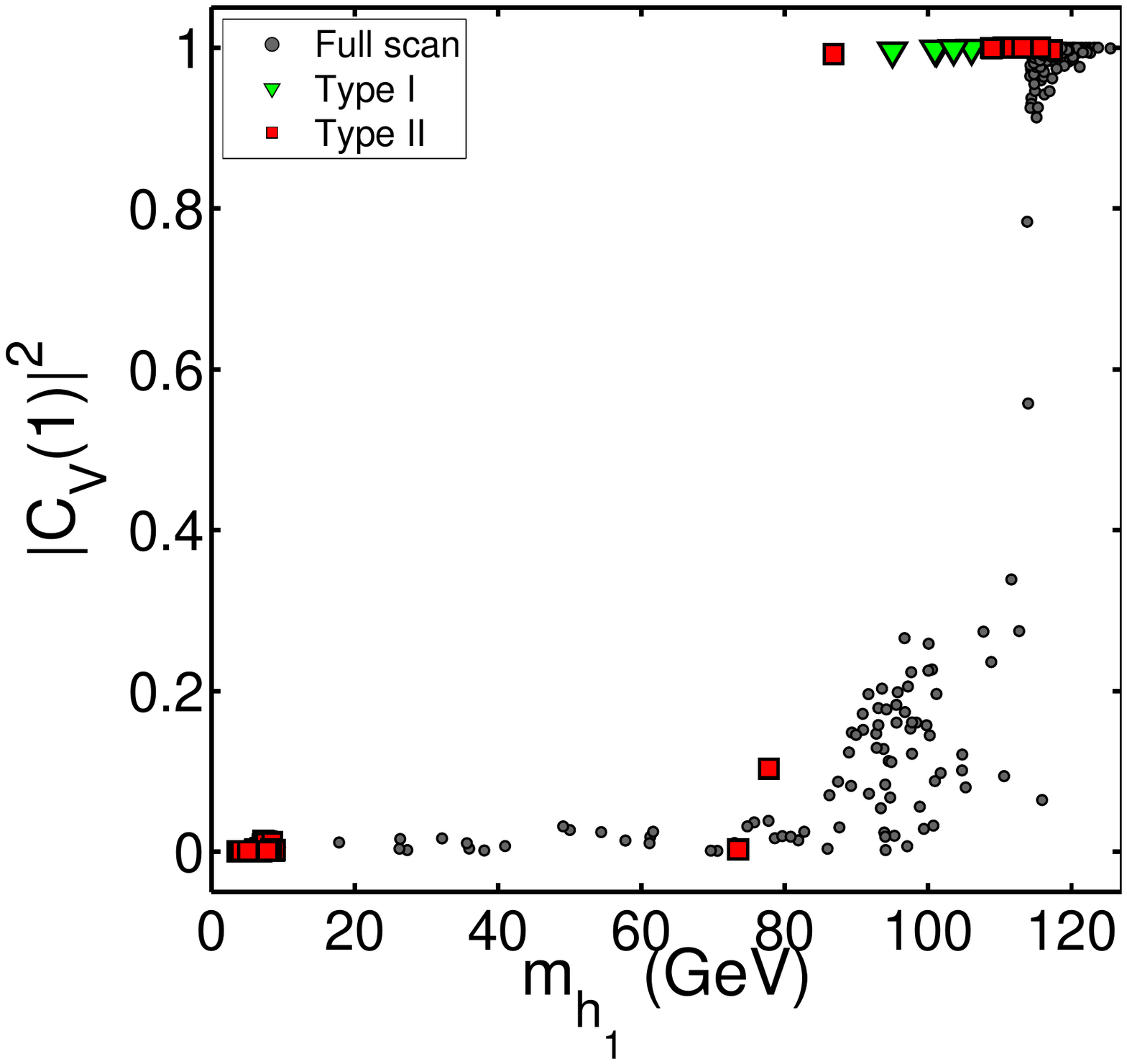}
\vspace*{-1.5in}\\
 \includegraphics[width=0.6\textwidth]{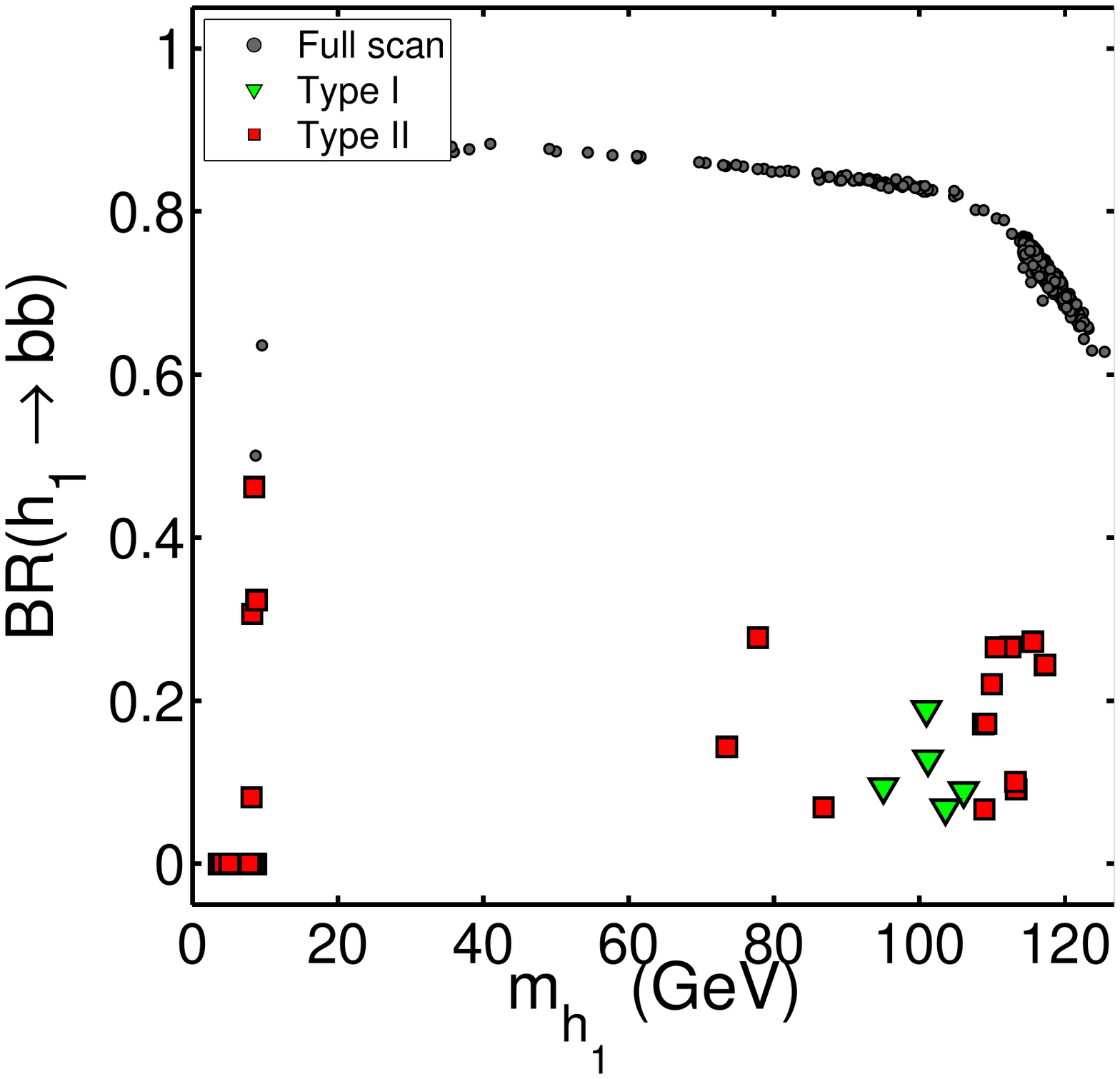} &
 \hspace*{-.5in}\includegraphics[ width=0.6\textwidth]{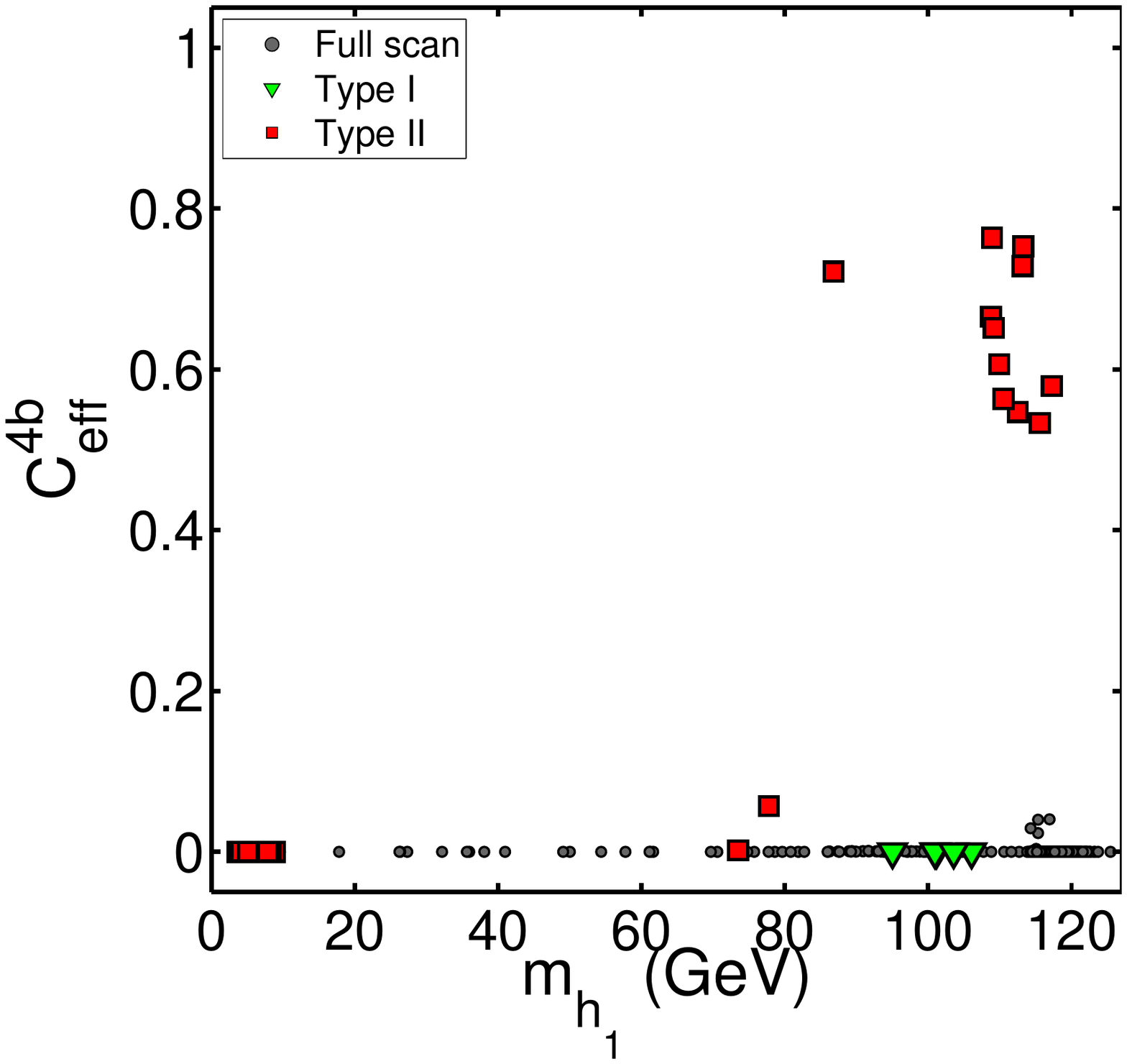} \\
 \end{tabular}
   \end{center}
\vspace*{-.8in}
 \caption[Plot of $\mhi$ against various phenomenological
   quantities.]{\label{fig:c2b} Plots of various phenomenological
     quantities --- see Eq.~(\ref{eqn:defs}) --- 
as a function of $\mhi$. Note that  all
plots and all figures to come use the color red for the Type~II
points, as opposed to the coloring in Fig.~\ref{fig:lepcomp}, where the color
blue was used for Type~II points.   }
   \end{figure}
   We display in Fig.~\ref{fig:c2b} an expanded look at this
   region. In the top left plot we put into context the correlation
   between $m_{h_{1}}$ and $\cbeffii$ for Type~I and Type~II points by
   showing ``background'' points from the full scan. There, we see
   that there are many background points with $\cbeffii$ of order 0.7
   to 0.8 but with large enough $\mhi$ to escape LEP limits, as
   characteristic of Type~IIIA points, as well as background points
   with very low $\mhi$ and $\cbeffii$, which include Type~IIIB
   points. Other plots in this figure show that we obtain the points
   with small $\cbeffii$ in two different ways. The first means is to
   suppress the branching ratio, $\hbb$, as is the case with most
   Type~I and Type~IIA points. The second is to suppress the
   squared-coupling $\cvisq$, as happens if the $\hi$ is sufficiently
   singlet in composition, which is the case in particular for
   Type~IIB points.  Indeed there can be a lot of points in a similar
   $\mhi$ region to that identified by Gunion \etal\ (\ie\
   $80\gev\lsim \mhi\lsim 100\gev$) that escape LEP limits via
   suppression of $\cvisq$ rather than via suppression of $\br(\hi\to
   b\anti b)$. Small $\cvisq$ implies that the $\hi$ cannot act as a
   ``ideal'' Higgs defined as having SM-like $WW,ZZ$ coupling but mass
   $\lsim 105\gev$.

   The bottom right figure in Fig.~\ref{fig:c2b} shows $\cbeffiv$
   against $\mhi$. This can be compared with a similar figure in
   \cite{gunion1}. In general it is clear that the Type~I points all
   have $\cbeffiv=0$ (since $\mai<2m_B$) while for Type~IIA points
   $\cbeffiv$ is quite significant and, for those points with
   $\mhi<114\gev$, is not far below the LEP limit.  As noted earlier,
   since the limits on $\cbeffii$ and $\cbeffiv$ are being applied
   individually and not in combination it could well be that the
   Type~IIA points with $\mhi<114\gev$, especially those with
   $\mhi\lsim 110\gev$, are in fact in contradiction
   with LEP. But, without a full LEP analysis, it can be instructive to
   leave them in with this caveat in mind. One thing to notice in
   general is that the ideal-Higgs-like Type~I and the Type~IIA points
   with $\mhi \lsim 110\gev$ and fairly large $|\cvi|\sim 1$ are very
   rare, even in the context of a scan looking for these regions; this
   could be an artifact of our scanning technique or it could be that
   these points are truly hard to find given the criteria and
   high-scale boundary conditions we have used.


\begin{figure}[tbh!]
\vspace*{-1.3in}
  \begin{center}\hspace*{-.4in}
\begin{tabular}{c c}
\includegraphics[width=0.6\textwidth]{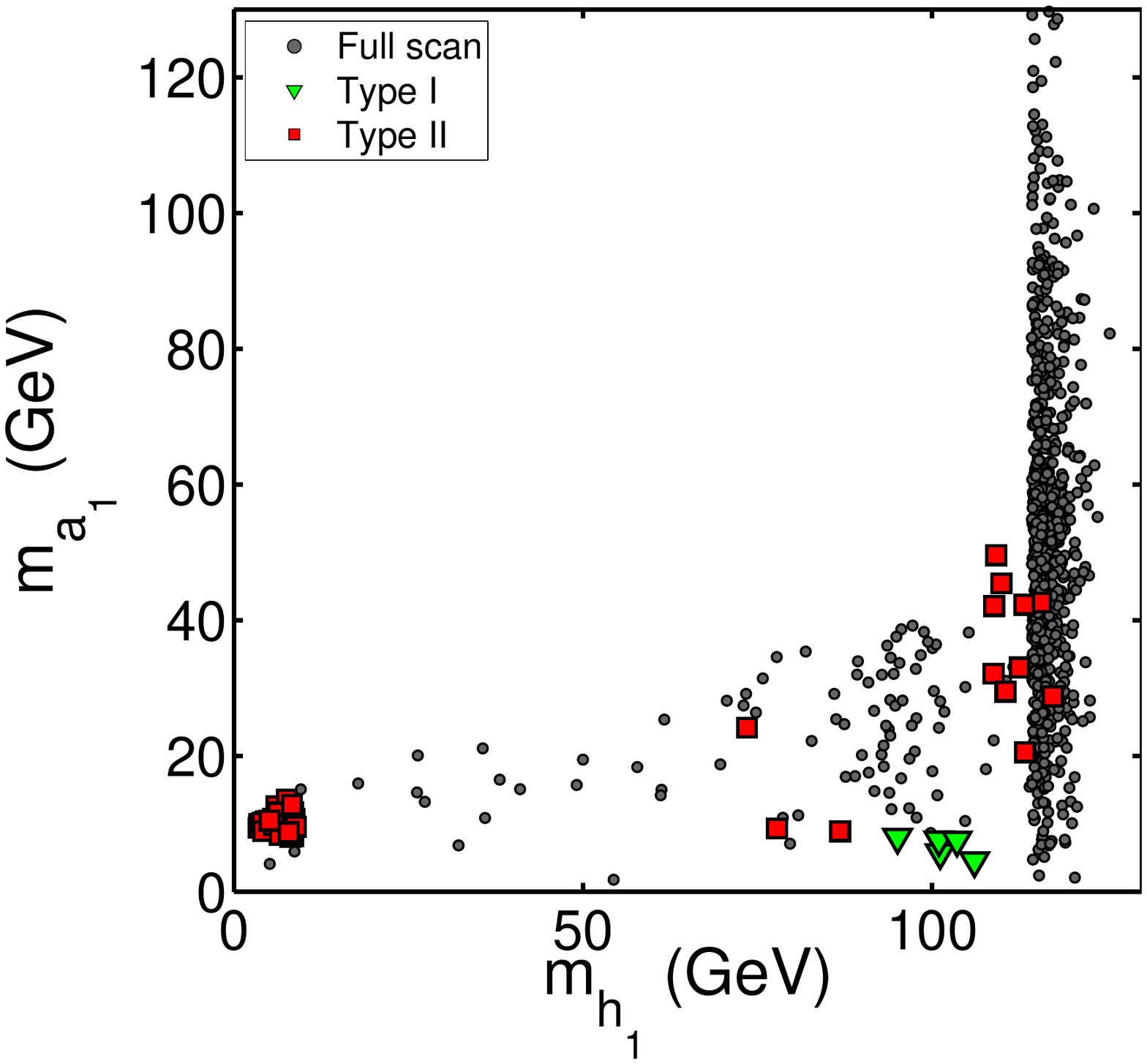} 
& \hspace*{-.5in}\includegraphics[width=0.6\textwidth]{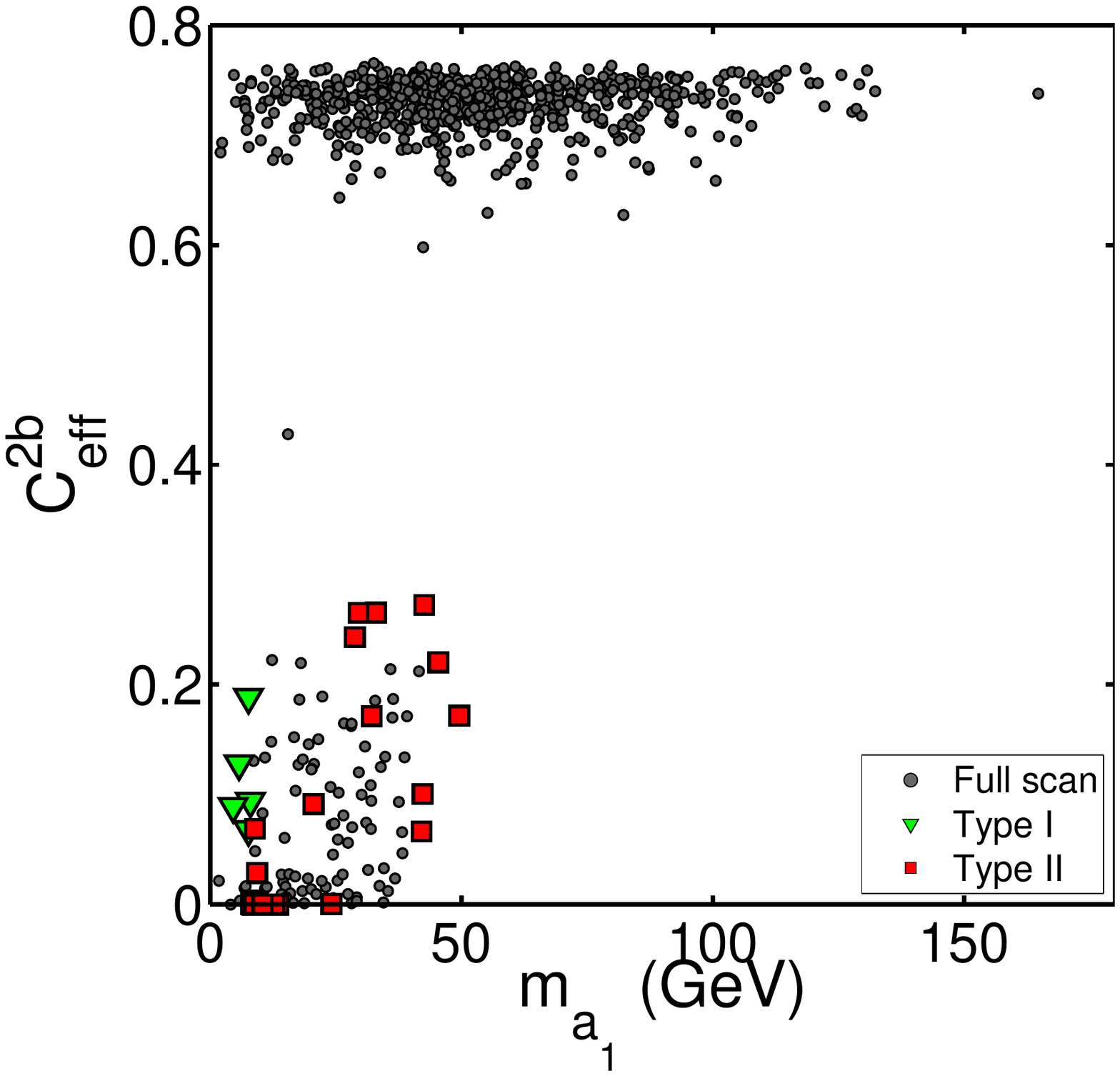} \\
   \end{tabular}
  \end{center}
\vspace*{-.9in}
\caption [Plots of $m_{h_{1}}$ and $\cbeffii$ against $m_{a_{1}}$
.]{\label{fig:tom2} Plots of $\mai$ as a function of $m_{h_{1}}$ and
  of $\cbeffii$ as a function of $m_{a_{1}}$.}
  \end{figure}
  Since $\mai$ is so crucial to whether or not a given point is ruled
  out by LEP data, it is useful to understand how $\mhi$ and
  $\cbeffii$ correlate with $\mai$. In the left plot of
  Fig.~\ref{fig:tom2} we display all the different types of points in
  the $m_{h_{1}}$--$m_{a_{1}}$ plane.  Note again the Type~IIA points with
  $\mai>2m_B$ and $109\gev\lsim \mhi<114\gev$ that escape LEP limits
  on $\cbeffiv$ despite having large $\br(\hi\to \ai\ai)$ and large
  $\br(\ai\to b\anti b)$. These also
  appeared in Fig.~\ref{fig:c2b}.

\begin{figure}[tbh!]
\vspace*{-.5in}
  \begin{center}\hspace*{-.5in}
    \begin{tabular}{c c c}
\includegraphics[width=0.41\textwidth]{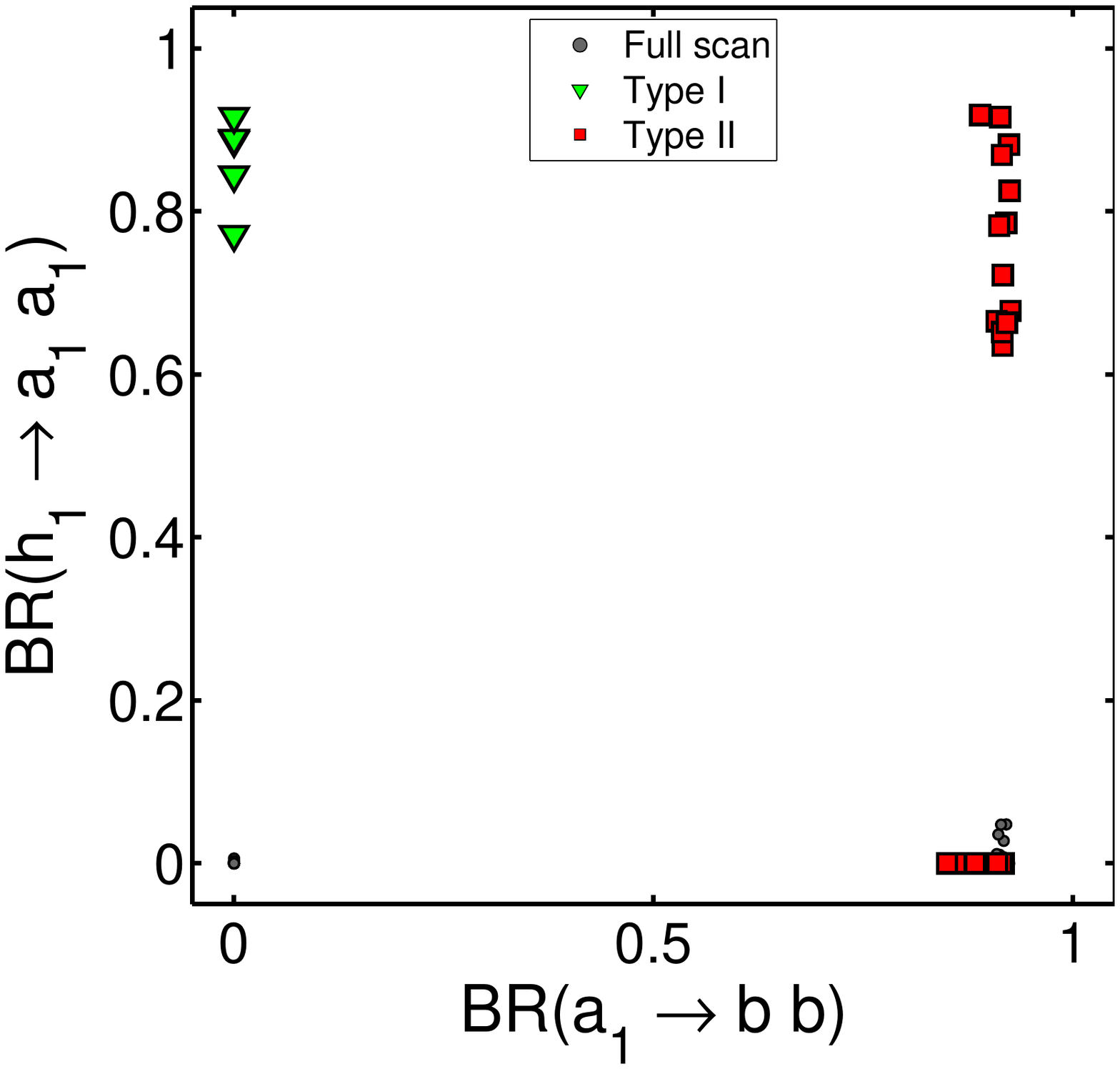} &
\hspace*{-.42in}\includegraphics[width=0.41\textwidth]{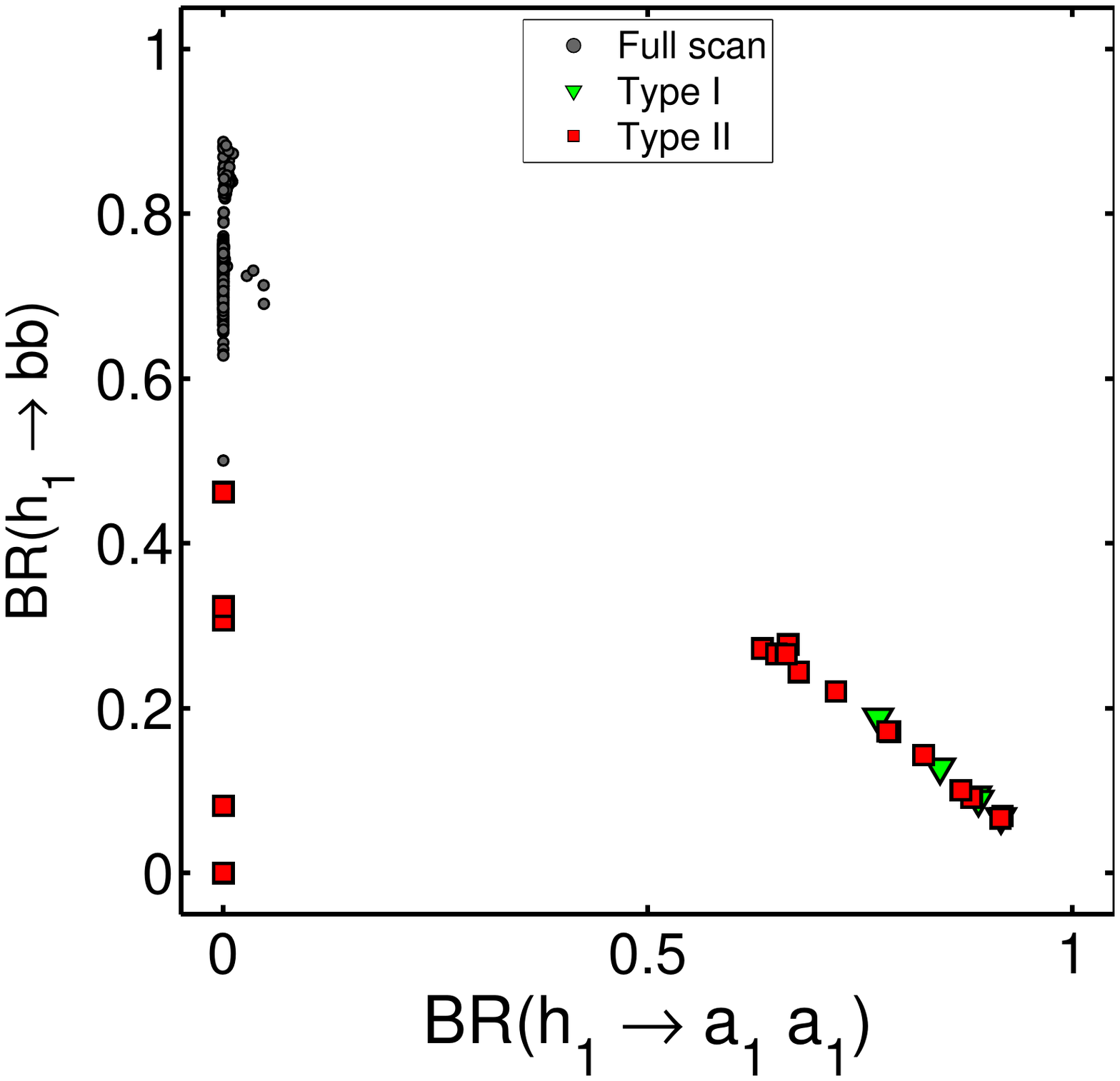} &
\hspace*{-.42in}\includegraphics[width=0.41\textwidth]{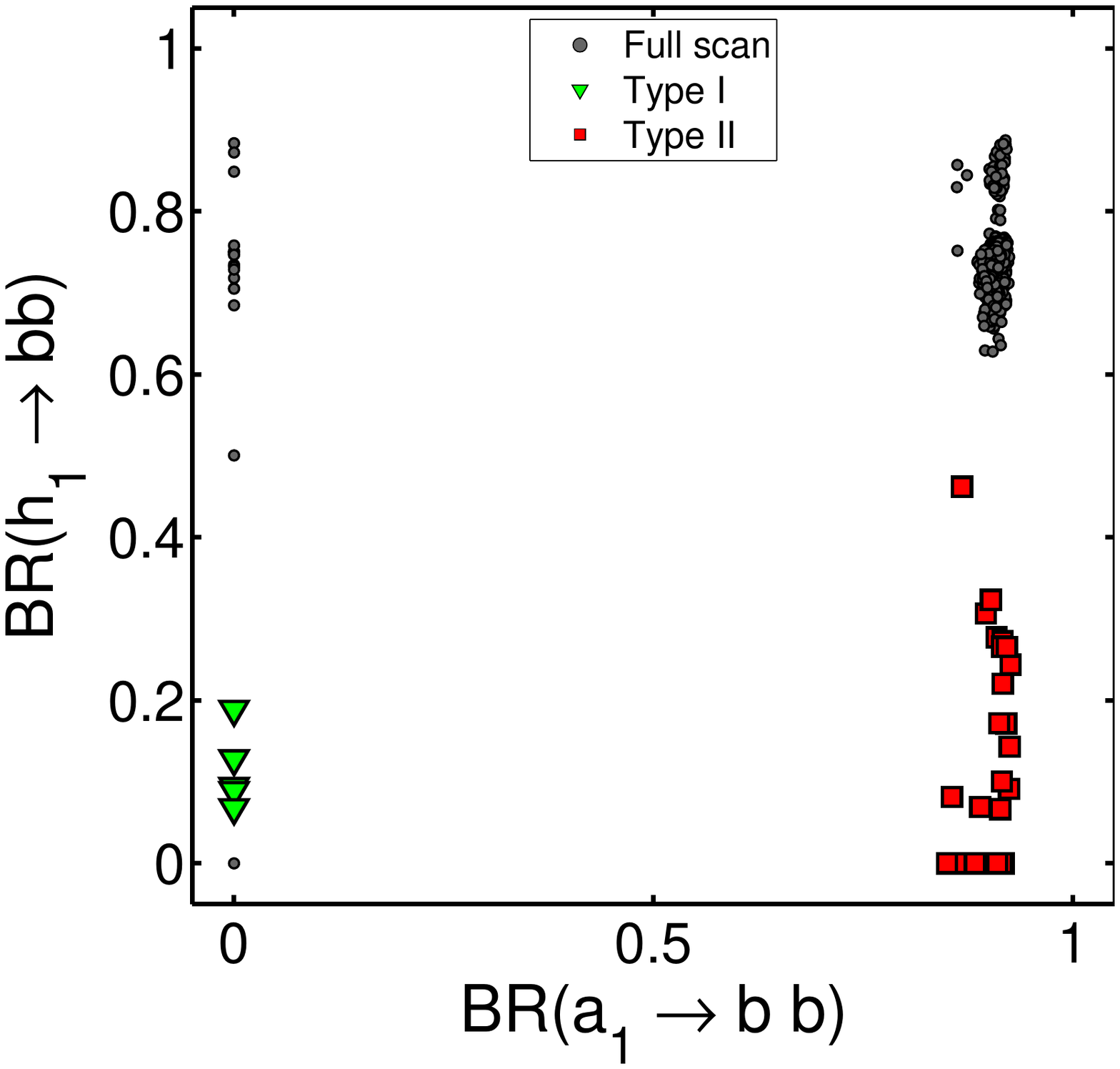}    \\
   \end{tabular}
  \end{center}
\vspace*{-.7in}
  \caption[Plots of the Higgs branching ratios of
  interest.]{\label{fig:brs} Plots of: $\br(\hi\to \ai\ai)$ vs.
    $\br(\ai\to b \anti b)$; $\br(\hi\to b \anti b)$ vs.
    $\br(\hi\to \ai\ai)$;  and $\br(\hi\to b\anti b)$
    vs. $\br(\ai\to b\anti b)$.}
  \end{figure}

  To better demonstrate the interplay between the various branching
  ratios needed to evade LEP constraints, some of the crucial ones are
  depicted in the same convention in Fig.~\ref{fig:brs}. One can see
  that the green Type~I points are clearly isolated, with the key
  discriminator from Type~II points being $\abb$. 
In Fig.~\ref{fig:brs}, some key differences between
  Type~IIA and Type~IIB points are apparent, the most notable being
  the very small $\br(\hi\to \ai\ai)$ for Type~IIB (singlet $\hi$)
  points. Note also that {\it all} Type~III points have very small
  $\br(\hi\to \ai\ai)$.

  We will shortly discuss whether or not the Type~I points escape the
  latest ALEPH limits on $\hi\to \ai\ai$ with $\ai\to \tau^+\tau^-$.
  Such escape is possible when $\tanb$ is small,
  since at small $\tanb$ one predicts that 
  $\br(\ai\to \tau^+\tau^-)$ is significantly suppressed due to
  substantial branching ratios for $\ai$ to $c\anti c$, $s\anti s$ and
  $gg$ and the resulting final states in $\hi\to\ai\ai$ are less
  strongly constrained than the $\hi\to\ai\ai\to 4\tau$ final
  state. This was discussed in \cite{Dermisek:2010mg}.

 \begin{figure}[tbh!]
\vspace*{-1.3in}
   \begin{center}\hspace*{-.5in}
 \begin{tabular}{c c}
 \includegraphics[width=0.6\textwidth]{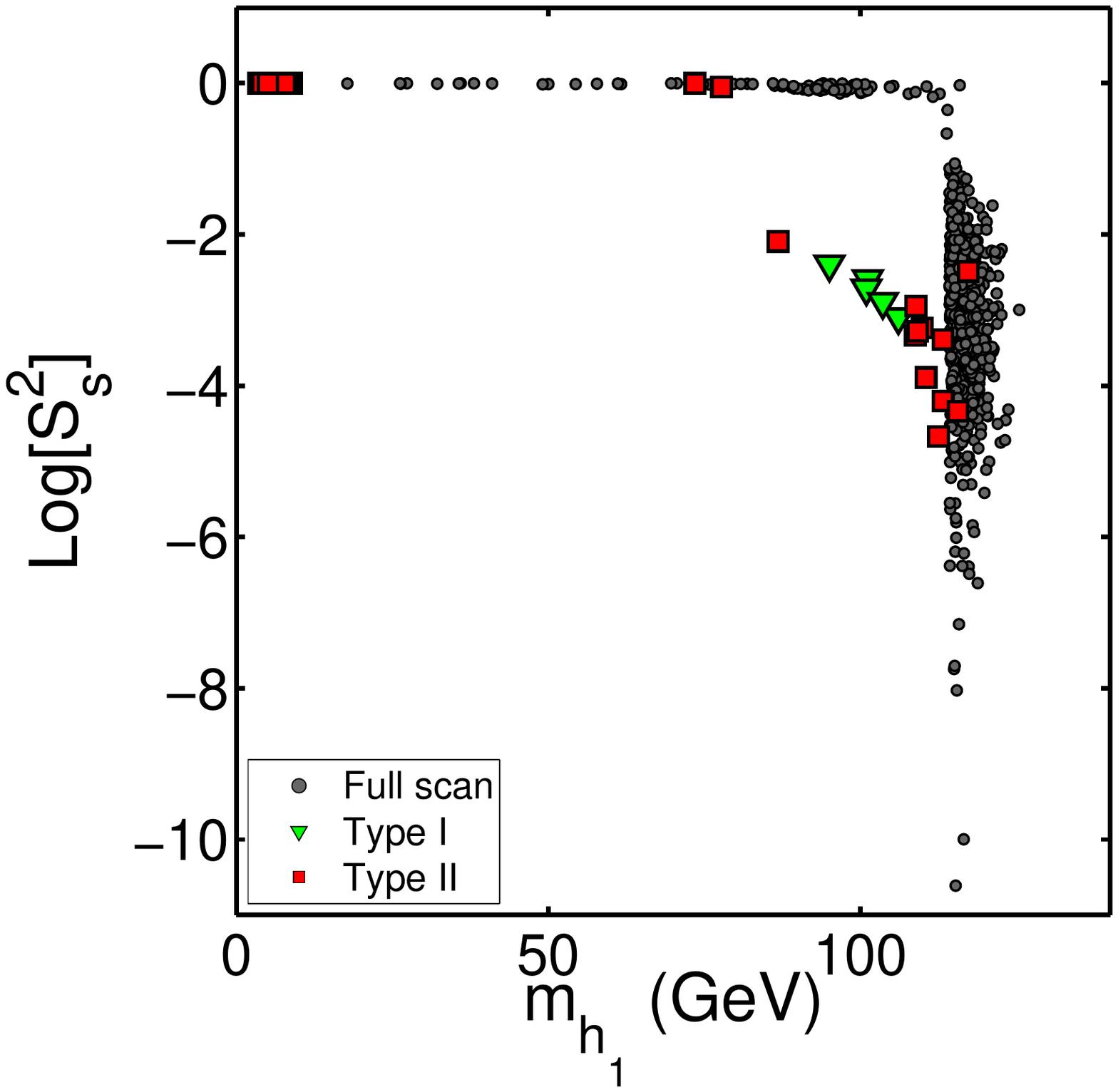}
 & \hspace*{-.5in}\includegraphics[width=0.6\textwidth]{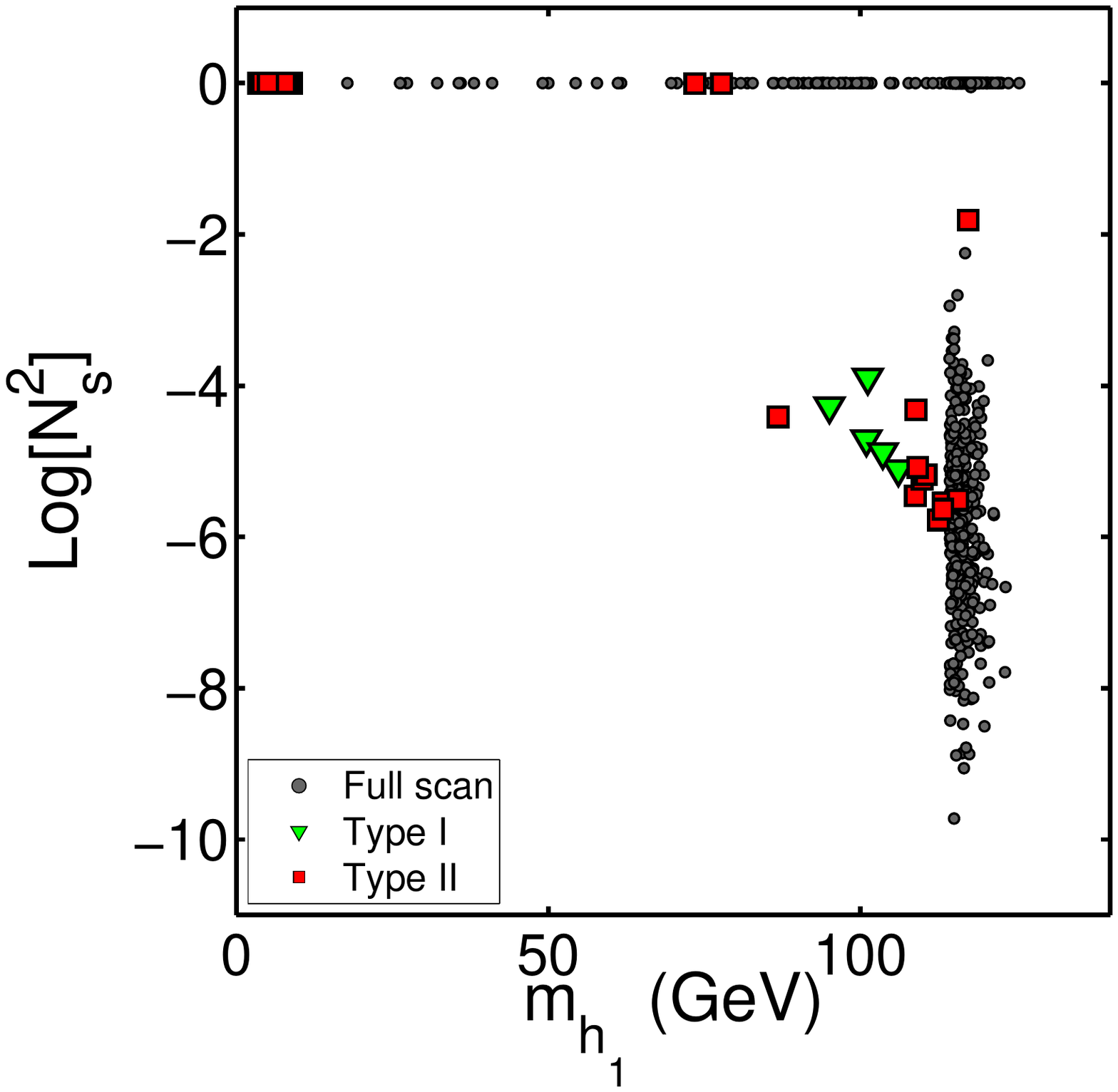}\\
    \end{tabular}
   \end{center}
\vspace*{-.9in}
\caption[Mass of the $\hi$ against singlet and singlino
compositions.]{\label{fig:comp} We plot the square of the singlet
  component of the $\hi$ and the square of the singlino component of
  $\cnone$ as functions of $\mhi$.}
   \end{figure}

   In Fig.~\ref{fig:comp} we show the square of the singlet component
   of the $\hi$, and the square of the singlino component of the
   lightest neutralino, the $\cnone$, as functions of
   $m_{h_{1}}$. These figures illustrate a number of things.  First
   note the large number of points with $N_s^2\sim 1$ and $S_s^2\sim
   1$, the latter implying that $\cvisq$ is greatly
   suppressed. Included in this set of points are the Type~IIB points
   with $\mhi<80\gev$ as well as the Type~IIIB points with low $\mhi$,
   large $\br(\hi\to \ai\ai)$ and non-zero $\br(\ai\to b\anti b)$ ---
   all these points escape LEP limits since the $\hi$ is very
   singlet-like. Second, we observe that the $S_s^2$ plot is closely
   related to the top right panel in Fig.~\ref{fig:c2b}. As noted in
   the discussion of the latter figure, it is the Type~I points and
   Type~IIA points with $\mhi\gsim 80\gev$ for which the $\hi$ is
   highly doublet-like whereas the Type~IIB points with $\mhi\lsim
   80\gev$ have a singlet-like $\hi$.  And, finally, there is the
   large collection of points with $\mhi>114\gev$ that are
   unconstrained by LEP data and typically are very doublet-like. 

   The right hand plot of Fig.~\ref{fig:comp} will be of more use
   below when we consider dark matter, but it does show an interesting
   correlation, namely that the Type~I points and the Type~IIA points
   with $\mhi\gsim 80\gev$ stand out by having a $\cnone$ that is
   bino-like instead of singlino-like, the latter being more typical
   of the majority of points found in our scans, including Type~IIB
   points. We further note that the Type~IIIB points that have
   $\mhi\lsim 110\gev$ and $S_s^2\sim 1$ also have $N_s^2\sim 1$.  In
   contrast, the Type~IIIA points which have $\mhi>114\gev$ and small
   $S_s^2$ (with $S_u^2$ being large instead) can have either
   large $N_s^2$ or large $N_B^2$ (\ie\ singlet-like $\chi$ or
   bino-like $\chi$).

%
%

\begin{table} 
\begin{center}
\resizebox{\textwidth}{!}{%
\begin{tabular}{|c | c | c | c | c | c | c | c | c |c |c |}
\hline
\hline
Point   & $\mzero$ (GeV) & $\mhalf$ (GeV)& $\mhu$ (GeV)& $\mhd$ (GeV) & ~~~~$\akappa$ (GeV)~~~~ &~~~~$\azero$ (GeV)~~~~ &~~~~$\lambda$~~~~&~~~~$\tanb$~~~~ &~~~~$G$~~~~ \\
\hline
1  & 452 & 223 & 3.54 & 543 &  5.69 &33.7   & 0.481 & 2.54 & 30.4 \\ 
\hline
2  & 10.9 & 287 & 710 & 180 &  7.89 &  51.7 &  0.436 & 3.61 & 30.9 \\ 
\hline
3  & 7.57& 467 &2.23 & 655 &  -4.65 & 28.1  &  0.408 & 2.15 & 17.6 \\ 
\hline
4  & 0.717 & 393 &  0.724& 622& 2.43 & 46.6   & 0.328 & 2.52 & 15.2 \\ 
\hline
5  & 0.804 & 387 & 42.0 & 526 &   7.17& 46.3   & 0.399 & 2.05 & 18.8 \\ 
\hline
\hline
\end{tabular}}
\end{center}
\begin{center}
\resizebox{\textwidth}{!}{%
\begin{tabular}{|c |c | c | c | c | c | c | c|}
\hline
\hline
Point  &~~~~ $\Omega~h^2$~~~~ & $\deltagmtwo$ & $\brbsgamma$ & $\br(\bsmumu)$ & $m_{h_1}$ (GeV) &$m_{a_1}$ (GeV)& $\chi^2$ \\
\hline
1  & 2.35 & $1.78\times 10^{-10}$ & $3.11\times 10^{-4}$ & $2.54\times 10^{-8}$ &  95.0 & 8.04 & 70.3 \\ 
\hline
2  & $0.276$ & $10.3\times 10^{-10}$ & $2.951\times 10^{-4}$  & $0.127\times 10^{-8}$ &  101 & 5.77 & 39.6\\ 
\hline
3  & 0.344 & $2.45\times 10^{-10}$ & $3.211\times 10^{-4}$  &  $4.30\times 10^{-8}$ &  106 & 4.58 & 30.1\\ 
\hline
4  & 0.341 & $3.90\times 10^{-10}$ & $3.191\times 10^{-4}$  & $3.30\times 10^{-8}$ & 101 & 7.66 & 29.9\\ 
\hline
5  & 0.245 & $3.42\times 10^{-10}$ & $3.391\times 10^{-4}$  & $3.22\times 10^{-8}$ &  104 & 7.63 & 25.7\\ 
\hline
\hline
\end{tabular}}
\end{center}
\begin{center}
\resizebox{\textwidth}{!}{%
\begin{tabular}{|c|c | c  |c | c | c | c | c | c | c | c|}
\hline
\hline
Point  &$cos\theta_{A}^{Max}$ & $cos\theta_{A}$  & $\caibb^{Max}$
&$\caibb$ &$\haa$ & $\hbb$  & $BR(a_{1} \to \tau^{+} \tau^{-})$ &
$BR(a_{1} \to \mu^{+} \mu^{-})$ &$(\xi^2)^{Max}_{ALEPH}$& $\xi^2$ \\
\hline
1  & 0.341& 0.0186 & 0.867 & 0.0472 & 0.887  & 0.0943  & 0.833 & 0.0034
& .2902 & 0.615 ! \\ 
\hline
2  &  0.199 &  0.0197& 0.719& 0.0711  &0.844 & 0.128  &  0.881 & 0.0042
&  .49809 & 0.655 !\\ 
\hline
3  &  0.309 & 0.00631 & 0.664& 0.0136  &0.890 & 0.0893  & 0.771 & 0.0047
& .75916 & 0.529\\ 
\hline
4  & 0.336 & 0.00527& 0.849&  0.0133  &0.772 & 0.189   & 0.837 & 0.0035
&  .52524 & 0.541 !\\ 
\hline
5  & 0.443 &  0.00716& 0.906&  0.0146  &0.916 & 0.0682  & 0.786 &
0.0034 & .67593 & 0.566\\ 
\hline
\hline
\end{tabular}}
\end{center}
\caption[Values of interest for Type~I points.]{\label{finetune}
  Displayed are some values of interest for the Type~I points found in
  our scans. In the upper table we show the base parameters that give
  us our population of interesting (Type~I) points. The final column,
  denotes G, defined in Eq.~(\ref{gfinetune}), a measure of the
  fine-tuning needed to obtain the (low) value of $m_{a_{1}}$. In the
  middle table are some of the phenomenological values for the points
  of interest. Notice the likelihood (to be precise the
  $-2\textrm{log}(\textrm{likelihood})=\chi^2$) in general is large
  reflecting a poor fit, and this is largely being driven by poor fits
  to $\abund$ \cite{tfhrr1}. The bottom table shows some of the
  key branching ratios of interest for Type~I points, and compares the
  ALEPH limits on $\xi^2\equiv |C_{V}(1)|^2 \times \haa \times
  \left[BR(a_{1} \to \tau^{+} \tau^{-} \right]^{2}$ with the predicted
  values. Points appended with an exclamation mark are excluded by the ALEPH analysis.  However,  as discussed later, by adjusting $A_\kappa$ by a very small amount they can be brought into agreement with the ALEPH limits without affecting any other phenomenology. }
\end{table}

Full details regarding Type~I points appear in
Table~\ref{finetune}. The upper table shows the input parameter values
for each of the Type~I points. The corresponding
``light-$\ai$'' fine-tuning measure, $G$, defined by: 
\beq 
G\equiv Min\left\{[Max(|F_{A_{\lambda}}|,|F_{A_{\kappa}}|)],
  |F_{A_{\lambda}} + F_{A_{\kappa}}|\right\},\eeq where, \beq
F_{A_{\lambda}}\equiv \frac{A_{\lambda}}{m^2_{a_1}}\frac{d
  m^2_{a_1}}{d A_{\lambda}}~~~~~~~~~~~~~~~~~~~~~~F_{A_{\kappa}}\equiv
\frac{A_{\kappa}}{m^2_{a_1}}\frac{d m^2_{a_1}}{d A_{\kappa}}
\label{gfinetune}
\eeq 
is also shown. One can see some common threads for all of the Type~I
points. Perhaps the most intriguing is the need for $\lambda$ to be
quite large and away from the decoupling limit. This suggests that
these points are in some sense specific to the NMSSM and are unlikely
to be found in similar parametrizations of the MSSM. 
It is also nice to see that despite the
light-$\ai$ fine-tuning measure $G$ not being used in the scans, the
resultant values for Type~I points are not wholly
unreasonable. Finally, we note that the values of $\tanb$ for which
Type~I points were found are relatively low. In the less constrained
scans of parameter space performed in \cite{gunion1,gunion2,gunion3}
Type~I (ideal-Higgs) points were found at large $\tanb$ as well.

\begin{figure}[h!]
\vspace*{-1.9in}
  \begin{center}
\includegraphics[width=0.78\textwidth]{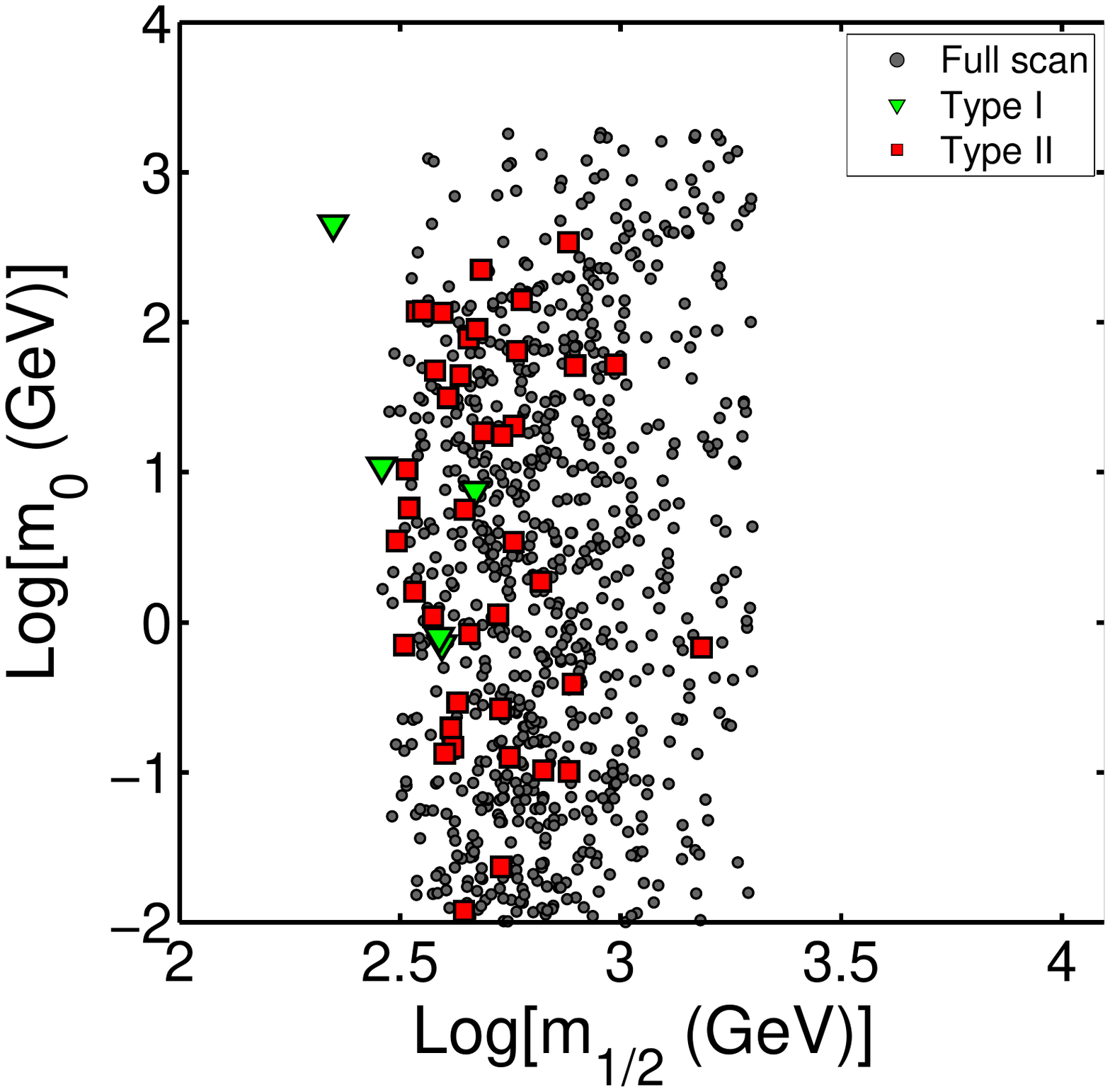}
  \end{center}
\vspace*{-1in}
\caption{\label{fig:m0m12} A plot of our points in the $m_0$ ---
  $m_{1/2}$ plane. }
  \end{figure}

In Fig.~\ref{fig:m0m12}, we show the values of $m_0$ and $m_{1/2}$ (at
the GUT scale) for the various different types of points.  We see that
many of the points of Type~II and Type~III have quite small values of
$m_0$ and that most Type~I points have quite modest $m_0$.  As regards
$m_{1/2}$, it is typically of order $250-300\gev$ for Type~I points but ranges from $\sim 250\gev$ up to
$2\tev$ for Type~II and Type~III points.  Regardless, the resulting gluino and (non-stop)
squark masses are always at least as large as $650\gev$ and often
significantly larger.  Such values are above the limits currently
being set by LHC data, which limits are typically of order
$500-600\gev$ (assuming universality for the gaugino masses and for
the non-Higgs scalar masses at the GUT scale). Of course, the LHC will probe
gluino and squark masses of order $1\tev$ after another year or two of
running.  In common with other models employing universality at the
GUT scale, the parameter points typical of our study will then start
to be ruled out.

  The middle table in Table~\ref{finetune} gives some corresponding
  experimental values for the Type~I points. It is interesting to see
  that the points provide phenomenologically viable results for
  $\br(\anti B\to X_s \gamma)$ and $\br(\anti B_s \to
  \mu^+\mu^-)$. However, the lower $2\sigma$ boundary for the observed
  $\deltagmtwo$ is $\sim 10\times 10^{-10}$ and only Type~I point 2
  barely predicts this high a value, the other Type~I points
  predicting values in the range $(1.78-3.90)\times 10^{-10}$. The
  relic density is equally problematical, with the best value barely
  getting to within $2\sigma$ of the WMAP value. The likelihood is
  dominated by this contribution as in general the relic density is
  the strongest constraint on the parameter space
  \cite{tfhrr1}. Hence, the points with best likelihood correspond to
  $\Omega h^2$ closest to its experimental value. As discussed
  shortly, the Type~IIIB points (\ie\ points $\mhi\lsim 104\gev$ that
  escape LEP limits by virtue of the $\hi$ being mainly singlet) quite
  readily achieve an $\Omega h^2$ near the WMAP value; as a result,
  the $\chi^2$ for Type~IIIB points ranges from a low of $\sim 1.9$ to
  a high of $\sim 6$, vs. the best value of $\sim 26$ found for Type I
  points in our scans.  Type~IIIA points have $\chi^2$ values only
  slightly larger than Type~IIIB points and definitely below $26$, as
  consistent with our requirement that Type~III points be consistent
  with the observed $\abund$ within $\pm 2\sigma$.  Note that small
  $\chi^2$ can be achieved within the other defining characteristics
  for Type~IIIA and Type~IIIB points because small $\abund$ is
  possible despite the singlino or bino nature of the $\chi$ by virtue
  of near mass degeneracy of the $\chi$ and $\widetilde \tau_1$.

  The bottom table of Table~\ref{finetune} gives the values of the
  coupling $\caibb\equiv \tanb\cta$ in comparison to the maximum
  absolute value allowed by BaBar data in the
  $\Upsilon_{3S}\to\gamma\tau+\tau^-$ channel. Here, $\cta$ is the
  doublet component of $\ai$ as defined by $\ai=\cta a_{MSSM}+\sta
  a_S$. This bottom table also gives the value of $\xi^2\equiv
  |C_{V}(1)|^2 \times \haa \times \left[BR(a_{1} \to \tau^{+} \tau^{-}
  \right]^{2}$ in comparison to the upper limit for each point from
  the recent ALEPH analysis.  We observe that the Type I points have
  no problem obeying the limits from BaBar but that the ALEPH limits
  are very problematical for three out of five of the Type I
  points. However, we show below that a very small change in
  $A_\kappa$ will bring these points into agreement with the ALEPH
  limits without affecting any other phenomenology. We also wish to
  note that the ALEPH constraints are much stronger than what was
  expected on the basis of Monte Carlo and so, in our opinion, some
  relaxation of the ALEPH bounds could be considered.  If $\sim
  1\sigma$ relaxation is allowed, then all our Type~I points survive
  ``as is''. We also wish to note that $\caibb^2\times \br(\ai\to
  \mu^+\mu^-)$ roughly determines the ability to detect $gg\to \ai \to
  \mu^+\mu^-$ at hadron colliders~\cite{Dermisek:2009fd}.  Very
  roughly, in the $\mai<2m_B$ region where $\br(\ai\to \mupmum)\sim
  (0.003-0.005)$ detection will only be ``easy'' if $|\caibb|\geq
  1$. Unfortunately, for our Type~I points, $|\caibb|$ is always
  small, with point 2 providing the largest value of $\sim 0.07$.

\begin{figure}[tbh!]
\vspace*{-1.3in}
  \begin{center}\hspace*{-.4in}
\begin{tabular}{c c}  
  \includegraphics[width=0.6\textwidth]{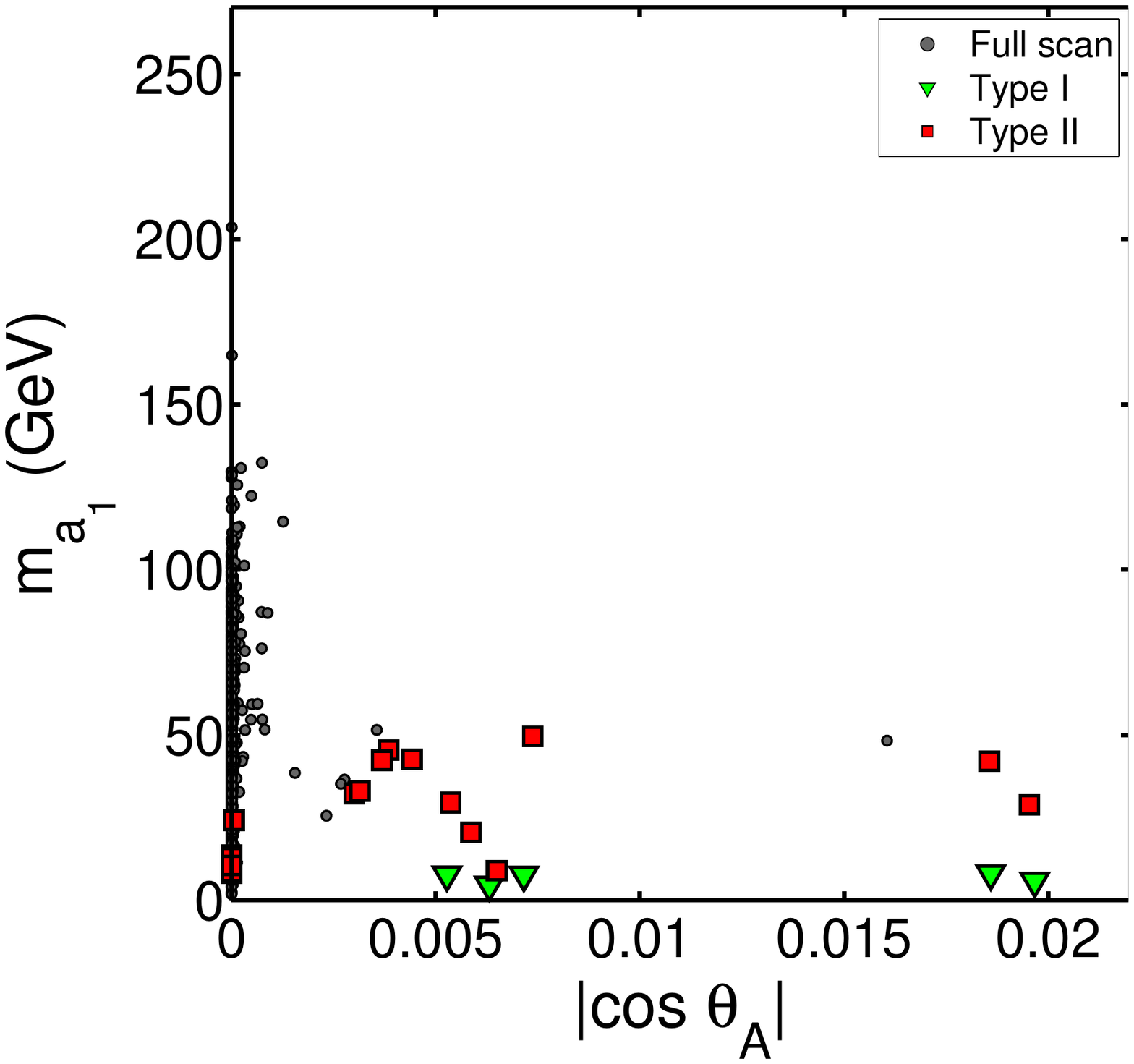}
  & 
\hspace*{-.6in} \includegraphics[width=0.6\textwidth]{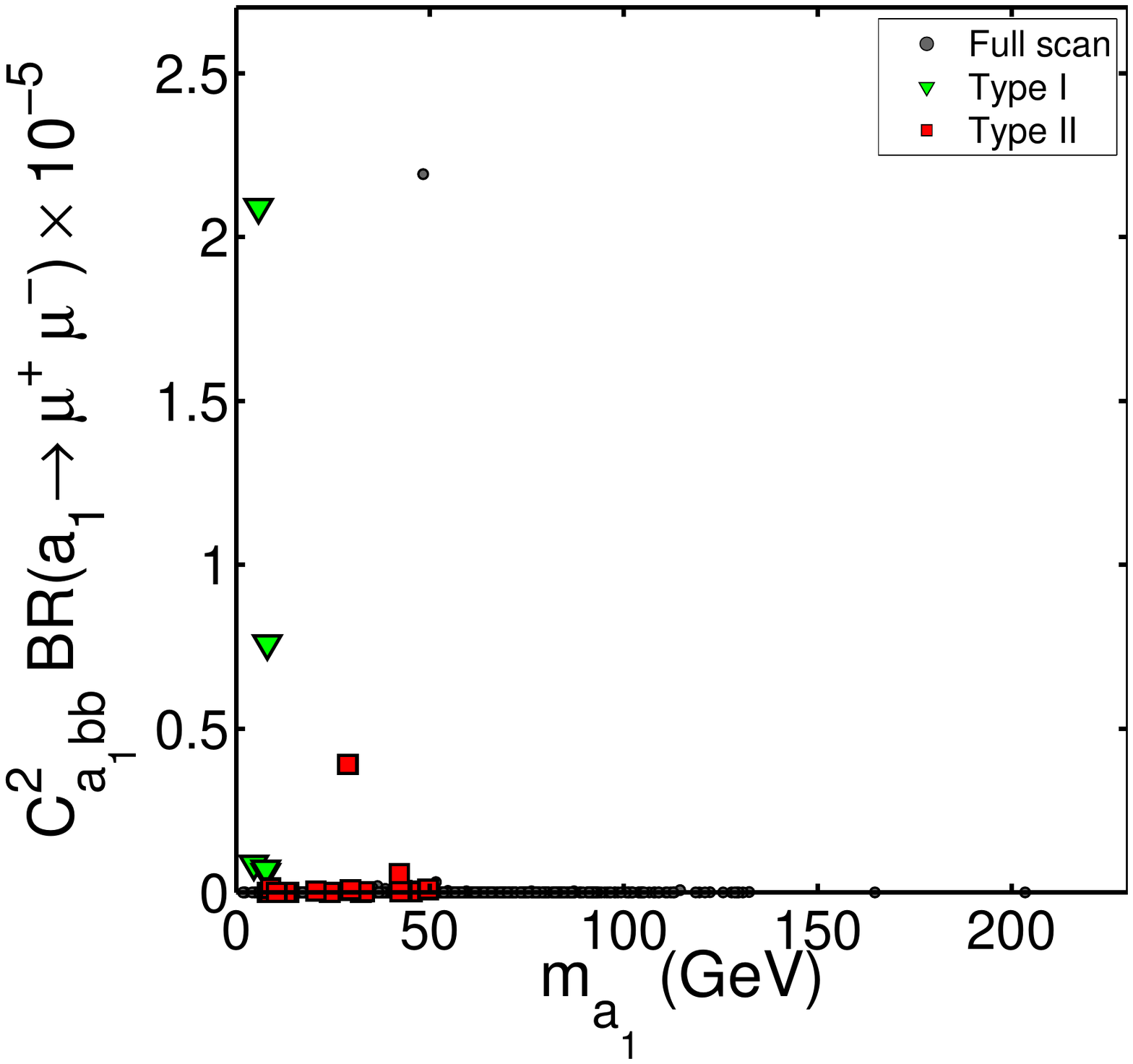} \\
   \end{tabular}
  \end{center}
\vspace*{-.6in}
\caption[]{\label{fig:caibb} We plot $\mai$ as a function of its
  singlet component, $|\cos \theta_A|$ and
  $\caibb^2\br(\ai\to\mu^+\mu^-)$ as a function of $\mai$. }
\end{figure}

A more global picture of $|\cta|$ and $\caibb^2 \br(\ai\to
\mu^+\mu^-)$ is provided by Fig.~\ref{fig:caibb}. In the left plot, we give $\mai$ as a function of
the magnitude of its singlet component, $|\cta|$, for all points in
order to show more generally how singlet the $\ai$ is for the
different classes of points. We observe that the $\ai$ is extremely
singlet for the vast bulk of points, including the Type~IIB and Type~IIIB
points.  This correlates with the fact that  the $\hi$ is
mainly singlet for these same two types of points.

In the right plot of Fig.~\ref{fig:caibb} we show
$\caibb^2\br(\ai\to\mu^+\mu^-)$ as a function of $\mai$ to indicate
which points have a reasonable probability that the production/decay
channel $gg\to \ai \to \mu^+\mu^-$ could be detected.  Points for
which this product is $\gsim 0.001$ would have a viable signal at the
LHC for accumulated luminosities of order
$10\fbi$~\cite{Dermisek:2009fd} (more being required if $\mai$ is in
the region of the $\Upsilon$ resonances). We see that none of our points
are even close to allowing such detection.

\begin{figure}[tbh!]
\vspace*{-1.3in}
  \begin{center}\hspace*{-.4in}
\begin{tabular}{c c}
  \includegraphics[width=0.6\textwidth]{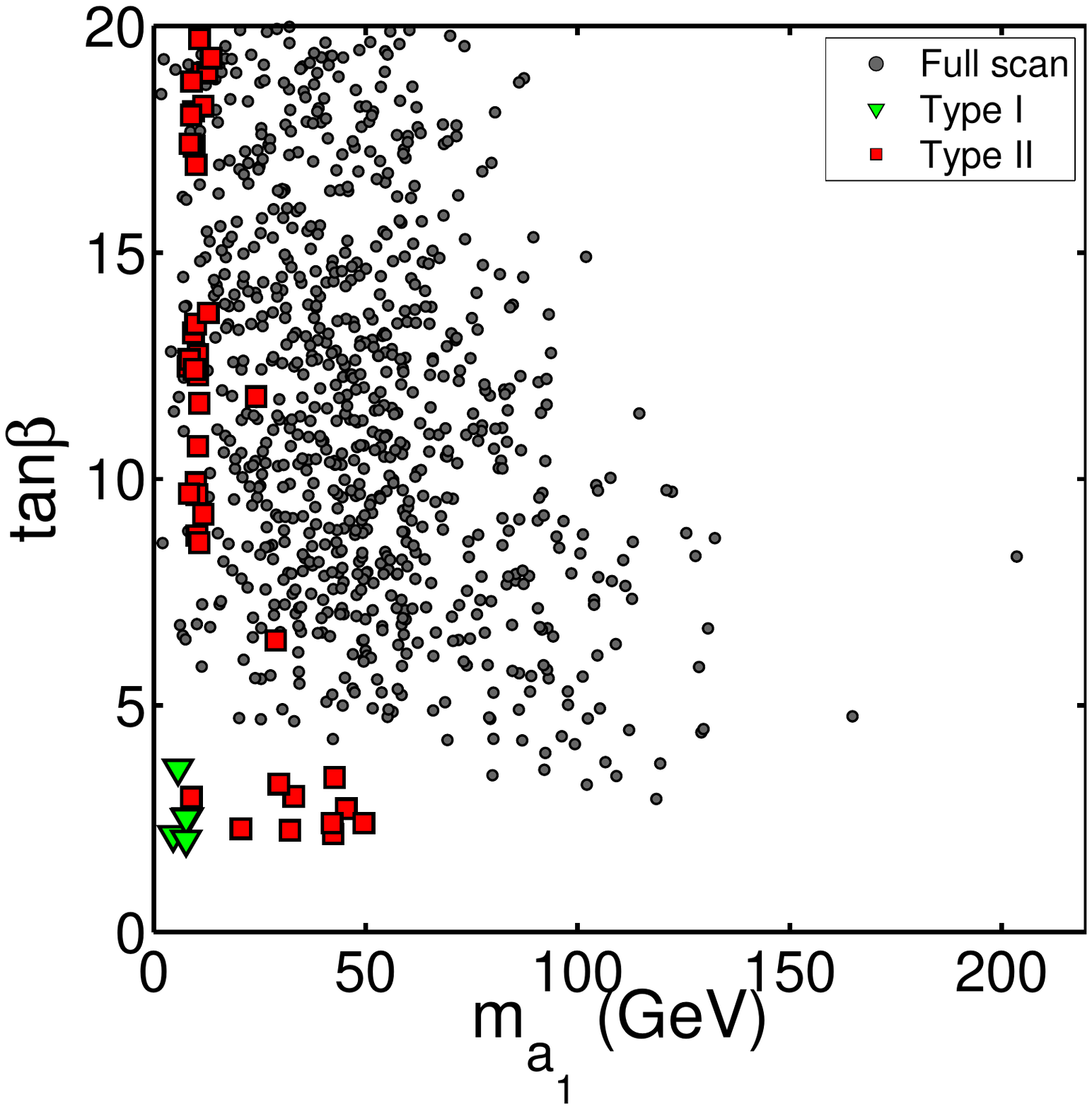} &
\hspace*{-.6in} \includegraphics[width=0.6\textwidth]{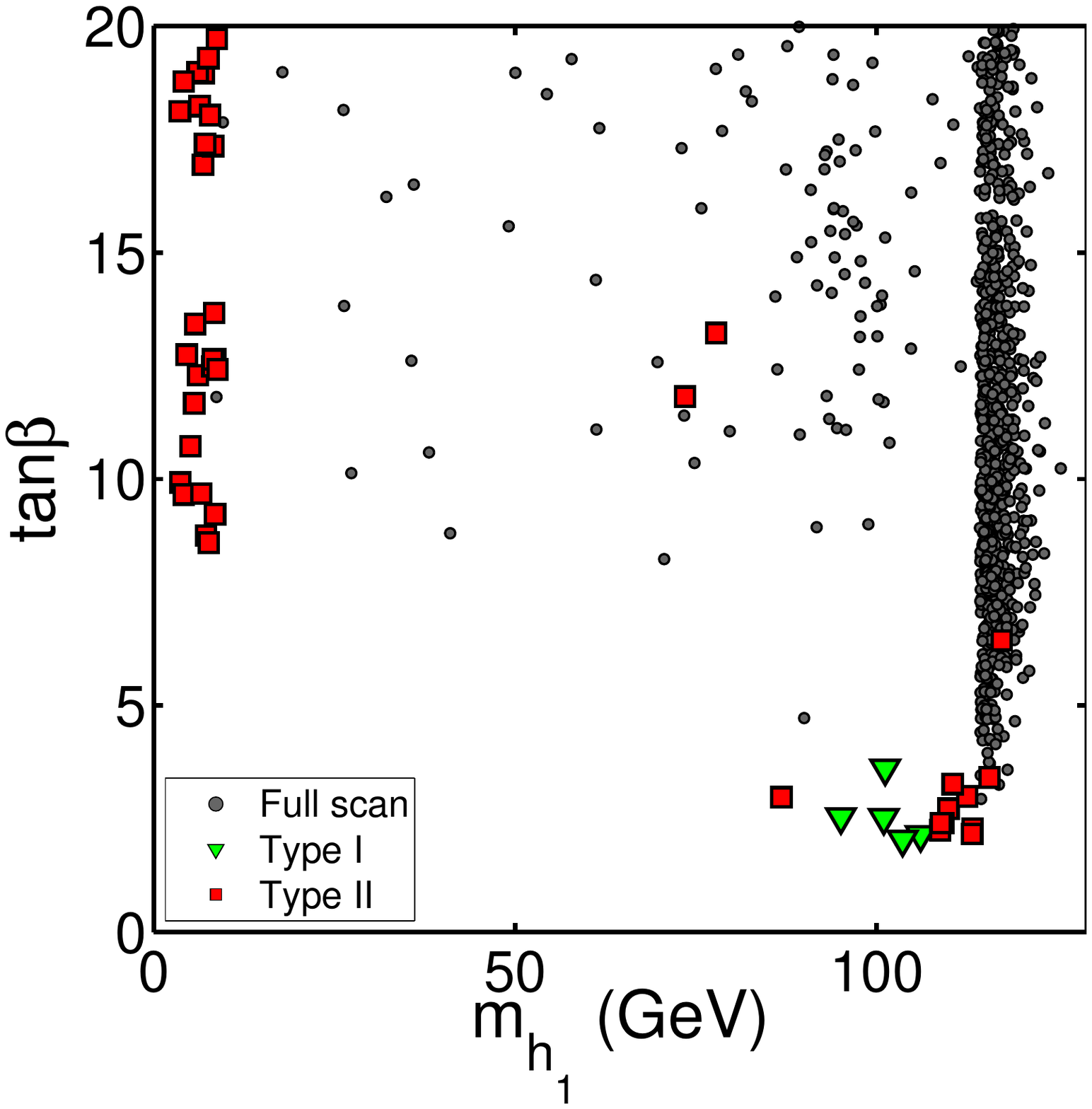} \\
   \end{tabular}
  \end{center}
\vspace*{-.9in}
\caption[]{\label{fig:tanbview} We plot $\tanb$ vs. $\mai$ and
  vs. $\mhi$.}
\end{figure}

\begin{figure}[tbh!]
\vspace*{-1.6in}
  \begin{center}\hspace*{-.4in}
\begin{tabular}{c c}\includegraphics[width=0.6\textwidth]{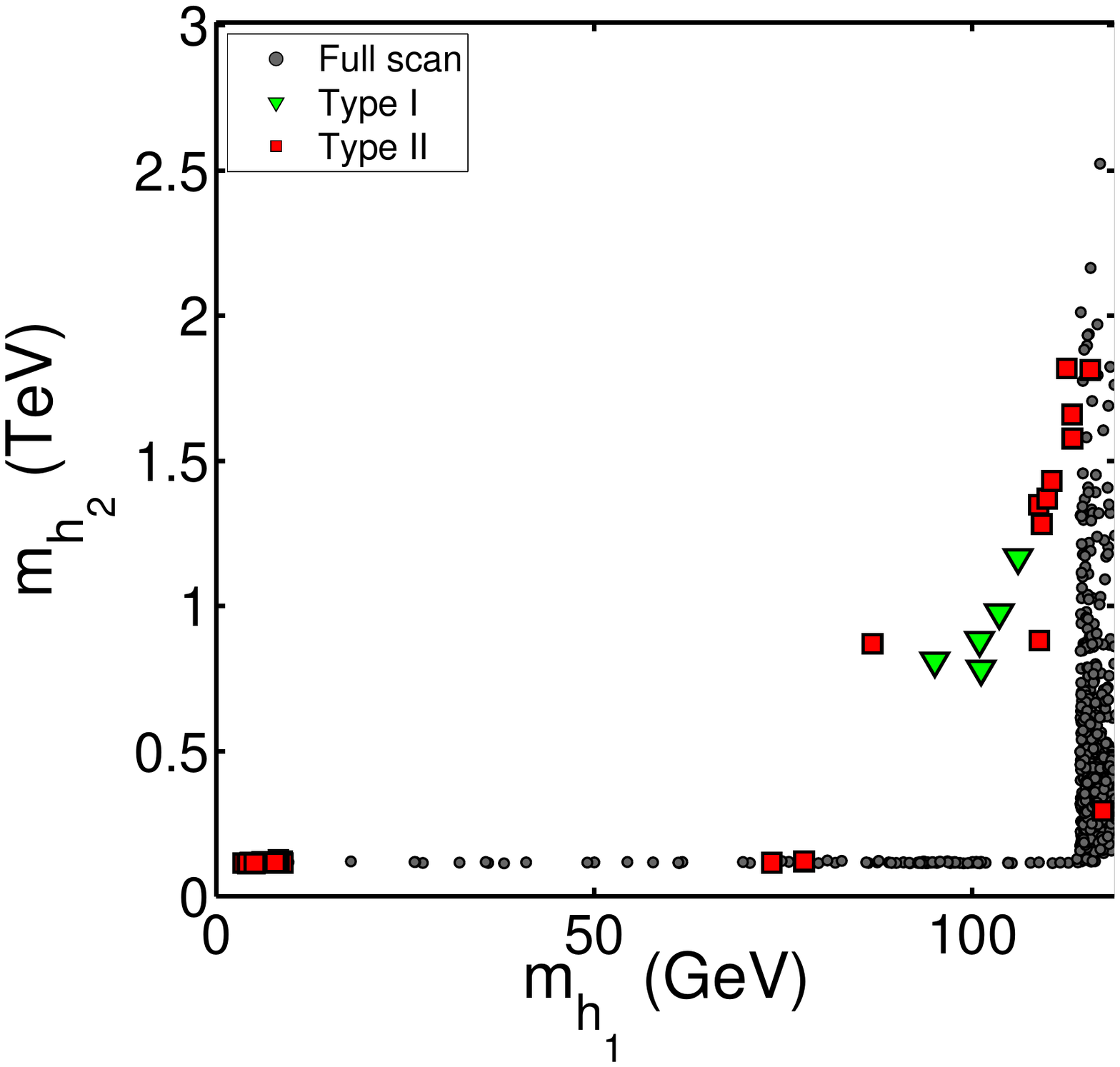}
  &
\hspace*{-.6in}  \includegraphics[width=0.6\textwidth]{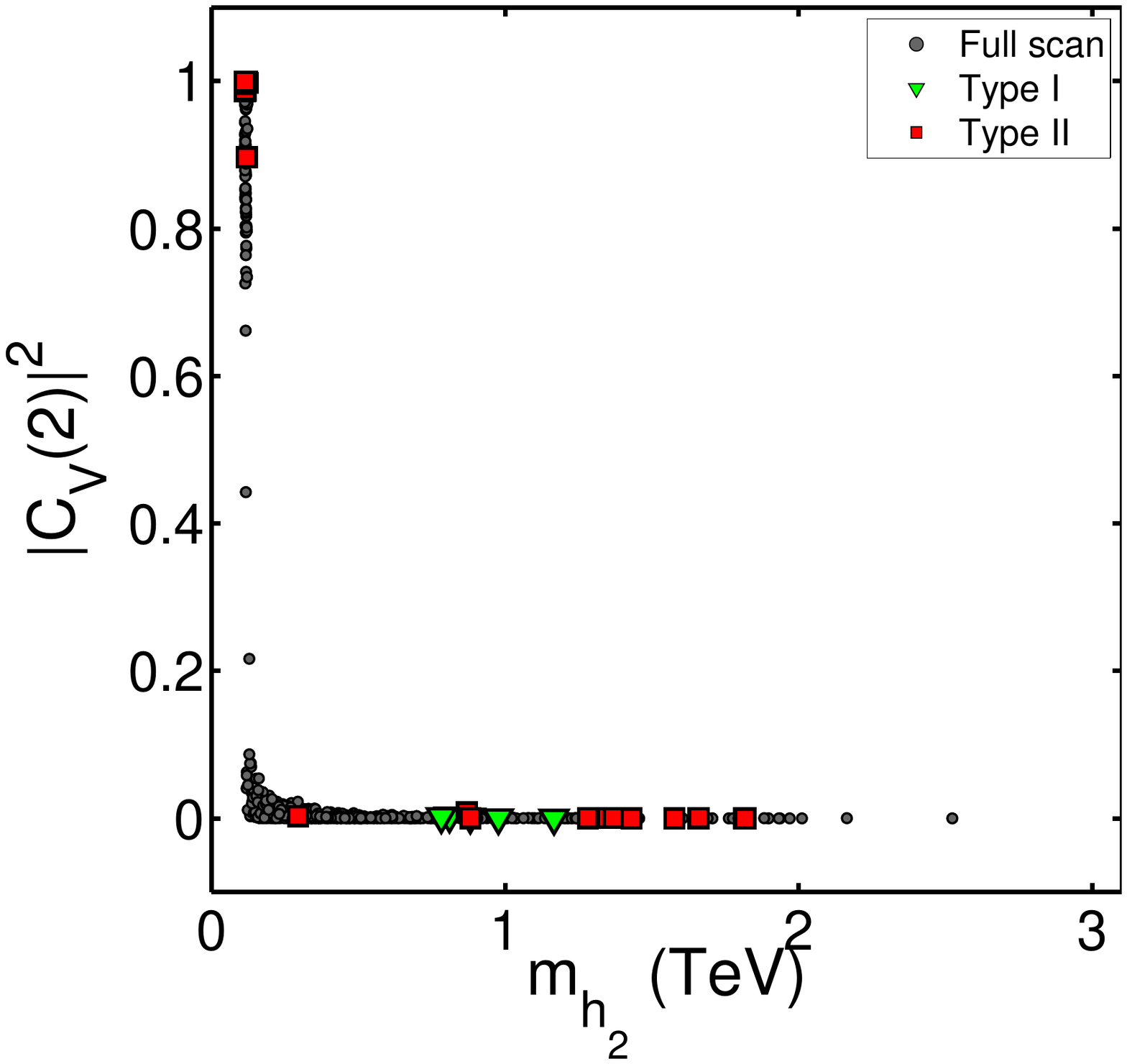} 
 \\
   \end{tabular}
\vspace*{-.9in}
   \end{center}
\caption[]{\label{fig:mhiivsmhi} We plot our points in the $\mhii$ --
  $\mhi$ and show $\cviisq$ as a function of $\mhii$.}
\end{figure}

Of course, the size of $\caibb$ derives both from $\cta$ and
$\tanb$. Thus, it is perhaps useful to display $\tanb$ as a function
of $\mai$ and $\mhi$.  This is done in Fig.~\ref{fig:tanbview}. We see
that only the Type~I and the Type~IIA points with $\mhi\gsim 80\gev$
(for which the $\ai$ and $\hi$ have at least a modest doublet
component) are forced into the low $\tanb$ region. In contrast,
Type~IIB and Type~IIIB points all have $\tanb>8$.  Type~IIIA points
can have any $\tanb$ above $\sim 3$. In the left plot of
Fig.~\ref{fig:mhiivsmhi} we show how all the points are distributed in
the $\mhii$ -- $\mhi$ plane. This plot shows very clearly two branches
for all the points that are neither Type~I nor Type~IIA points.
The vertical branch corresponds to Type~IIIA points where
$\mhi\gsim 114\gev$ (thereby escaping LEP limits) with $\cvisq\sim 1$
(see Fig.~\ref{fig:c2b}). The horizontal branch encompasses the
Type~IIB and Type~IIIB points for which the  $\hi$ is
very singlet and it is instead the $\hii$ that is SM-like with $\mhii\gsim 114\gev$ and
$\cviisq\sim 1$, as displayed in the right plot of
Fig.~\ref{fig:mhiivsmhi}. Finally, we remark that $\br(\hii\to
\ai\ai)$ is very small for all points --- this is perfectly OK since
LEP limits for the $\hii$ are evaded either because it is very singlet
or because $\mhii>114\gev$ --- the extra $\hii\to \ai\ai$ decay
channel is not needed.

 \begin{figure}[tbh!]
\vspace*{-1.9in}
   \begin{center}\hspace*{-.5in}
 \begin{tabular}{c c}
 \includegraphics[width=0.6\textwidth]{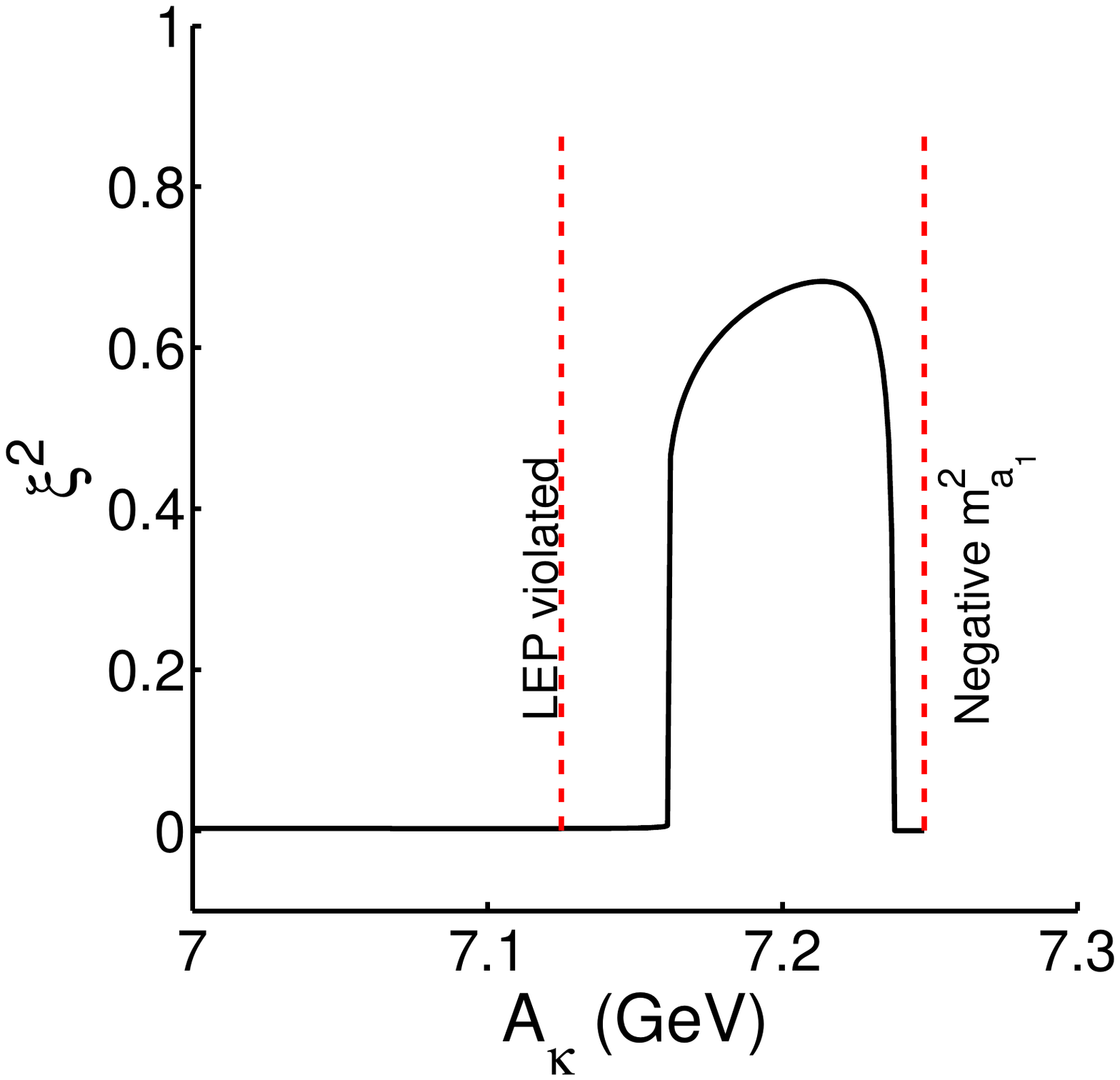}
 &\hspace*{-.5in} \includegraphics[width=0.6\textwidth]{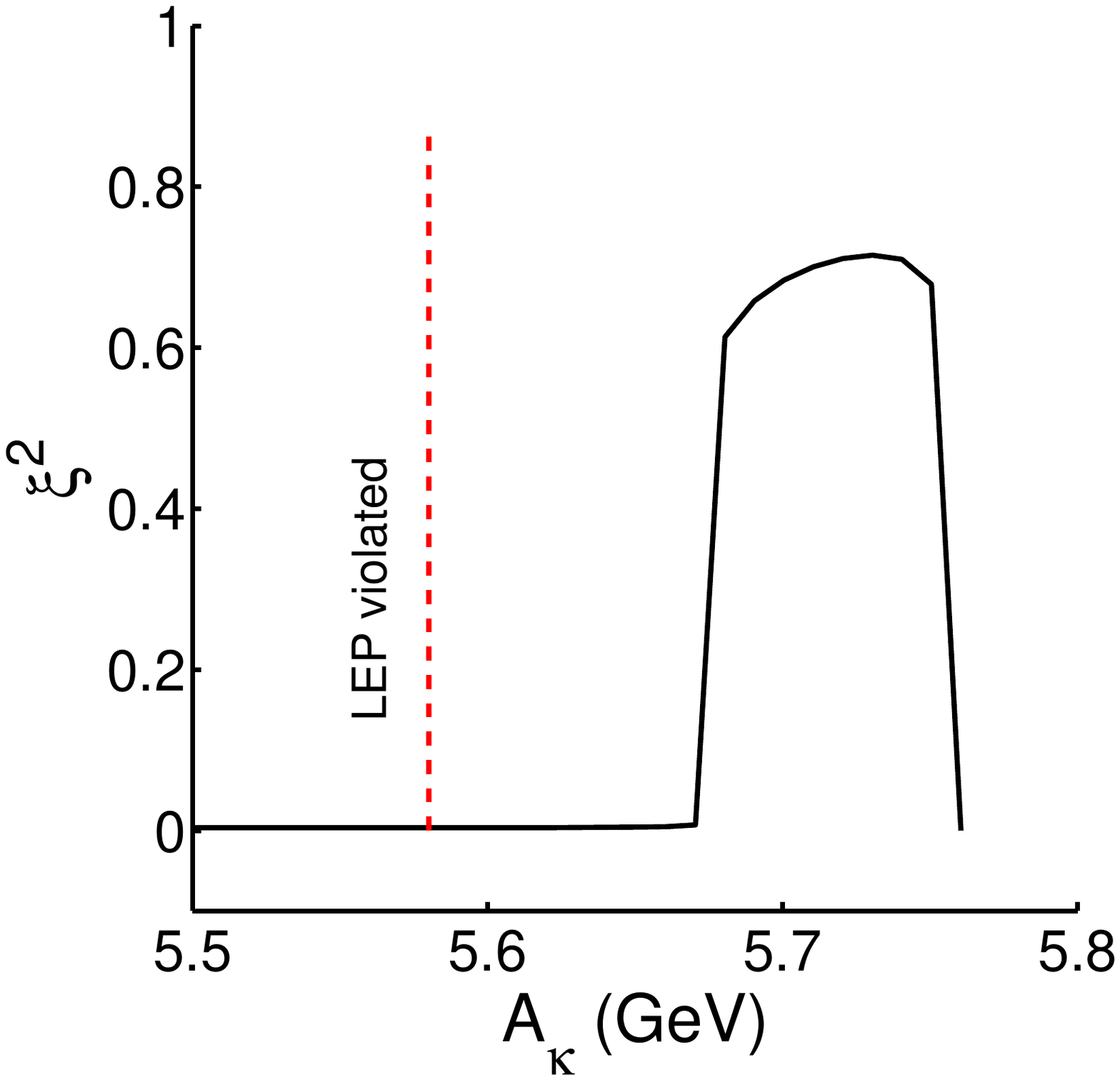}\\
    \end{tabular}
   \end{center}\vspace*{-.4in}
 \caption[The effect of varying the $\akappa$ parameter.]{\label{fig:akvary2}
The effect of varying the $\akappa$ parameter of point 5 (left) and point 1
(right) in table \ref{finetune}.}
   \end{figure}

Let us now return to the ALEPH limits on $\xi^2= |C_{V} (1)|^2
BR(h_{1} \to a_1 a_1) [BR(a_1\to \tau^+\tau^-)]^2$.  In
Fig. ~\ref{fig:akvary2}, we show specifically the effect of varying
$A_\kappa$ on $\xi^2$.  One can see that there exist two
"discontinuities" as $A_\kappa$ is varied.  These arise because
$m_{a_1}$ decreases with increasing $A_\kappa$ . The first abrupt
change occurs at the upper range of $A_\kappa$ plotted when $m_{a_1}$
passes below $2m_\tau$, and $\xi^2$ goes to zero because $BR(a_1\to
\tau^+\tau^-)=0$. The 2nd abrupt change arises as $A_\kappa$ is
decreased and $m_{a_1}$ becomes larger than $2m_B$.  In this region,
the dominant decay channel for the lightest Higgs is $h_1\to a_1
a_1\to 4b$ and $BR(h_1\to a_1 a_1\to 4\tau)$ is very small.  As
$A_\kappa$ decreases just a little bit more, the model point in
question will start to exceed LEP bounds on the $h_1\to 4b$ final
state.  As seen in the left hand plot of Fig.~\ref{fig:akvary2}, in
the case of point 5 the LEP bound on
$C_{eff}^{4b}$ comes into play quite quickly as $A_\kappa$ is
decreased.  In comparison, the right hand plot shows that in the case
of point 1 there is a
larger range of $A_\kappa$ for which the LEP bound on $C_{eff}^{4b}$
is satisfied {\it and} $\xi^2=0$ so that the ALEPH bound is
automatically satisfied. Of course, the nominal value of $A_\kappa$
from Table V for point 5 (the left-hand plot) is such that the ALEPH
bound is satisfied and no adjustment of $A_\kappa$ is required.  We
presented the plot to show how sensitive the LEP phenomenology is to
$A_\kappa$. In this case, the value of $A_\kappa$ can be changed
somewhat from its nominal Table V value without immediately
encountering a problem with either the LEP bound on the $4b$ final
state or the ALEPH bound on $\xi^2$.  In contrast, the nominal value
of $A_\kappa$ for point 1 in Table V gives a value for $\xi^2$ that is
considerably too large in comparison to the $\xi^2\leq 0.29$ ALEPH
limit.  In this case, we must lower $A_\kappa$ in order to satisfy the
ALEPH bound.  Roughly, any $A_\kappa$ above $5.58$ GeV, the value at
which the LEP $4b$ bound (dashed line) becomes relevant, but below
about $5.67$ GeV (vs. the nominal value of $5.69$ GeV) would be
allowed. 

The above discussion illustrates that acceptable $\xi^2$ can be obtained for
the type I points 1, 2 and 4 by very small shifts in $A_\kappa$ that
do not any way affect the remainder of the phenomenology of these
points. Of course, it must be acknowledged that there is a certain
level of fine-tuning of $A_\kappa$ involved in getting $m_{a_1}$ into
an allowed range.  This is, in fact, already reflected in the somewhat
large $G$ values of points 1 and 2 of Table V, these points being ones
where the nominal $\xi^2$ is substantially above the ALEPH limit.
Point 4 has a much more modest $G$ value and correspondingly a broader
range of $A_\kappa$ would allow it to satisfy the ALEPH limit that is
only slightly below the value of $\xi^2$ predicted by the nominal
$A_\kappa$ value tabulated in Table V.
%
%
%
%

\begin{figure}[tbh!]
\vspace*{-1.3in}
  \begin{center}\hspace*{-.5in}
\begin{tabular}{c c}
\includegraphics[width=0.6\textwidth]{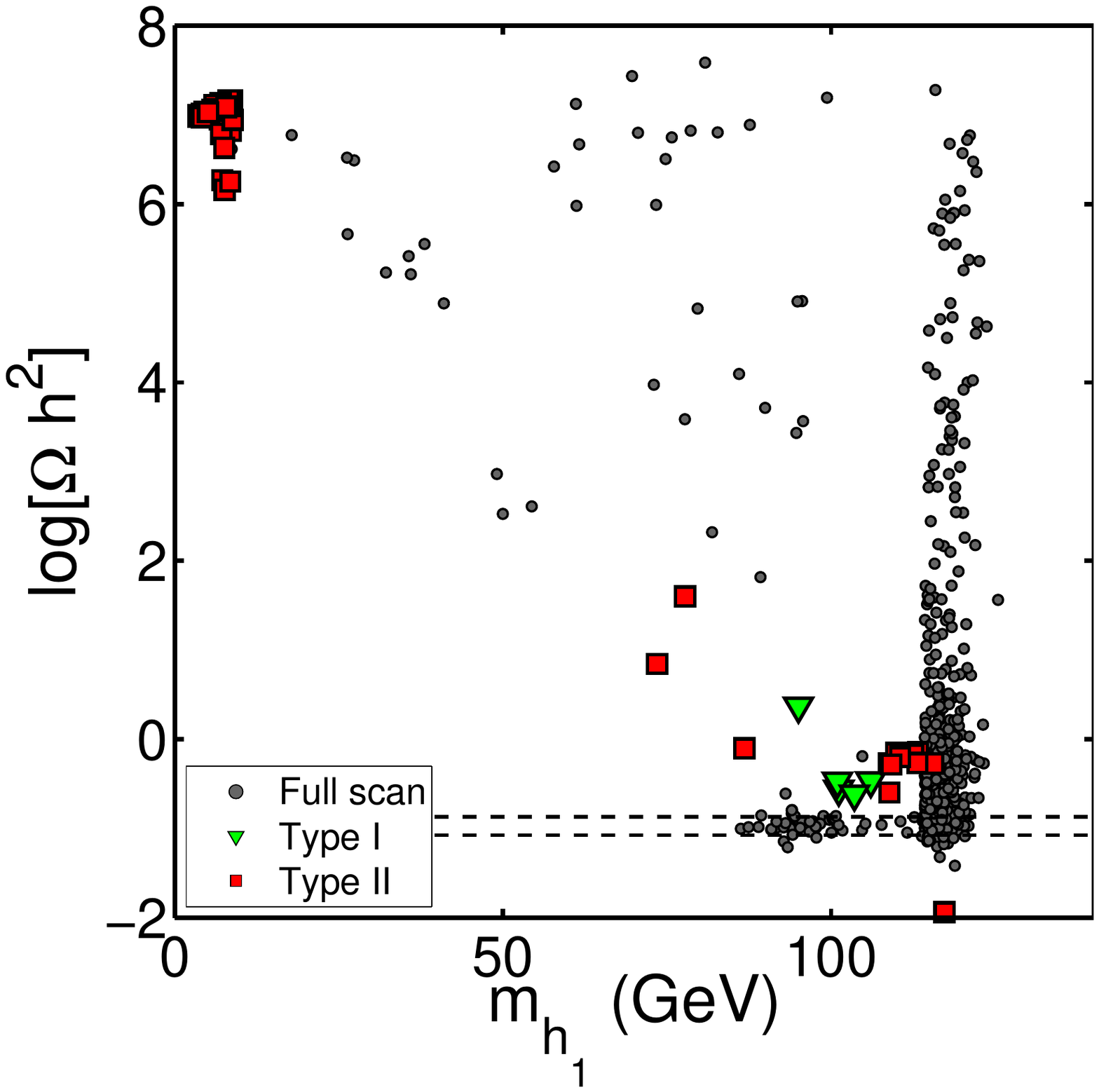}
& \hspace*{-.5in}\includegraphics[width=0.6\textwidth]{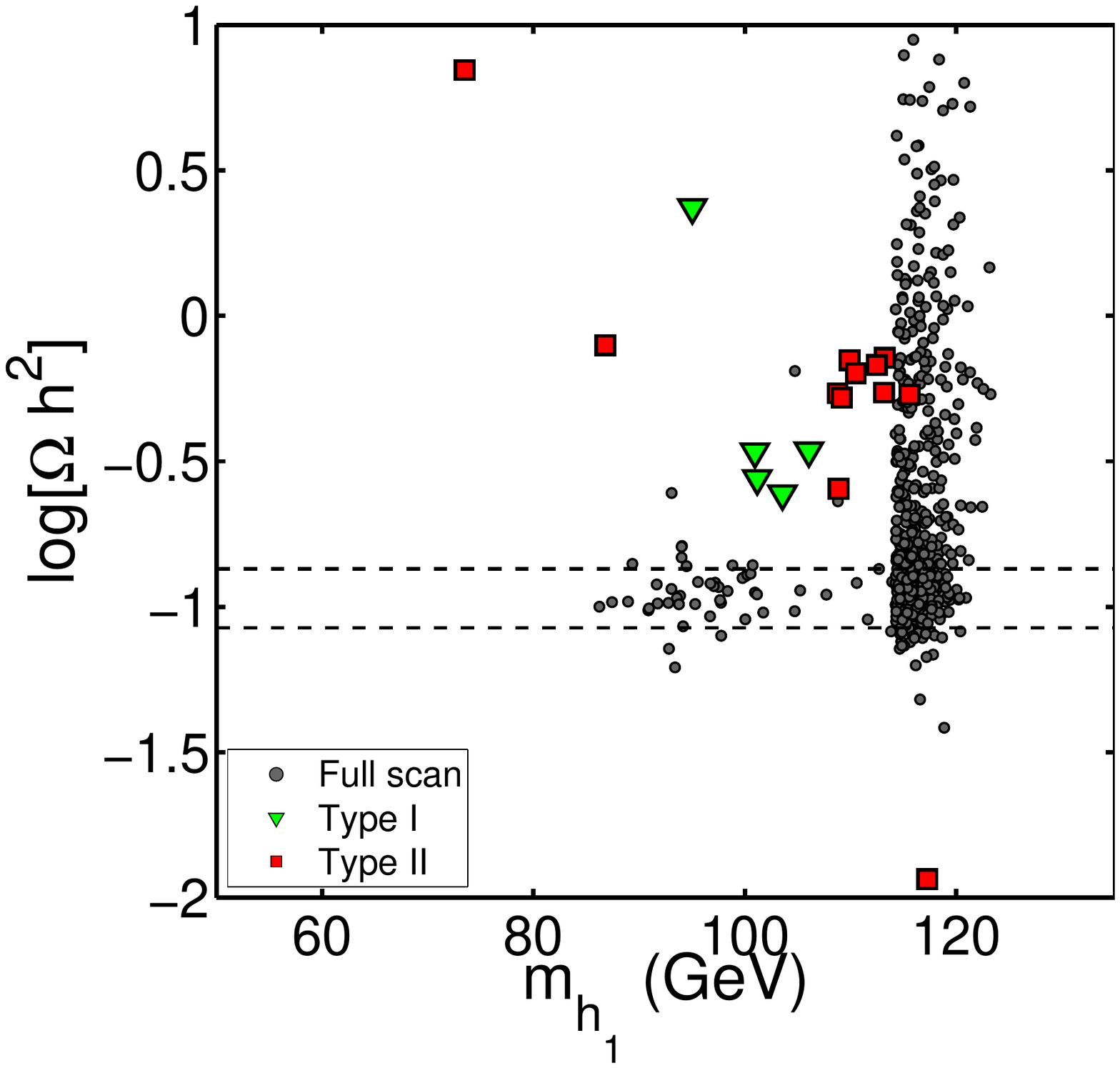}\\
   \end{tabular}
  \end{center}
\vspace*{-.9in}
\caption[The relic density, $\log(\Omega \, h^2)$ versus $m_{h_1}$]{\label{fig:relic} The relic density, $\log(\Omega \, h^2)$, versus $m_{h_1}$, }
  \end{figure}

As stated earlier, $\omhsq$ plays a pivotal role in determining the
likelihood of a given point in parameter space.  Fig.~\ref{fig:relic}
shows us that the range of values for $\Omega h^2$ is huge, with many
points having relic densities that are too large by orders of
magnitude. In this context, the fact that Type~I points tend to
achieve the right order of magnitude seems quite remarkable.  That
said, despite the relic density constraint pushing our scanning quite
strongly towards the WMAP value, we did not find Type~I points with a
relic density that is less than two sigma away from the observed
value. Similar remarks apply to the Type~IIA points.  In contrast,
Type~IIB points (for which the $\cnone$ is very singlino-like) have
much too large $\abund$ as a result of too small an annihilation cross
section.  

As noted already, Type~III points were defined by requiring not only
$\br(\hi\to b\anti b)>0.5$ and $\mai>2m_B$ (so that $\br(\ai\to b\anti
b)\neq 0$) but also by demanding that
$\abund$ is within $\pm 2\sigma$ of the observed value. Thus, it is mainly the
Type~III points that populate the band in the right-hand expanded plot
of Fig.~\ref{fig:relic}. It turns out that for all the Type~III points
the dominant process responsible for getting correct $\abund$ is
coannihilation of the $\cnone$ with $\stauone$ --- they are quite
closely degenerate in mass for all the Type~IIIA and Type~IIIB
points.  In the case of points where the $\cnone$ is very
singlino-like, which comprises all Type~IIIB points and a sizable
fraction of Type~IIIA points, the mass difference between
$\cnone$ and $\stauone$ is at most about $4\gev$ and the common mass
is typically of order $120\gev$. The $\cnone$ has just enough gaugino
and higgsino components (of order $10^{-6}$ at the probability level)
to couple effectively and coannihilate with the $\stauone$ to get the right relic
density. In the case of the Type~IIIA points,
for which the $\cnone$ is mainly bino-like, the common mass is most often
$>250\gev$ (but not always) and for such points the mass difference is
more typically of order $10\gev$.  The very smallest $\chi^2$ values
are achieved for the Type~IIIB points for which both the $\hi$ and the
$\chi$ are mainly singlet and singlino, respectively, the
``singlet-singlino'' (or SS) scenario.  The ease with which such low $\chi^2$
points were found in our scans suggest that the SS scenario for dark
matter should be taken quite seriously as possibly being the correct
paradigm for dark matter.

One question is whether small parameter changes for the Type~I points
could bring the predicted $\abund$ into close agreement with observation. 
After all, our scans might just be slightly missing Type~I points with the
right relic density.  To examine this, an attempt to vary the gaugino masses
independently at the GUT scale was made to see if this could lead to
the right amount of dark matter. In general it is not difficult to get
to $\abund \sim 0.1$, but in all the cases studied, it comes at the
price of exceeding LEP limits on several channels, including
Higgstrahlung processes such as $Z\hi$ production with $\hi\to b\anti b$.

\begin{figure}[tbh!]
\vspace*{-1.4in}
   \begin{center}\hspace*{-.5in}
 \begin{tabular}{c c}
 \includegraphics[width=0.6\textwidth]{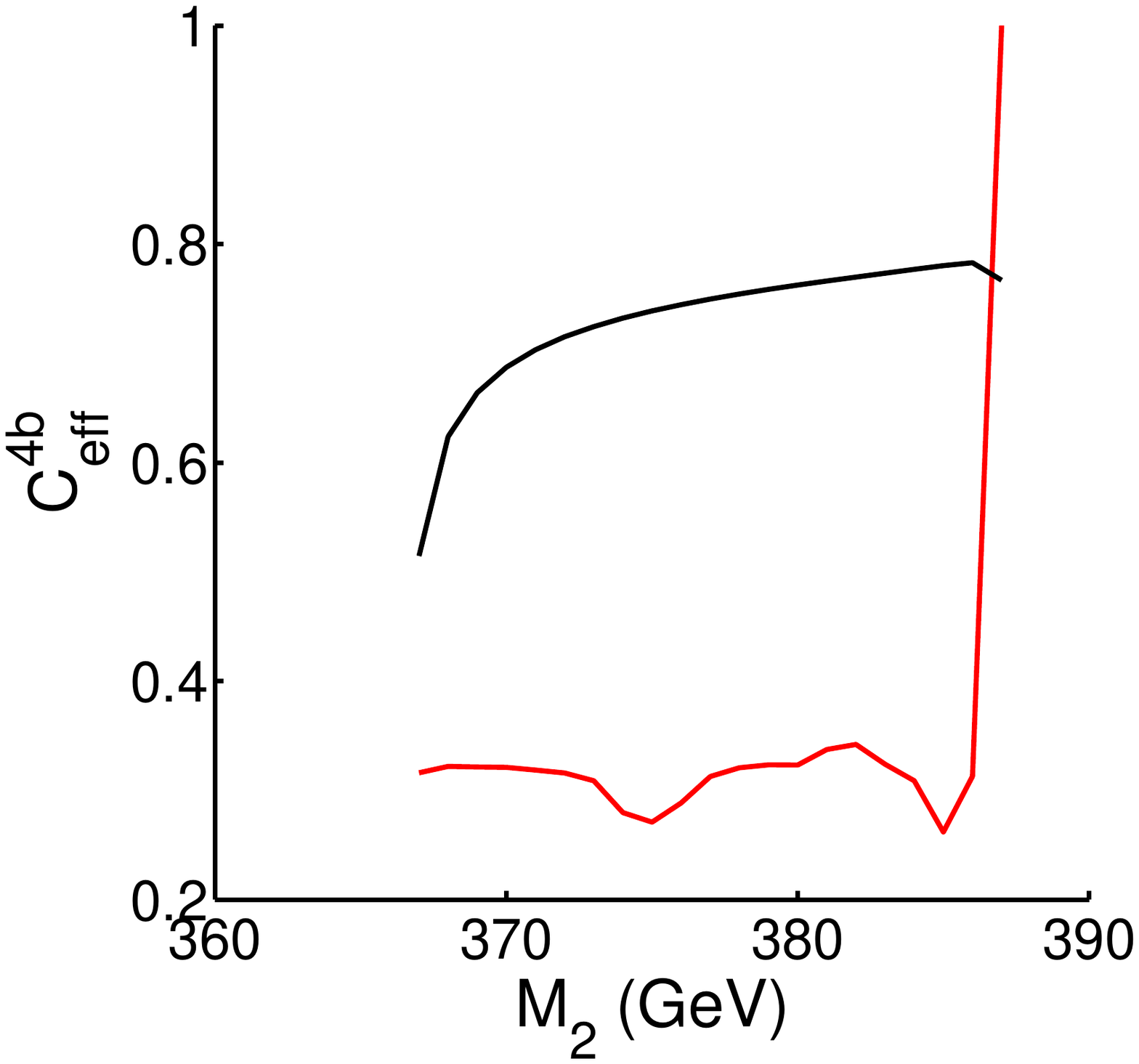}
 & \hspace*{-.5in}\includegraphics[width=0.6\textwidth]{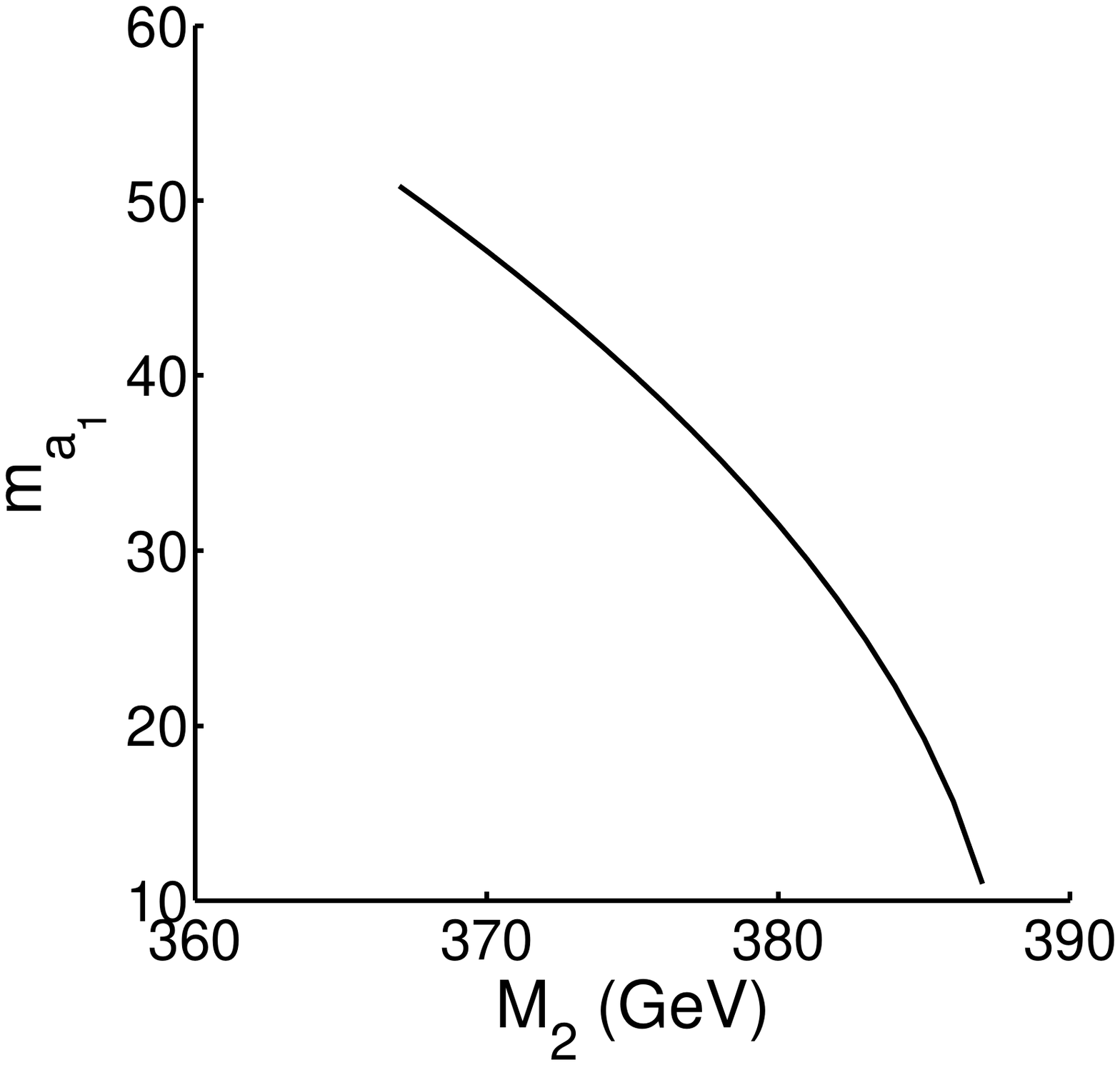}\\
    \end{tabular}
   \end{center}\vspace*{-.4in}
 \caption[The effect of varying the $\mtwo$ parameter.]{\label{fig:m2_vary} The effect of varying the $M_2$ parameter of point 5 in table \ref{finetune}. The left-hand plot gives the values for $C_{eff}^{4b}$ as defined in Eq.\eqref{eqn:defs}
with red denoting the experimental limit and black the NMSSM predicted value as a function of $M_2$. The black curve terminates as $M_2$ increases when $m_{a_1}^2$ becomes negative.  The sharp rise in the experimental limit occurs as $m_{a_1}$ approaches and then falls below $2m_B$, the point at which the $a_1\to 2b$ decay mode becomes kinematically forbidden. On the right, we show the rather dramatic change of $m_{a_1}$ with $M_2$. In contrast, $m_{h_1}$ remains roughly at 102 GeV over this range of $M_2$.}
   \end{figure}

\begin{figure}[h!]
\vspace*{-2.1in}
   \begin{center}
 \begin{tabular}{c}
 \includegraphics[width=0.65\textwidth]{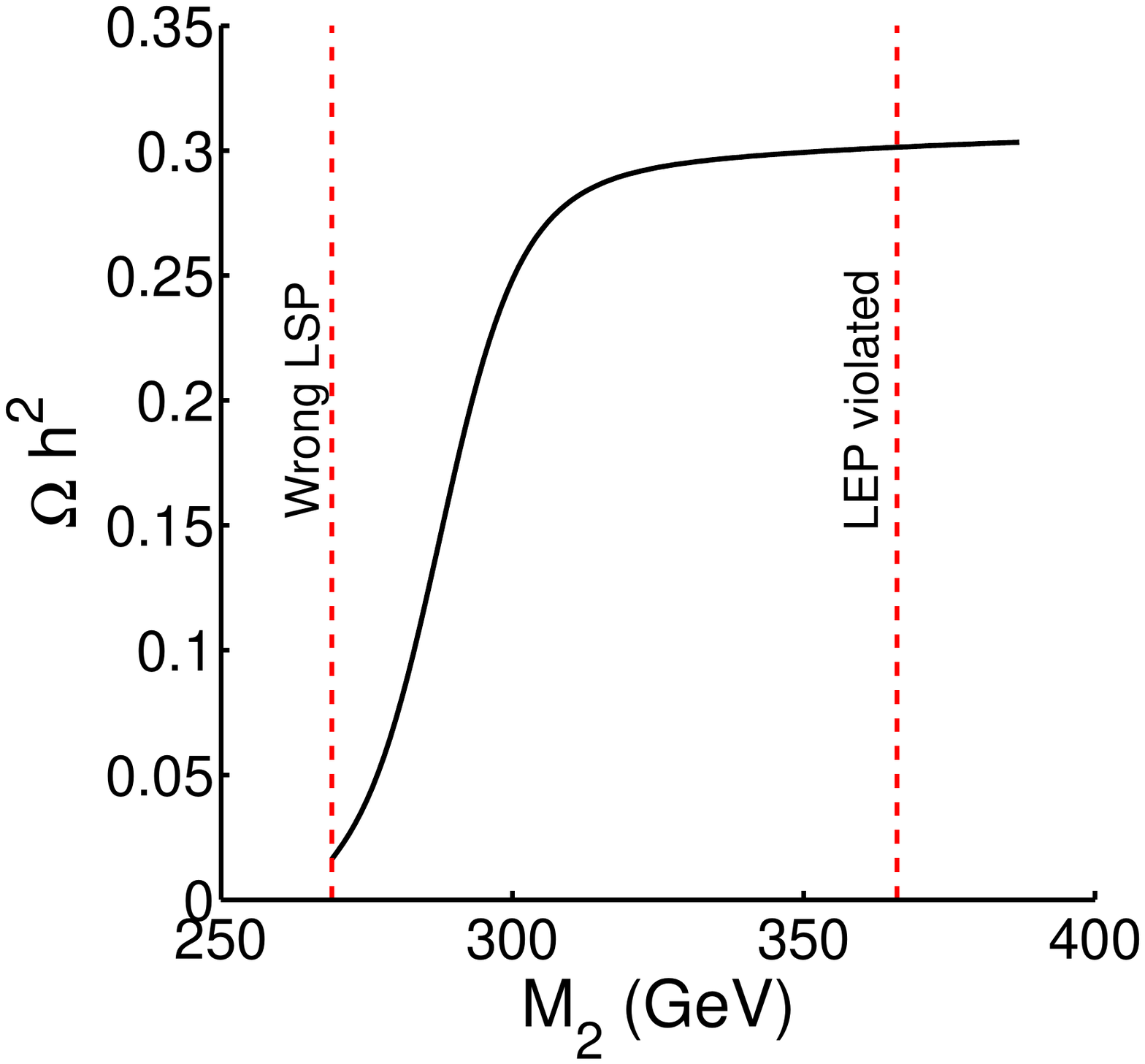}
    \end{tabular}
   \end{center}\vspace*{-.4in}
 \caption[The effect of varying the $\mtwo$ parameter on
 $\abund$]{\label{fig:m2_varydm} The effect of varying the $\mtwo$
   parameter on point 5 in table \ref{finetune}. Here we show the
   change in the relic density, $\abund$. It is possible to bring DM
   in line with experiment, but in doing so one violates LEP limits on
   several channels, below $\mtwo\sim 360\gev$ in this case, which is
   the reason that the curves in the previous figures don't go below this value.}
   \end{figure}

  In order to do this perturbation, we took the most promising Type~I
  point (\ie~the one with the best relic density value, point 5 in
  table~\ref{finetune}~) and perturbed $\mtwo$ away from universality
  at the GUT scale. The results are shown in Figs.~\ref{fig:m2_vary}
  and~\ref{fig:m2_varydm}. What seems to be happening here is that as
  $\mtwo$ changes in this region, $m_{a_{1}}$ also changes
  dramatically, so much so that perturbing $M_2$ by only a few $\gev$ gives
  us a point that is ruled out by the $h \to aa \to 4b's$ limit from
  LEP. As $\mtwo$ gets bigger, eventually we get to the crucial point
  where $\abb$ is kinematically suppressed as $m_{a_{1}}$ is
  sufficiently light.

   The curves in Fig.~\ref{fig:m2_vary} do not extend below $\mtwo
   \sim 367\gev$ or so, since, as seen in Fig.~\ref{fig:m2_varydm}, in
   this region one exceeds the LEP limits on $\cbeffii$. In addition,
   the sensitivity of $m_{a_{1}}$ means that soon after getting to a
   point where LEP limits are respected, (the exact point being
   unclear due to the resolution of the exploration done) $m_{a_{1}}$
   is driven tachyonic. Looking at this admittedly very specific case,
   one can perhaps begin to understand why we are obtaining so few
   points that are Type~I and anywhere near the right relic density,
   given that $m_{a_{1}}$ is very sensitive to changes and is the
   crucial element here. This is by no means a blanket statement about
   points being ruled out, merely an observation for this specific
   situation. For further study, a scan with all the gaugino masses
   disunified might be useful to shed light on this.

\begin{figure}[tbh!]
\vspace*{-1.3in}
  \begin{center}\hspace*{-.5in}
\begin{tabular}{c c}
\includegraphics[width=0.6\textwidth]{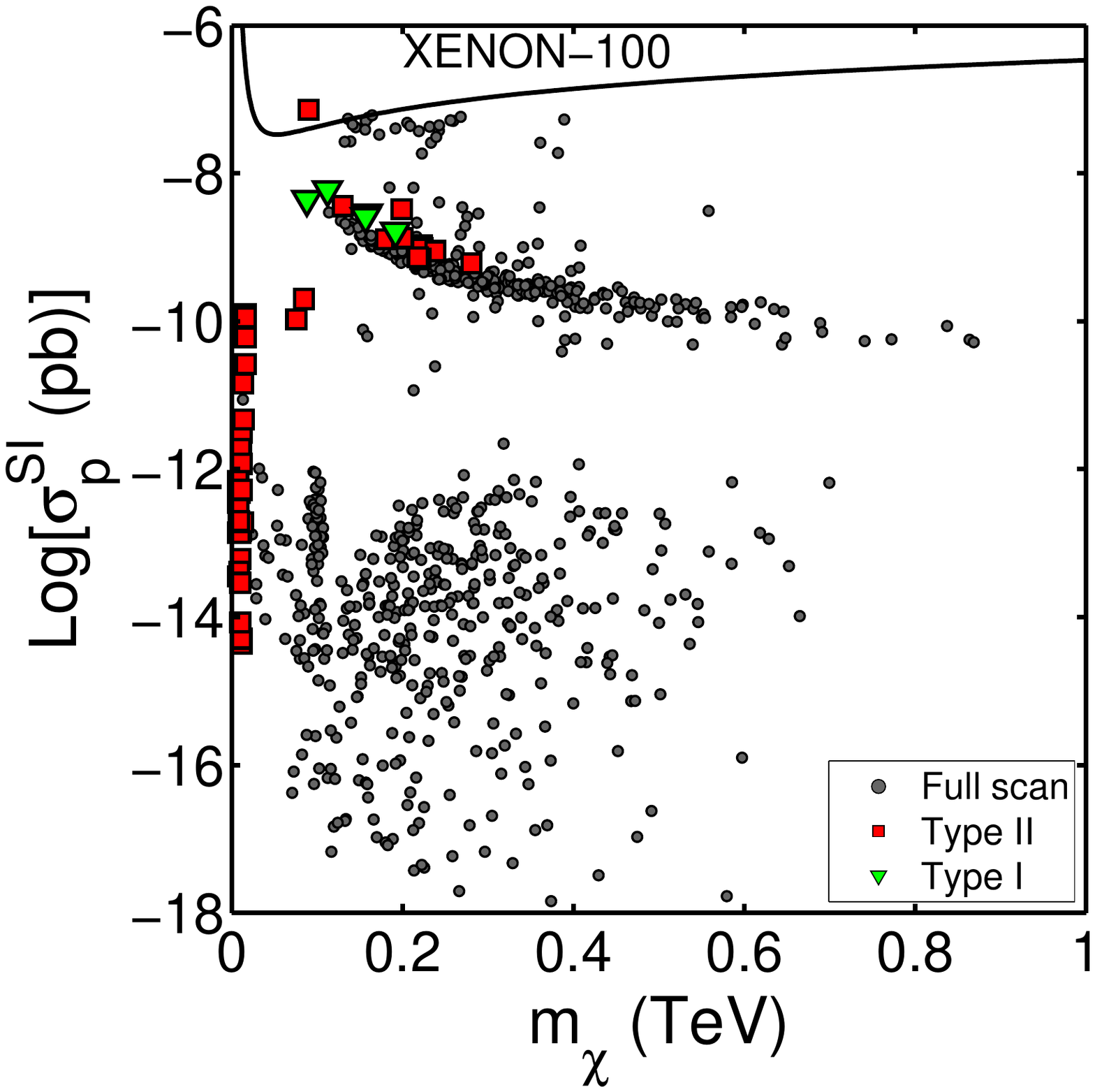} &
\hspace*{-.5in}\includegraphics[width=0.6\textwidth]{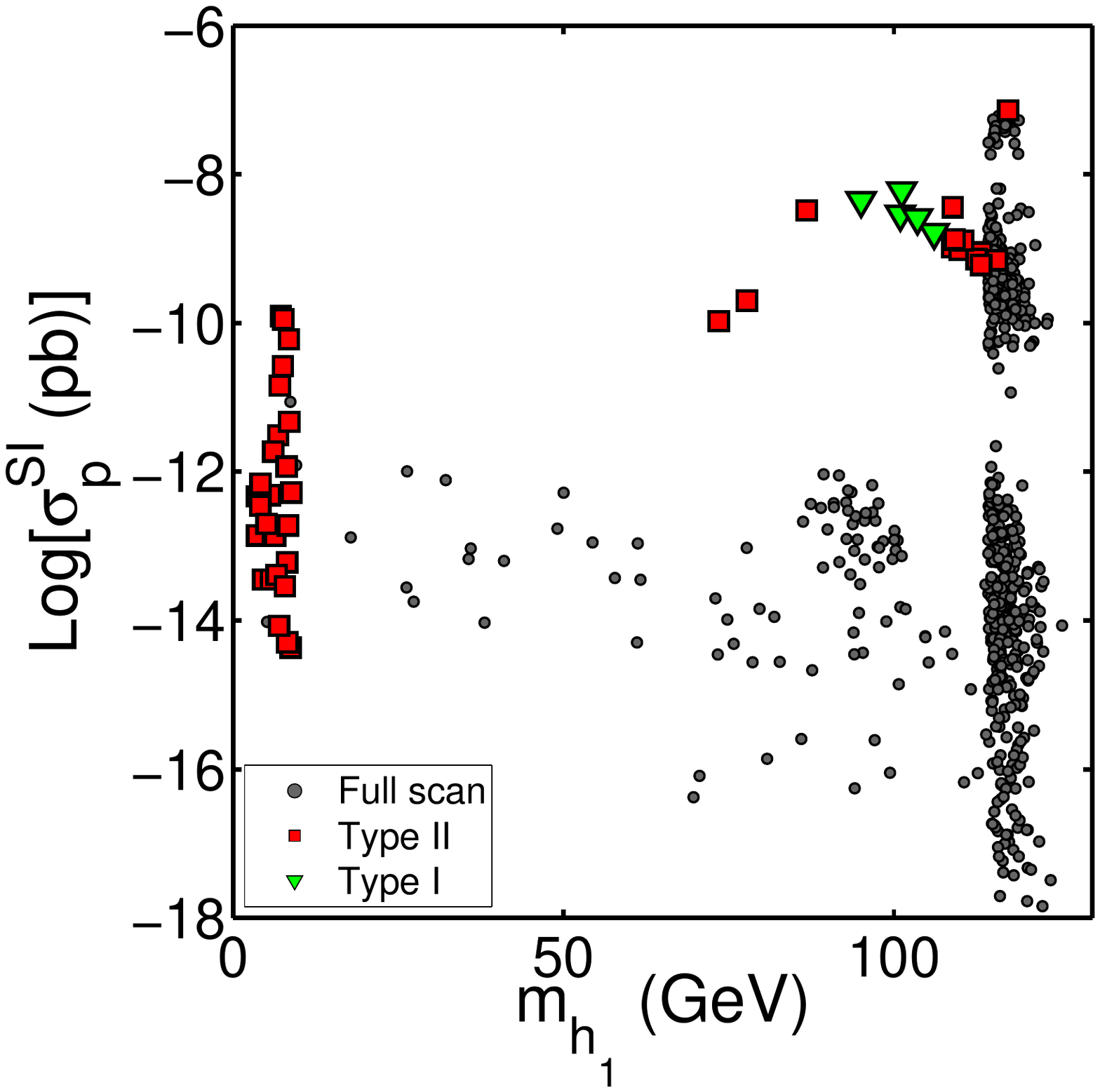} \\
   \end{tabular}
  \end{center}
\vspace*{-.9in}
\caption[The dark matter spin-independent cross
section.]{\label{fig:sigmachi} The dark matter spin-independent cross
  section $\sigma^{SI}_P$ as a function of, on the left, $\mcnone$
  and, on the right, $m_{h_{1}}$. Included on the left hand side is an illustrative limit from the Xenon-100 direct detection experiment \cite{xenon100}. }
\end{figure}

\begin{figure}[tbh!]
\vspace*{-1.5in}
  \begin{center}\hspace*{-.5in}
\begin{tabular}{c c}
\includegraphics[width=0.6\textwidth]{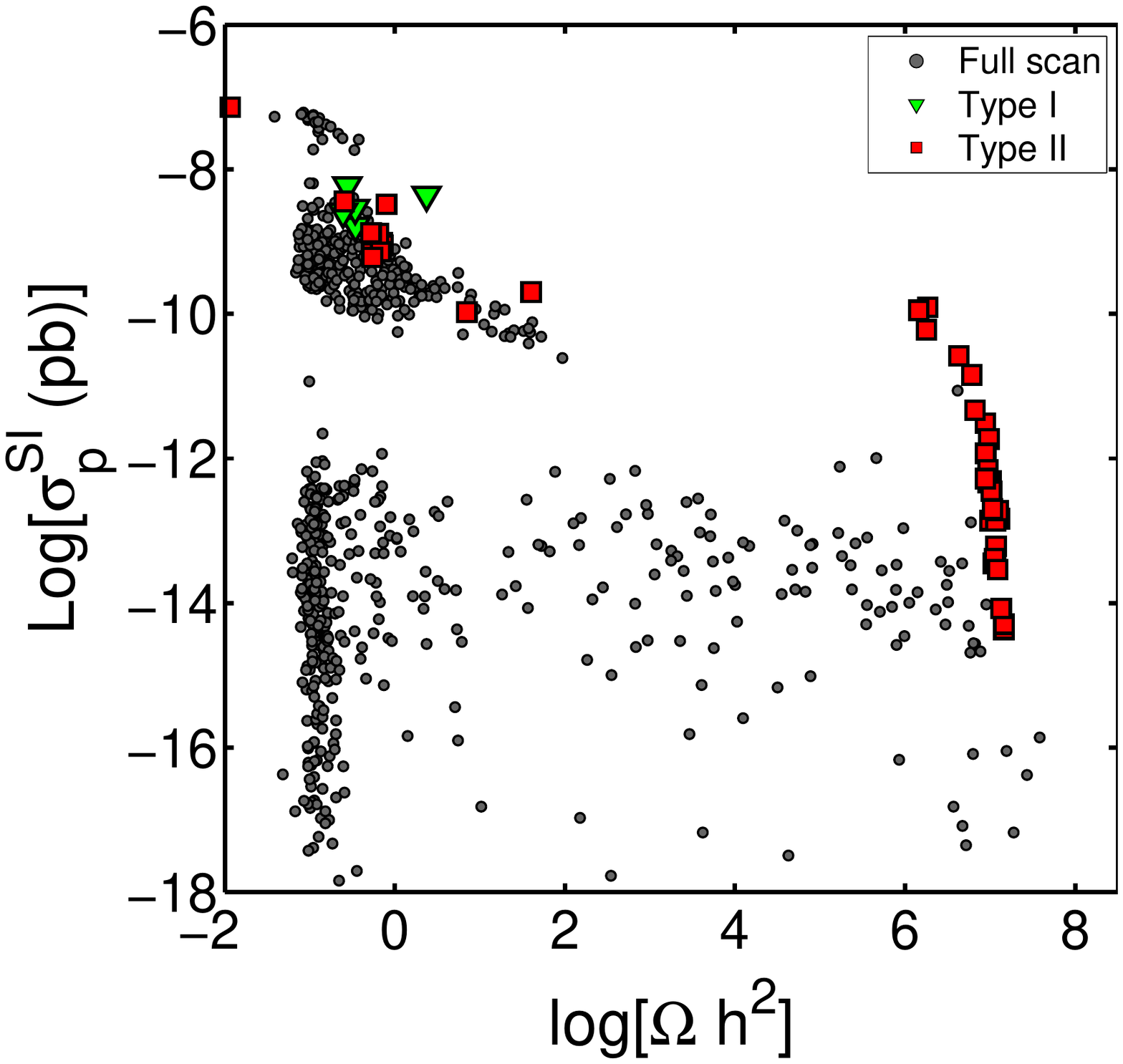}
& \hspace*{-.5in}\includegraphics[width=0.6\textwidth]{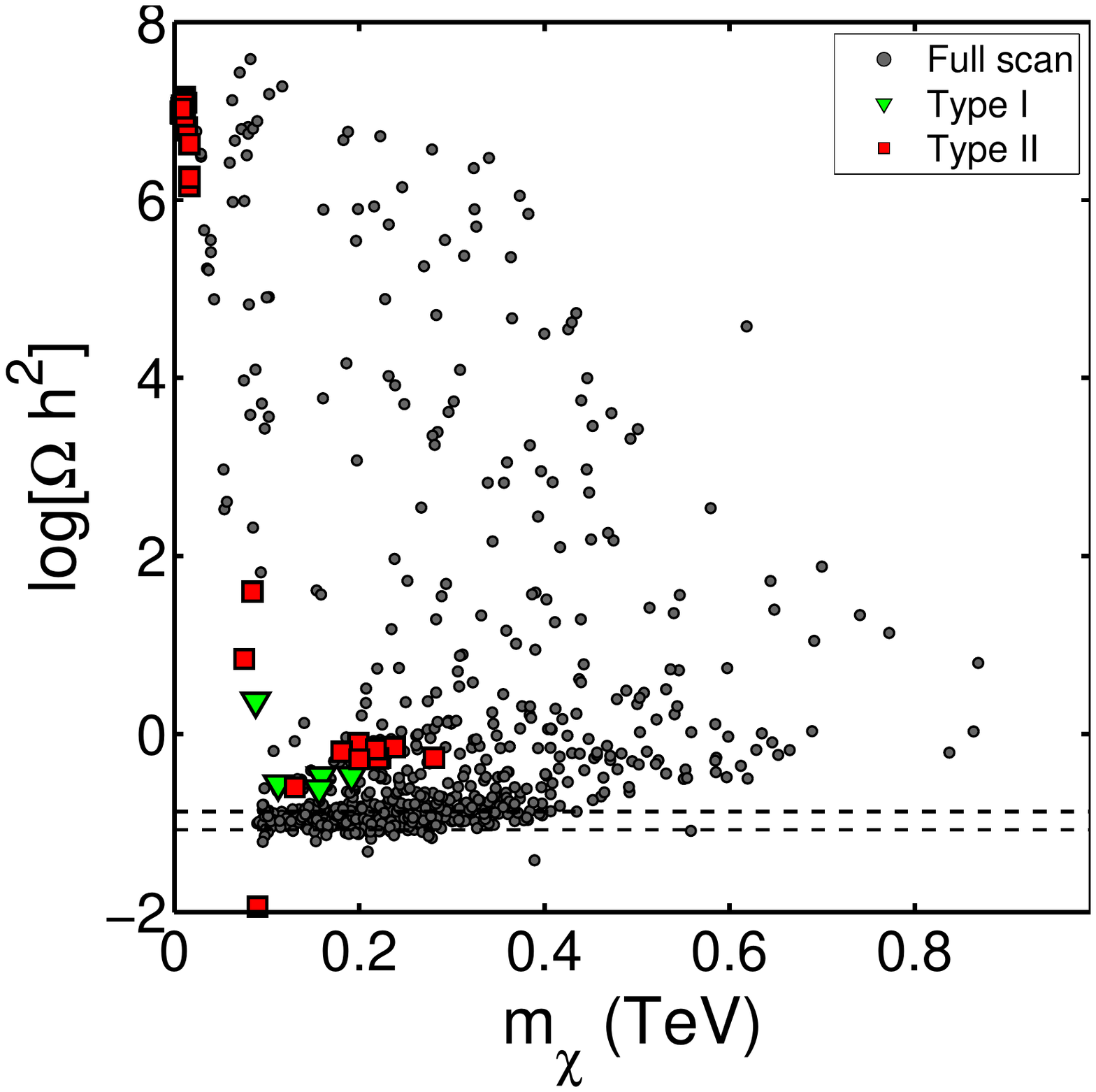}\\
   \end{tabular}
  \end{center}
\vspace*{-.7in}
\caption[]{\label{fig:relic2} Left: The spin-independent cross
  section, $\log(\sigsip)$, vs. the relic density, $\log(\Omega \,
  h^2)$. Right: $\log(\Omega \, h^2)$ vs. $\mcnone$. }
  \end{figure}

Let us now turn to predictions for direct detection of the
neutralino via scattering on nucleons.
In Fig.~\ref{fig:sigmachi} we show the spin-independent cross section
as a function of both $\mcnone$ and $m_{h_{1}}$, and see something not
entirely unexpected: the large swathe of singlino points, which
comprise Type~IIB, Type~IIIB and a sizable fraction of Type~IIIA
points, will be nearly undetectable in any upcoming direct detection
experiment, having cross sections of at most $10^{-12}\pb$ or
so. On the other hand, about half the Type~IIIA points have
$\sigpsi>10^{-10}\pb$, with some having $\sigpsi\sim 10^{-7}\pb$ (and
good $\abund$). And, we should again note that all Type~IIIA points
have $\mhi>114\gev$. In contrast, the Type~I and
Type~IIA points have $\mhi<114\gev$ and often $\mhi\sim
100\gev$, \ie\ in the $\mhi$ region of interest for explaining the LEP
excess near $100\gev$.  As apparent from Fig.~\ref{fig:sigmachi} these
points for which the $\hi$ is doublet-like seem to have a large, or at
least measurable, direct detection cross section. As a result, if a
largish value of $\sigpsi$ is eventually observed, the value of $\mhi$ 
could be used to distinguish between the Type~IIIA and Type~I/IIA
regions of parameter space.  Note
that {\it a priori} there is no reason for Type~I and Type~IIA points
to have a direct detection cross section that is both high enough to
be tested by future experiments and low enough to avoid current
constraints, and as such it is interesting to note that many such
points do appear.  

On another note, by comparing between the two
figures in Fig.~\ref{fig:sigmachi}, we can again detect the
correlation between singlet-like Higgses and singlino-like neutralinos
that was apparent in Fig.~\ref{fig:comp} for values of the Higgs mass
in the range $m_{h_{1}} < 90\gev$. As already apparent in
Fig.~\ref{fig:comp}, for $114>m_{h_{1}}>90\gev$ the situation is more
complicated and there is no clear correlation between the singlet
nature of the $\hi$ and the singlino nature of the $\cnone$.

It is interesting to comment on results for $\sigpsi$ for points with
very low $\mcnone\lsim 15\gev$.  We see in Fig.~\ref{fig:sigmachi}
that the Type~IIB points are those that populate the very low
$\mcnone$ region. However, these points all have a singlino-like
$\cnone$ and, correspondingly, the largest cross section is
of order $\sigpsi\sim 10^{-10}\pb$, \ie\ far below the region that is
needed to explain the possible CoGeNT~\cite{Aalseth:2010vx} and
DAMA~\cite{Bernabei:2010mq} excesses in the $6\gev\lsim \mcnone\lsim
9\gev$ region for which $\sigpsi\gsim 10^{-4}\pb$ is
required~\cite{Hooper:2010uy}. In \cite{Gunion:2010dy}, it is shown
that if one ignores GUT-scale unification then values of $\sigpsi$
within a factor of 10 of the above range are possible while still
having a Higgs with SM-like $WW,ZZ$ couplings that is sufficiently
light to achieve ``ideal-like'' agreement with precision data.  From the study presented in the
present paper, it seems that such large values of $\sigpsi$ cannot be
achieved in the context of the relaxed-CNMSSM boundary conditions
employed here.  In particular, very low $\mcnone$ values for a
bino-like $\cnone$ (which allowed for the largest $\sigpsi$ values in
\cite{Gunion:2010dy}) require small $M_1$ values, a region that is
quite inaccessible in the strict CNMSSM context.

Before concluding, we provide a tabular summary of the most important characteristics of the
five different classes of points that we have particularly focused on:
Type~I, Type~IIA, Type~IIB, Type~IIIA and Type~IIIB.

\begin{table}
\label{sumtable}
\caption{Important characteristics of the different classes of
  points.  The top entries in the Table list the requirements imposed
  in defining the classes.  The bottom entries list the resulting
  properties of points in the different classes. Three additional
  requirements are imposed for Type~III points (only): (i) that $\omhsq$ be within
  $2\sigma$ of the observed value (roughly $\log[\abund]\sim -1$); 
  (ii) ALEPH limits on $\hi\to\ai\ai \to 4\tau$ are satisfied; and
  (iii) BaBar limits on $\Upsilon(3S)\to \gamma\ai$ are satisfied.}
\begin{tabular}{|c|c|c|c|c|c|}
\hline
\ & Type~I & Type~IIA & Type~IIB & Type~IIIA & Type~IIIB \cr
\hline
$\br(\hi\to b\anti b)$ & $<0.3$ & $<0.5$ & $<0.5$ &$>0.5$ & $>0.5$ \cr
$\br(\ai\to b\anti b)$ & $0$ & $\neq 0$ & $\neq 0$ & $\neq 0$ &
$\neq0$ \cr
$S_s^2$ & $\sim 0$ & $\sim 0$ & $\sim 1$ & $\sim 0$ &
$\sim 1$ \cr
\hline
\hline
$|\cvi|$ & $\sim 1$ & $\sim 1$ &$\ll 1$ & $\sim 1$ & $\ll 1$ \cr
$\mhi$ & $\in[95,108]$ & $>75$ & $<90$, mostly $<20$ & $\gsim 114$ &$<110$ \cr
$\mai$ & $<2m_B$ & $\in[2m_B,50]$ & $\in[2m_B,20]$ & $>2m_B$ &
$\in[2m_B,40]$ \cr
$\br(\hi\to \ai\ai)$ & $> 0.7$ & $>0.6$ & $\sim 0$ & $<0.1$ & $\sim 0$
\cr
$N_s^2$ & $\ll 1$ & $\ll 1$ & $\sim 1$ & $\ll 1$ or $\sim 1$  & $\sim 1$
\cr
$N_B^2$ & $\sim 1$ & $\sim 1$ & $\ll 1$ & $\sim 1$ or $\ll 1$ & $\ll
1$ \cr
$\tanb$ & $<3.5$ & $<6$ & $>8$ & $\in[2,20]$ & $>8$ \cr
$|\cta|$ & $\in[0.005,0.02]$  & $\in[0.003,0.02]$ & $\sim 0$ & $\in[0.002,0.017]$ & $\sim 0$ \cr
$\mchi$ & $\in[80,200]$ & $\in[70,300]$ & $<15$ & 
$\in[113,400]$ & $\in[91,110]$ \cr 
$\log[\abund]$ & $\in[-0.65,0.4]$ & $\in[-1.9,1.8]$ & $\in[6.3,7.3]$ &
$\sim -1$ & $\sim -1$ \cr
$\log[\sigpsi(\pb)]$ & $\in[-8.7,-8.2]$ & $\in[-10,-7]$ & $\in[-14.5,-9.8]$
& $\in[-10,-7]$ & $<-12$ \cr
$\chi^2$ & $\in[26,70]$ & $>10$ & $>10$ & $\in[2.7,26]$ &
$\in[1.9,6.0]$ \cr
\hline
\end{tabular}
\end{table}


\section{Conclusion and summary}\label{sec:gunion_summary}

In the version of the NMSSM studied in this paper, we have relaxed the
unification condition of $A_\kappa$ and of the Higgs soft masses
$m_{H_u}$ and $m_{H_d}$ (at the GUT scale) with respect to the CNMSSM
in order to explore the extent to which the absence of the
ideal-Higgs-like scenarios in the CNMSSM scans depended upon these
particular (rather unmotivated) universality assumptions. Allowing for
non-universal $\akap$ and non-universal Higgs soft masses, five
parameter space points corresponding to the so called ``ideal Higgs''
scenario were indeed found, although these were far outnumbered by
other points. The phenomenology of the ideal-Higgs points was studied,
and in the context of this particular scan these points were seemingly
acceptable in terms of flavour observables like $\brbsgamma$ and
$\brbsmumu$ (with $\gmtwo$, in common with every other scan we have
done, struggling to fit the observed $3\sigma$ difference relative to
the SM). However, only two of the five ideal-Higgs-like points we
found are strictly consistent with the latest ALEPH limits on the
Higgs to four tau mode, with a third being very close to
consistency. However, we have shown that by changing $A_\kappa$ by a
very small amount compared to the nominal value found in the scan
(which did not use the ALEPH limit on the four tau final state as an
input to the chi-squared employed) will allow consistency with the
ALEPH limit without altering any other phenomenology.

As regards the relic density, the $\abund$ values of the Type~I points
were not within the two sigma range of the observations, but four out
of the five were within a factor of 2 or 3 of $\abund\sim 0.1$.
Perhaps not too much should be read into this as the relative scarcity
of these points in our relaxed-CNMSSM scan could mean our
understanding of these parameter points is incomplete. Correct $\Omega
h^2$ can be achieved for the Type I points by varying the most
relevant gaugino mass parameter ($M_2$) slightly.  However, we found
that for $M_2$ such that the relic density was correct one or more of
the LEP Higgs limits was not satisfied.

There was a another very interesting class of points, denoted
Type~III, that appeared in our relaxed-CNMSSM scan. Type~III points
are, first of all, characterized by $\br(\hi\to b\anti b)>0.5$ (\ie\
the normal SM decay is dominant) and by $\mai>2m_B$ (\ie\ $\ai\to
b\anti b$ is dominant). Points satisfying this criterion are, as one
might expect, very numerous. Further, we found that it was very easy
to find points satisfying the above criteria that gave $\abund$ values
in close agreement ($\pm 2\sigma$) with the observed value (something
that we included in our final definition of Type~III points, along
with the requirement that they obey the ALEPH limits on
$\hi\to\ai\ai\to 4\tau$ decays and the BaBar limits on
$\Upsilon(3S)\to\gamma\ai$ decays).  Within the Type~III class, as
finally defined, the very best predictions for $\abund$ were obtained
for cases in which the lightest Higgs is very singlet-like with
$85<\mhi<110\gev$ and the lightest neutralino is very singlino-like
with $91<\mchi<110\gev$.  Such scenarios are dubbed SS scenarios. For
all such SS scenarios, sufficiently small $\abund$ in agreement with
experiment is achieved via $\chi-\stauone$ coannihilation. We believe
that one should take these SS scenarios seriously. It will then be the
second lightest Higgs boson (predicted to have mass $\mhii$ close to
$114\gev$) that has SM-like couplings to WW,ZZ and its decays will
also be SM-like.  Unfortunately, in such scenarios the
spin-independent cross section for direct dark matter detection is
predicted to be very small, $\sigpsi<10^{-12}\pb$. Rates for collider
production of the singlet-like $\hi$ will be very low.  The $\chi$
will appear in chain decays at the LHC and its roughly $100\gev$ mass
should be measurable with reasonable accuracy.  However, to determine
how singlet it is would require observation of a displaced vertex.
Unfortunately, the predicted non-singlet content of the $\chi$ for the
SS scenarios is of order $1-N_s^2\sim few\times 10^{-6}-10^{-4}$,
sufficient to make the decays to the $\chi$ prompt.

Overall, it is clear that even a slight extension of the strongly
constrained CNMSSM to allow non-universality for the Higgs
soft-masses-squared and for $A_\kappa$ opens up the phenomenological
possibilities very considerably. One finds fairly good
ideal-Higgs-like scenarios. In addition, the very intriguing
singlet-scenarios that are consistent with all experimental
constraints and give excellent $\abund$ become quite prominent.



\section{Acknowledgements}

JFG is supported by US DOE grant DE-FG03-91ER40674.  JFG acknowledges
support by the Aspen Center for Physics during a portion of this
project. The work of R. RdA has been supported in part by MEC (Spain)
under grant FPA2007-60323, by Generalitat Valenciana under grant
PROMETEO/2008/069 and by the Spanish Consolider Ingenio-2010 program
PAU (CSD2007-00060).  R. RdA would like to thank the support of the
Spanish MICINN's Consolider-Ingenio 2010 Programme under the grant
MULTIDARK CSD2209-00064. DEL-F is supported by the French ANR TAPDMS
ANR-09-JCJC-0146 and would like to thank the Science Technology and
Facilities Council for its support at the beginning of this
collaboration. LR is supported by the EC 6th Framework Proramme
MRTN-CT-2006-035505 and by the Foundation for Polish Science. TV would
like to thank the Science Technology and Facilities Council. The use
of the Iceberg cluster is gratefully acknowledged.

\end{document}